\documentclass[12pt,a4paper]{JHEP}
\usepackage{epsfig}
\usepackage{array}
\usepackage{cite}
\usepackage{amsmath}
\usepackage{amssymb}
\usepackage{epic}
\usepackage{afterpage}
\newcommand{\newc}{\newcommand}
\newc{\definmath}[2] {\def#1{\ifmmode#2\else$#2$\fi}}

\definmath\gsim{\,\,\rlap{\raise 3pt\hbox{$>$}}{\lower 3pt\hbox{$\sim$}}\,\,}
\definmath\lsim{\,\,\rlap{\raise 3pt\hbox{$<$}}{\lower 3pt\hbox{$\sim$}}\,\,}

\def\ie{{\it{i.e.}}}

\def\eg{{\it e.g.}}

\def\compProg{\tt}
\def\herwig{{\compProg HERWIG}}
\def\herwigv#1{{\compProg HERWIG-#1}}
\def\isajet{{\compProg ISAJET}}
\def\isajetv#1{{\compProg ISAJET-#1}}

\def\atlfast{{\compProg ATLFAST}}
\def\atlfastv#1{{\compProg ATLFAST-#1}}

\newc{\barr}{\begin{eqnarray}}
\newc{\earr}{\end{eqnarray}}
\newc{\beq}{\begin{equation}}
\newc{\eeq}{\end{equation}}

\newc{\voidcol}{\phantom{m}\begin{rotate}{270}Not
reconstructed\end{rotate}\phantom{mn}}

\definmath\half{{\frac 1 2}}
\definmath\threehalfs{{\frac 3 2}}
\definmath\quarter{{\frac 1 4}}
\definmath\sixth{{\frac 1 6}}
\definmath\third{{\frac 1 3}}
\definmath\twothirds{{\frac 2 3}}
\definmath\fourthirds{{\frac 4 3}}
\definmath{\mPl}{M_\mathrm{Pl}}
\definmath{\invfb}{\mathrm{fb}^{-1}}
\definmath{\Omegadm}{\Omega_\mathrm{CDM}}
\definmath{\omegadm}{\omega_\mathrm{CDM}}

\definmath\amin{\mathrm{min}}
\definmath\amax{\mathrm{max}}

\newc{\mr}{\mathrm}

\def\slashchar#1{\setbox0=\hbox{$#1$}           
   \dimen0=\wd0                                 
   \setbox1=\hbox{/} \dimen1=\wd1               
   \ifdim\dimen0>\dimen1                        
      \rlap{\hbox to \dimen0{\hfil/\hfil}}#1 
   \else                                        
      \rlap{\hbox to \dimen1{\hfil$#1$\hfil}}/                                    \fi}

\definmath{\etmiss}{\slashchar{E}_T}
\definmath{\pmiss}{\slashchar{p}}
\definmath{\ptmiss}{\slashchar{p}_T}
\definmath{\Ptmiss}{\slashchar{{\bf p}}_T}
\definmath{\pt}{p_T}
\definmath{\pthat}{{\hat{p}_t}}
\definmath{\qth}{Q_\mr{thr}}

\newc{\appref}[1]{appendix~\ref{#1}}
\newc{\chref}[1]{chapter~\ref{#1}}
\newc{\Chref}[1]{Chapter~\ref{#1}}
\newc{\secref}[1]{sec.~\ref{#1}}
\newc{\tabref}[1]{table~\ref{#1}}
\newc{\figref}[1]{fig.~\ref{#1}}
\newc{\partref}[1]{part~\ref{#1}}
\newc{\Secref}[1]{Sec.~\ref{#1}}
\newc{\eqnref}[1]{eq.~\ref{#1}}
\newc{\Eqnref}[1]{Eq.~\ref{#1}}
\newc{\Tabref}[1]{Table~\ref{#1}}
\newc{\Figref}[1]{Fig.~\ref{#1}}
\newc{\Partref}[1]{Part~\ref{#1}}

\definmath{\z}  {\mathrm{Z}^{0}}
\definmath\tbar{{\bar t}}
\definmath\ttbar{{t \tbar}}

\definmath{\cht}{{\tilde{\chi}}}
\definmath{\chgone}{{\cht^+_1}}
\definmath{\chgtwo}{{\cht^+_2}}
\definmath{\chgonem}{\cht^-_1}
\definmath{\chgonepm}{\cht^{\pm}_1}
\definmath{\chgall}{\cht^{\pm}_{1,2}}
\definmath{\ntlone}{{{\cht^0_1}}}
\definmath{\ntltwo}{{{\cht^0_2}}}
\definmath{\ntlthree}{\cht^0_3}
\definmath{\ntlfour}{\cht^0_4}
\definmath{\ntlall} {\tilde{\chi}_{1,2,3,4}^{0}}
\definmath{\gluino}{{\tilde{g}}}

\definmath{\ssul} {{\tilde{u}_{L}}}
\definmath{\ssdl} {{\tilde{d}_{L}}}
\definmath{\sscl} {{\tilde{c}_{L}}}
\definmath{\sssl} {{\tilde{s}_{L}}}
\definmath{\sstone} {\tilde{t}_{1}}
\definmath{\ssbone} {\tilde{b}_{1}}
\definmath{\ssur} {\tilde{u}_{R}}
\definmath{\ssdr} {\tilde{d}_{R}}
\definmath{\sscr} {\tilde{c}_{R}}
\definmath{\sssr} {\tilde{s}_{R}}
\definmath{\ssttwo} {\tilde{t}_{2}}
\definmath{\ssbtwo} {\tilde{b}_{2}}
\definmath{\sqr} {\tilde{q}_{R}}
\definmath{\sql} {\tilde{q}_{L}}
\definmath{\squark} {{\tilde{q}}}
\definmath{\ssulbr} {\bar{\tilde{u}_{L}}}
\definmath{\ssdlbr} {\bar{\tilde{d}_{L}}}
\definmath{\ssclbr} {\bar{\tilde{c}_{L}}}
\definmath{\ssslbr} {\bar{\tilde{s}_{L}}}
\definmath{\sstonebr} {\bar{\tilde{t}_{1}}}
\definmath{\ssbonebr} {\bar{\tilde{b}_{1}}}

\definmath{\ssurbr} {\bar{\tilde{u}_{R}}}
\definmath{\ssdrbr} {\bar{\tilde{d}_{R}}}
\definmath{\sscrbr} {\bar{\tilde{c}_{R}}}
\definmath{\sssrbr} {\bar{\tilde{s}_{R}}}
\definmath{\ssttwobr} {\bar{\tilde{t}_{2}}}
\definmath{\ssbtwobr} {\bar{\tilde{b}_{2}}}

\definmath{\ssel} {\tilde{e}_{L}}
\definmath{\ssellp} {\tilde{e}_{L}^{+}}
\definmath{\ssellm} {\tilde{e}_{L}^{-}}
\definmath{\ssellpm} {\tilde{e}_{L}^{\pm}}

\definmath{\sser} {\tilde{e}_{R}}
\definmath{\sselrp} {\tilde{e}_{R}^{+}}
\definmath{\sselrm} {\tilde{e}_{R}^{-}}
\definmath{\sselrpm} {\tilde{e}_{R}^{\pm}}

\definmath{\ssmulp} {\tilde{\mu}_{L}^{+}}
\definmath{\ssmulm} {\tilde{\mu}_{L}^{-}}
\definmath{\ssmulpm} {\tilde{\mu}_{L}^{\pm}}

\definmath{\ssmurp} {\tilde{\mu}_{R}^{+}}
\definmath{\ssmurm} {\tilde{\mu}_{R}^{-}}
\definmath{\ssmurpm} {\tilde{\mu}_{R}^{\pm}}

\definmath{\sstauone} {{\tilde{\tau}_{1}}}
\definmath{\sstauonep} {\tilde{\tau}_{1}^{+}}
\definmath{\sstauonem} {\tilde{\tau}_{1}^{-}}
\definmath{\sstauonepm} {\tilde{\tau}_{1}^{\pm}}

\definmath{\sstautwop} {\tilde{\tau}_{2}^{+}}
\definmath{\sstautwom} {\tilde{\tau}_{2}^{-}}
\definmath{\sstautwopm} {\tilde{\tau}_{2}^{\pm}}

\definmath{\sslrpm} {\tilde{l}_{R}^{\pm}}
\definmath{\sslr} {\tilde{l}_{R}}
\definmath{\ssll} {\tilde{l}_{L}}

\definmath{\ssnu} {\tilde{\nu}}
\definmath{\ssnuel} {\tilde{\nu}_{e}}
\definmath{\ssnumul} {\tilde{\nu}_{\mu}}
\definmath{\ssnutl} {\tilde{\nu}_{\tau}}

\definmath{\mnought}{{m_0}}
\definmath{\mthreehalfs}{{m_{3/2}}}
\definmath{\DeltaMChi}{{\Delta M_{\cht_1}}}
\definmath\mtx{{m_{TX}}}
\definmath\mttwo{{m_{T2}}}
\definmath\mttwosq{{m_{T2}^2}}

\definmath{\lamp}{\lambda^\prime}
\definmath{\lampp}{\lambda^{\prime \prime}}
\definmath{\rparity}{R_P}
\def\atlas{ATLAS}

\def\susy{SUSY}

\title{Discovering anomaly-mediated supersymmetry at the LHC}

\author{A.J. Barr, C.G. Lester, M.A. Parker\\
        Cavendish Laboratory, University of Cambridge, Madingley Road, Cambridge,
 	CB3\nolinebreak\ \nolinebreak{0HE,} UK.}
\author{B.C. Allanach\\ Theory Division, CERN, 1211 Geneva 23, Switzerland.}

\author{P. Richardson\\
	DAMTP, Silver Street, \mbox{Cambridge~CB3~9EW,~UK}, and\\
	Cavendish Laboratory, University of Cambridge, Madingley Road, Cambridge,
 	CB3\nolinebreak\ \nolinebreak{0HE,} UK.}

\abstract{The discovery potential of the LHC is investigated for the minimal 
anomaly-mediated supersymmetry breaking (mAMSB) scenario,
using the ATLAS fast detector simulator, 
including track reconstruction and particle identification.
Generic supersymmetry search cuts are used to map the 5~$\sigma$ (and $\geq 10$~event) 
discovery contours in the \mnought--\mthreehalfs\ plane.
With 100~\invfb of integrated luminosity the search will reach up to 
2.8~TeV in the squark mass and 2.1~TeV in the gluino mass.
We generalise a kinematical variable and demonstrate that it is sensitive 
to the small chargino--LSP mass splitting characteristic of AMSB models. 
By identifying tracks from chargino decays we show that the Wino-like nature of the LSP 
can be determined for a wide range of chargino lifetimes.
}

\preprint{ATLAS-COM-PHYS-2002-034 \\ 
	Cavendish HEP-2002-02/11\\ 
	CERN-TH/2002-190\\ 
	DAMTP-2002-104 }

\begin{document}

\section{Introduction}
\label{AMSB:INTRO}

Weak scale supersymmetry (\susy) is the most promising solution to the technical 
hierarchy problem of the Standard Model (SM). However supersymmetric partners of known particles 
have not yet been discovered by experiment so if \susy\ exists it must be broken.
Dynamical supersymmetry breaking is usually assumed to occur in a `hidden sector'
which is isolated from ordinary particles and interactions. 
It is then communicated to the fields of the minimal supersymmetric standard model (MSSM)
by some messenger(s).  This must be achieved while maintaining both the theoretically 
appealing aspects such as naturalness \cite{Dimopoulos:1995mi,Anderson:1995tr} 
and respecting experimental constraints on masses, 
flavour-changing neutral currents and CP violation.

\susy\ breaking must be communicated to the observed particles 
by some interaction felt by both the visible and hidden sectors.
The most commonly discussed methods have been flavour-blind 
gravitational (SUGRA), or gauge (GMSB) interactions.

More recently an alternative mechanism known as anomaly-mediation (AMSB) has been 
proposed\cite{Randall:1998uk,Giudice:1998xp} 
in which a conformal anomaly in the auxiliary field of the supergravity multiplet 
transmits SUSY-breaking to the observable sector.
There will be an anomaly-mediated contribution to the gaugino masses in any hidden-sector model, 
but where no other direct contribution is present AMSB will be the leading effect.

Anomaly mediation provides a potential solution to the SUSY flavour problem,
in a highly predictive model.
One undesirable feature of pure anomaly-mediation is that the slepton has a negative mass-squared. 
There are various ways to solve this problem\cite{Randall:1998uk,Pomarol:1999ie,Chacko:1999am,Katz:1999uw,Carena:2000ad,Jack:2000cd,Arkani-Hamed:2000xj,Chacko:2000wq,Kaplan:2000jz,Nelson:2001ji,Luty:2001zv,Harnik:2002et,Jack:2002pn},
the simplest of which is the addition of a universal scalar mass at the GUT scale.
The unmeasured parameters of the minimal model (mAMSB) are then \mthreehalfs\ -- the gravitino mass; \mnought\ -- the
universal scalar mass; $\tan \beta$ -- the ratio of the vacuum expectation value of the Higgs fields;
and the sign of the $\mu$ parameter multiplying $H_1 H_2$ in the superpotential,
the magnitude of which is fixed from the condition of correct electroweak symmetry breaking.


The sparticle spectra for mAMSB have been calculated in \cite{Gherghetta:1999sw,Feng:1999hg,Huitu:2002fg,Feng:1999fu}.
With pure anomaly-mediation, (\ie\ $\mnought=0$) the gaugino masses are proportional to their beta functions:
\begin{equation}M_i=\frac{\beta_{g_i}}{g_i} \mthreehalfs \label{AMSB:GAUGINOMASS}\end{equation}
where $g_i$ are the gauge coupling constants with $i=1,2,3$ indicating the gauge group, 
$\beta_{g_i}$ are their corresponding renormalisation group beta-functions, and \mthreehalfs\ is the gravitino mass.
AMSB therefore predicts that the gaugino masses are in the approximate ratios $M_1 : M_2 : M_3 \approx 3 : 1 : 7$
so that the Wino (rather than the more conventional Bino) is the lightest supersymmetric particle
(LSP), and the gluino is nearly an order of magnitude heavier than the LSP.

The prediction in AMSB of a Wino-like LSP has interesting phenomenological consequences.
The most striking of these is that the lightest chargino is nearly mass-degenerate with the lightest neutralino.
Near-degenerate particles are not unusual in SUSY phenomenology, 
but with AMSB one of these particles is the LSP. Since R-parity is assumed to be conserved in our 
analysis (unlike \eg\ \cite{Allanach:2000gu,DeCampos:2001wq}) the \chgone\ may only decay into the
\ntlone\ so the small mass difference 
\begin{equation}\DeltaMChi \equiv m(\chgone) - m(\ntlone)\end{equation}
means that the lightest chargino may have a lifetime long enough to be detected at collider 
experiments.

In the present paper we examine the AMSB discovery potential of the LHC
general-purpose detectors from several points of view. We concentrate 
on the ATLAS detector but similar considerations will also apply to CMS.
In the next section we review the properties of the recently proposed benchmark points 
for minimal anomaly mediation.
In \secref{AMSB:ATLAS} we highlight some of the important features of the ATLAS detector.
We explore the potential reach of generic SUSY search techniques in detecting 
minimal anomaly mediated supersymmetry in \secref{AMSB:SCAN}.
While the generic search reach is dependent on the mAMSB spectrum, 
the same is not true for the classic anomaly-mediated signature -- the Wino-like  LSP -- 
which will apply beyond the minimal model.
The signatures for identifying Wino-like LSPs at the LHC 
are explored in \secref{AMSB:DIST:ANAL}.
Finally in \secref{AMSB:OTHER} we examine constraints on mAMSB from 
measurements of the cosmological relic density, the muon anomalous magnetic moment, and the 
branching ratio $B\to X_s\gamma$.

\section{Benchmark points}
\label{AMSB:BENCH}
\EPSFIGURE[t]{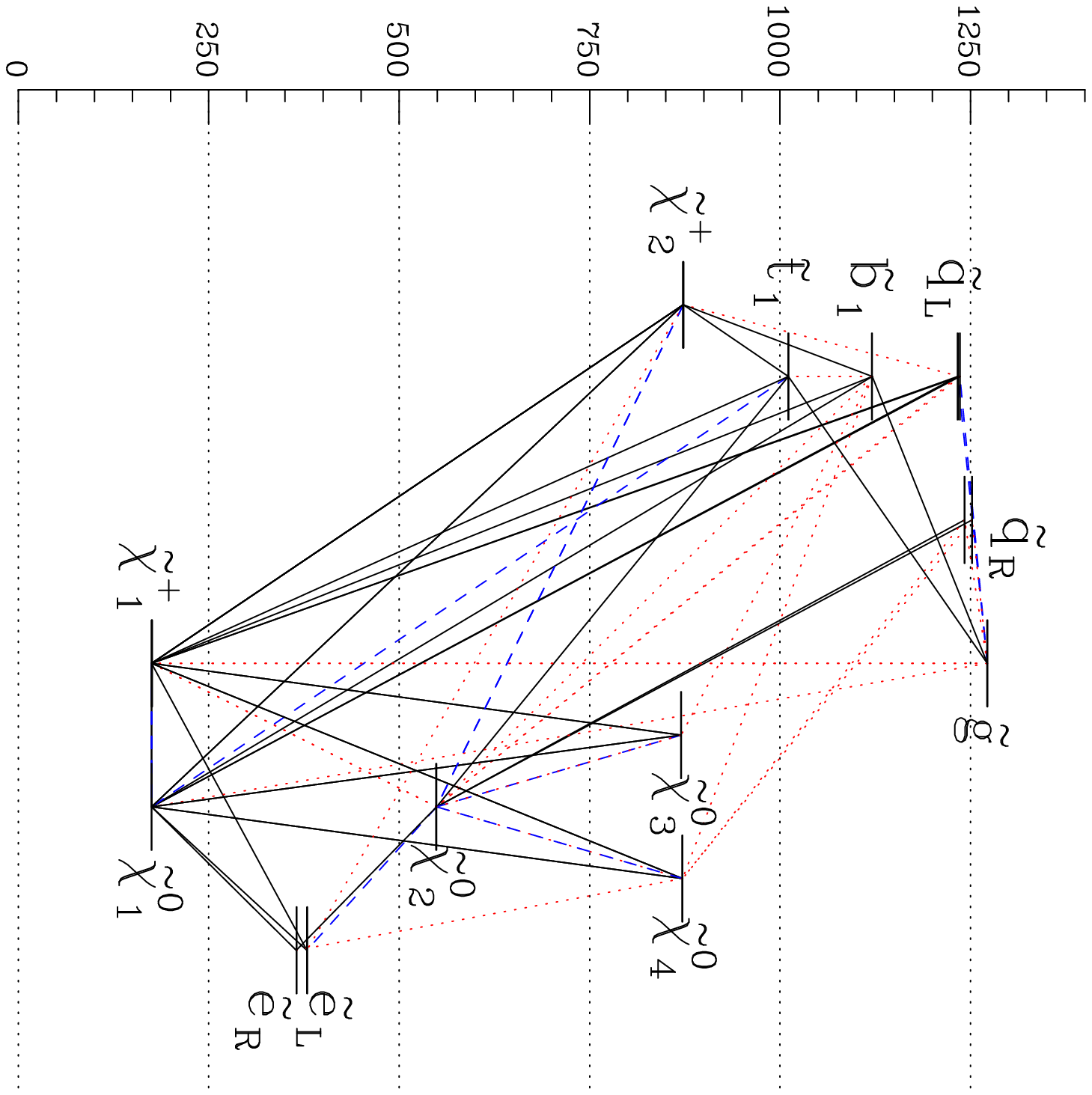, width=10.5cm, angle=90}{
Part of the sparticle spectrum at the Snowmass point {\bf SPS~9} 
which has $\mnought=450$~GeV, $\mthreehalfs=60$~TeV and $\tan\beta=10$, $\mu>0$.
Solid black lines indicate branching ratios (BR's) greater than 10\%,
dashed blue lines show BR's in the range $1\% \to 10\%$, while
red dotted lines show BR's in the range $0.1 \to 1\%$.
The sparticles are displaced horizontally for clarity.
\label{AMSB:SNOSPEC} }

A set of benchmark points and `slopes' or model lines was suggested 
for study at Snowmass\cite{Allanach:2002nj}.
Of the eleven points, only one ({\bf SPS~9}) applies to mAMSB.
That point has the parameters: $\mnought=450$~GeV,~$\mthreehalfs=60$~TeV, $\tan\beta=10$, with $\mu>0$,
and lies on the model line ``slope'' $\mnought = 0.0075\times\mthreehalfs$,
where $\mthreehalfs$ can vary.
Part of the sparticle spectrum for {\bf SPS~9} is shown in \figref{AMSB:SNOSPEC}.
The lightest sparticles --- the \chgone\ and \ntlone\ --- have masses about 170~GeV
while squark and gluino masses are about 1.25~TeV,
so one would expect copious sparticle production at the LHC. 
The Wino-like character of the lightest chargino and neutralino
increases the relative cross-section to non-coloured sparticles as compared to
\eg\ SUGRA models.
Indeed the \herwigv{6.3}\cite{Corcella:2000bw,Moretti:2002eu,Corcella:2001pi} 
Monte-Carlo event generator gives the inclusive SUSY cross-section as 3.9~pb, 
of which about 0.5~pb is to squarks and gluinos.
The chain $\tilde{q}\to\ntltwo\to\tilde{l}\to\ntlone$ is
available, \footnote{Note 
that SM particles are omitted from decays when this can be done without ambiguity.}
and has a large branching ratio, so good information
about the squark, slepton, and at least two neutralino masses could be
extracted from kinematic edges in the $ll$, $lq$ and $llq$ 
invariant masses\cite{phystdr,Allanach:2000kt,Paige:1999ui}.

\EPSFIGURE[t]{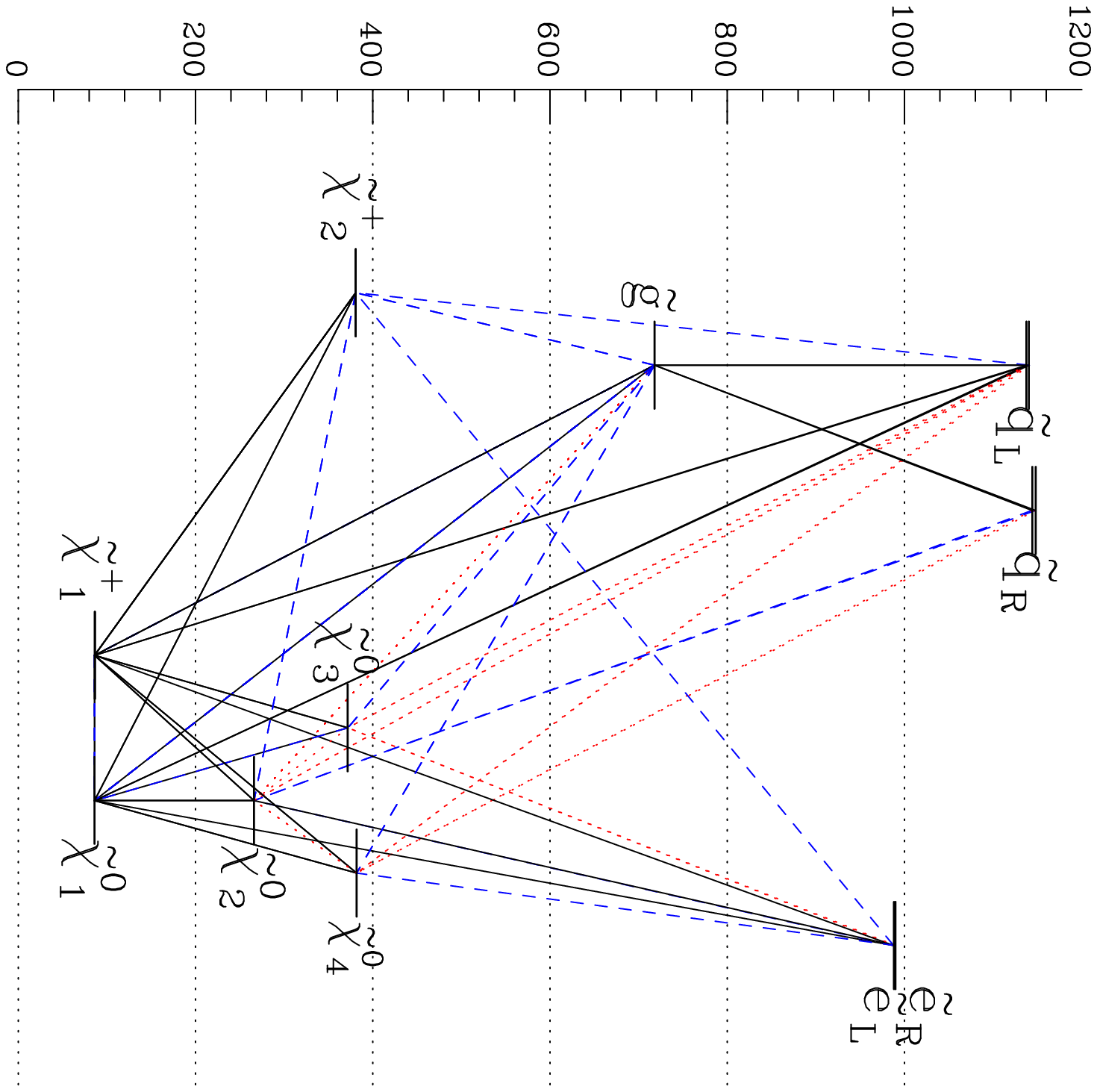, width=10.5cm, angle=90}{
Part of the sparticle spectrum at the Point d'Aix: 
$\mnought=1000$~GeV, $\mthreehalfs=30$~TeV, $\tan\beta=30$, $\mu>0$.
Branching ratios as for \figref{AMSB:SNOSPEC}.
The sparticles are displaced horizontally for clarity.
\label{AMSB:PDXSPEC}}

The \sser\ and \ssel\ masses are approximately equal (\figref{AMSB:SNOSPEC})
because in mAMSB the slepton masses are principally 
determined by \mnought. A more detailed analysis would be required to determine
if the \sser\ and \ssel\ masses could be separately measured 
from two nearby edges in the $ll$ invariant mass 
distribution, as was noted in \cite{Paige:1999ui} for a similar point.
The decay of the right-sleptons to the Wino-like LSP is suppressed,
but not sufficiently to produce a measurable displaced vertex from slepton decay ($\tau\sim 10^{-16}$~s).
In fact even if the Bino-like component of the LSP becomes 
{\em extremely} small, the decay $\sser\to\sstauone\to\ntlone$ will remain unsupressed.
This is particularly true at high $\tan\beta$ since the mixing between the left- and 
right-handed staus increases with the tau Yukawa coupling.

\FIGURE[tp]{
  \begin{minipage}[b]{.48\linewidth}
    \begin{center}
     \epsfig{file=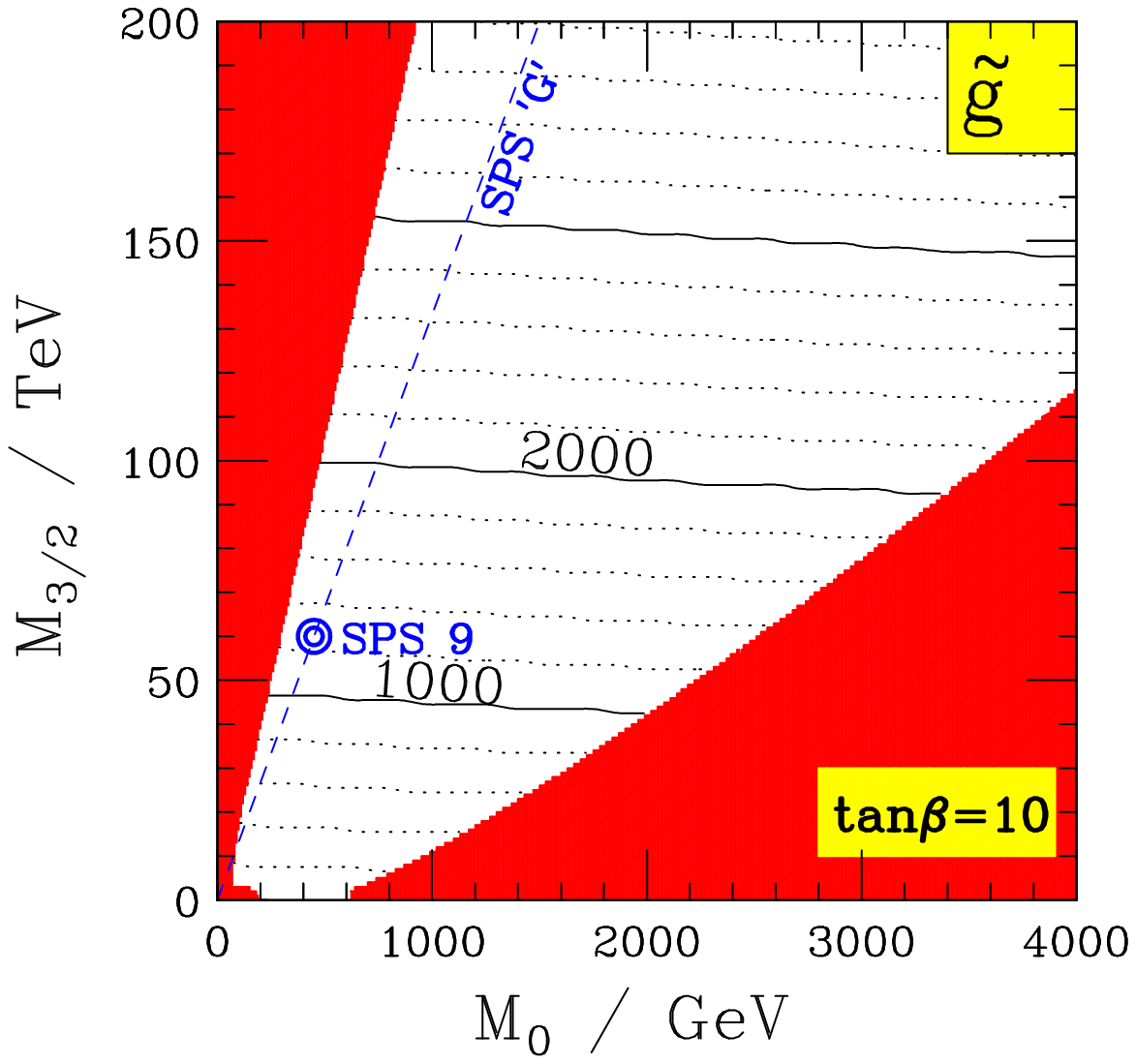, height=7.5cm,
      width=7cm, angle=90}\\
      \hspace*{0.5cm}{\bf (a)}\vspace*{1cm}\\
      \epsfig{file=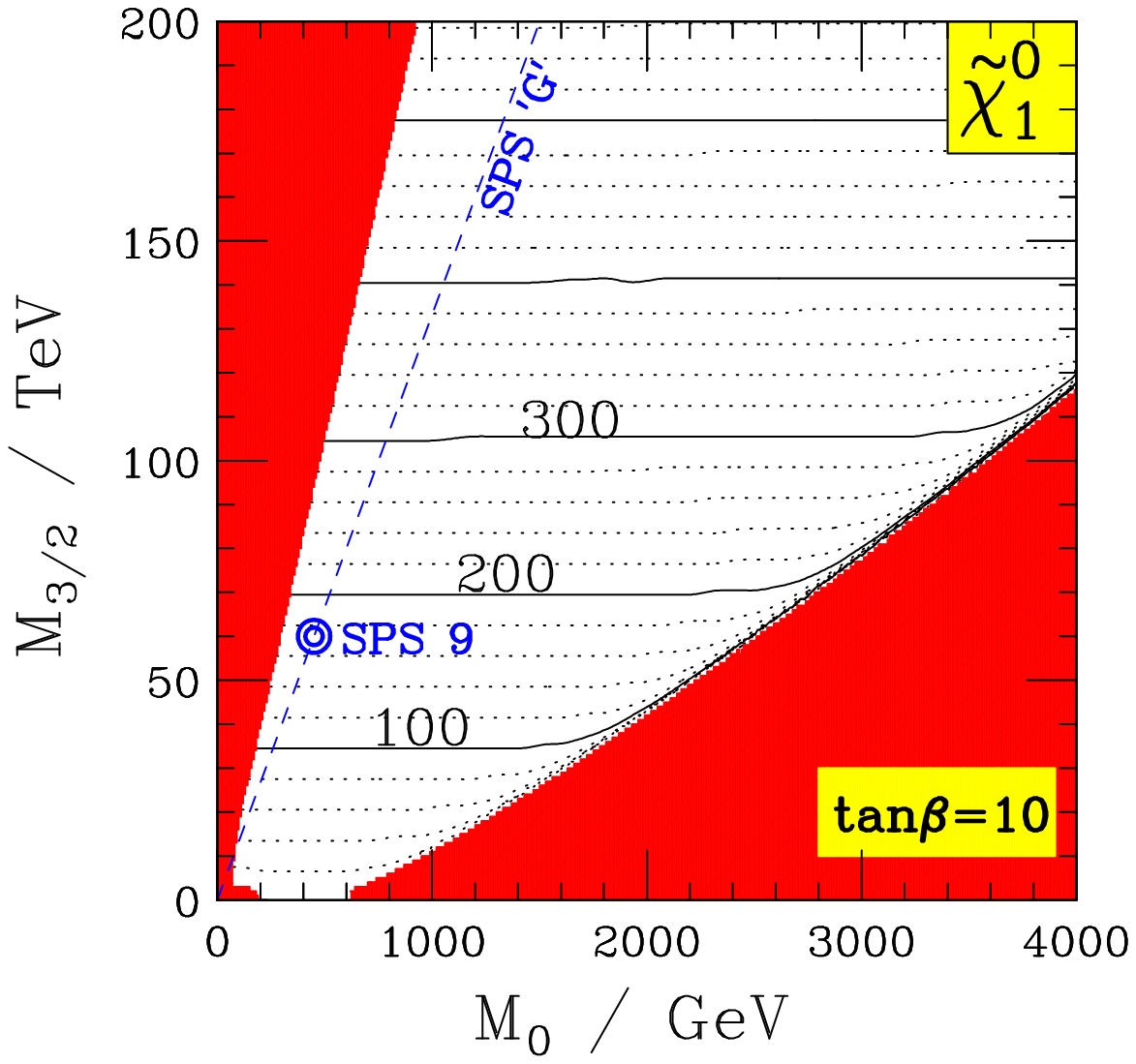, height=7.5cm,
       width=7cm, angle=90}\\
      \hspace*{0.5cm}{\bf (c)}
    \end{center}
  \end{minipage}\hfill
  \begin{minipage}[b]{.48\linewidth}
    \begin{center}
     \epsfig{file=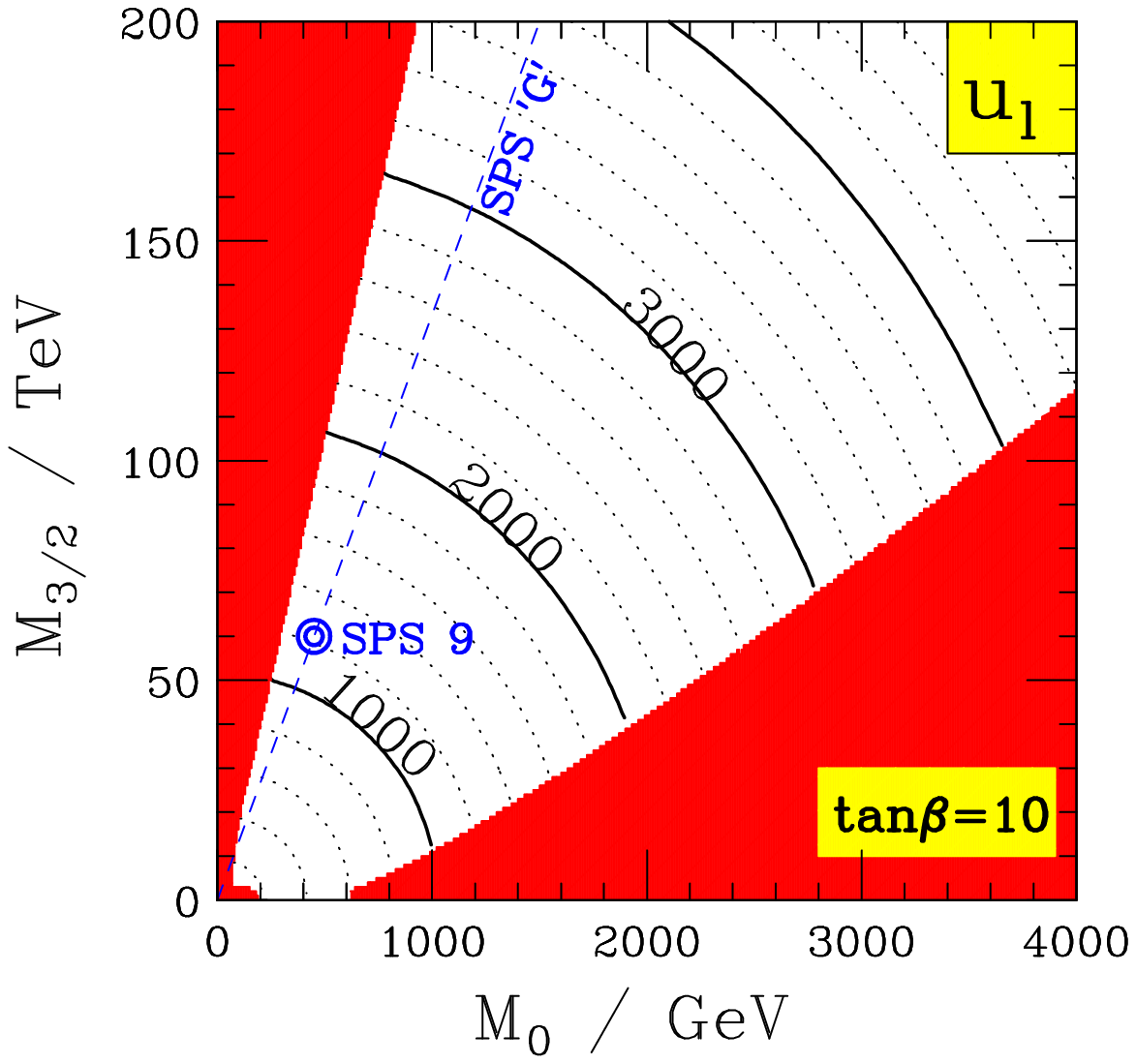, height=7.5cm,
       width=7cm, angle=90}\\
       \hspace*{0.5cm}{\bf (b)}\vspace{1cm}\\
      \epsfig{file=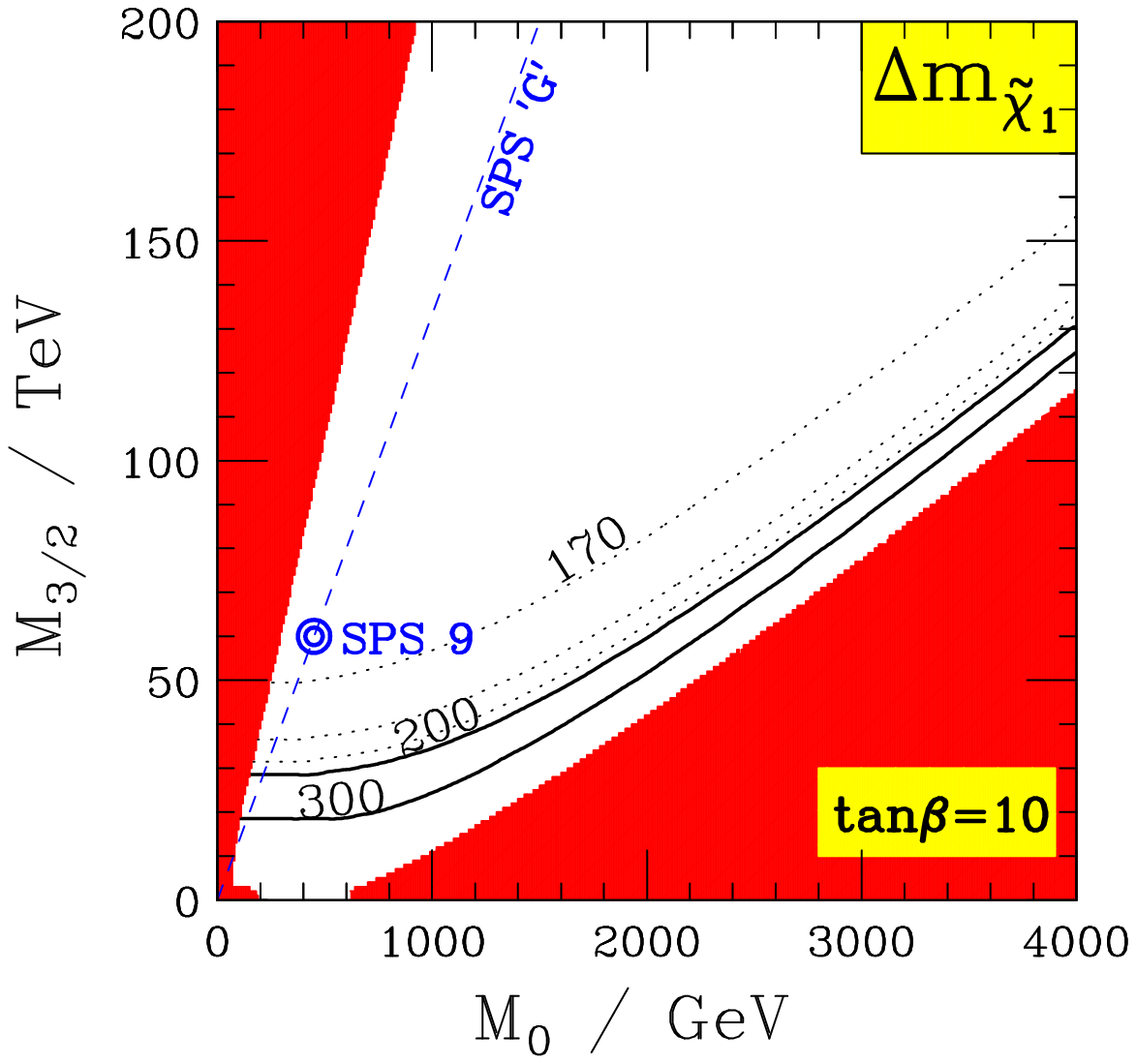, height=7.5cm,
       width=7cm, angle=90}\\
       \hspace*{0.5cm}{\bf (d)}
    \end{center}
  \end{minipage}\hfill
\caption{
The contours show the {\bf (a)}~gluino, {\bf (b)}~up left squark, and {\bf (c)}~lightest neutralino mass in GeV, and
{\bf (d)} the $\Delta m=\chgone-\ntlone$ mass difference in MeV
as a function of \mnought\ and \mthreehalfs.
The other parameters are: $\tan\beta=10$ and $\mu>0$.
The Snowmass point ($\mnought=450$~GeV,~$\mthreehalfs=60$~TeV) is marked as a pair of concentric circles,
through which passes the Snowmass slope ``model line G'' (dashed line).
The solid red regions are excluded because of lack of electroweak symmetry
breaking (bottom right) or because of charge- or colour-breaking
minima, or non-\ntlone\ LSP (left).
$\Delta m$ is greatest where $\mu$ is small, which occurs near to the 
region where electroweak symmetry is unbroken. 
\label{AMSB:MASSES} }
}

Because the gluino has a similar mass to the heavier squarks
at {\bf SPS~9}
it decays primarily to \sstone\ and \ssbone, meaning that
gluino production will lead to large numbers of $b$-quarks.
This is not a general (or unique) feature of mAMSB, and while vertex tagging 
of jets could help distinguish this particular point from the SM
it is not used in our analysis.

Another set of benchmarks, the `Points d'Aix'\cite{Djouadi:pdaix}
chooses the mAMSB point $\mnought=1000$~GeV, $\mthreehalfs=30$~TeV
also with $\mu>0$, and with the higher value $\tan\beta=$30.
With a \chgone\ mass of 85~GeV, this point violates the limits
from the LEP search\cite{Heister:2002mn} for $e^+ e^-\to \chgone\cht^-_1 \gamma$,
which places a lower limit on the \chgone\ mass of about 92~GeV
for almost all values of the mass difference \DeltaMChi.

The mAMSB Point d'Aix, despite being ruled out, still has a number of features which are 
of qualitative interest (\figref{AMSB:PDXSPEC}).
The light gluino means that the decays $\gluino\to\tilde{t}\bar{t}\ \mathrm{and}\ \tilde{b}\bar{b}$
are no longer kinematically allowed. This means that the 
major source of $b$ quarks is from $\cht^0_{2,3,4} \to \ntlone h$ followed by $h\to b \bar{b}$,
so fewer heavy quarks are produced than for the Snowmass point.
The heavy sleptons can no longer participate in the various chains $\tilde{q}\to\cht^0_x\to\tilde{l}\to\cht^0_y$
so slepton mass measurements would be extremely difficult at a hadron collider.
There will, however, be large numbers of events in which \chgonepm s are produced,
either directly or from $\gluino\to\chgonepm q \bar{q}$, 
so a signature based on the near-degeneracy of the \chgone\ and \ntlone\ is robust.

\section{The ATLAS detector}
\label{AMSB:ATLAS}
ATLAS is one of two general-purpose experiments which will explore a wide
variety of high \pt\ physics with proton on proton collisions at $\sqrt{s}=14$~TeV the LHC.
The LHC design luminosity is $(10^{34}$~$\mathrm{cm}^{-2} \mathrm{s}^{-1})$,
with a bunch-crossing period of 25~ns.

The detector design, performance and physics potential 
are described in \cite{phystdr}. The main components are:
\begin{itemize}
\item{An inner tracking detector contained within a 2~Tesla solenoidal field
which comprises three different elements, 
each of which has acceptance up to $|\eta|<2.5$ and $2\pi$ in $\phi$, where
$\eta=-\log[\tan(\theta/2)]$ is the pseudorapidity,
$\theta$ is the polar angle from one of the beams, and $\phi$ is the azimuth.
In order of increasing radius in the transverse direction ($r$ in mm) the
sub-detector components are: a pixel detector ($43<r<132$),
a silicon microstrip detector ($300<r<520$) and a straw-tube 
transition radiation tracker with  electron-identification ability ($560<r<1070$).}
\item{A highly segmented lead/liquid-argon electromagnetic calorimeter allows
electron and photon identification.
The hadronic calorimeter employs steel tiles with a scintillator in the 
barrel region $(|\eta|<1.7)$ with granularity $\Delta\eta\times\Delta\phi=0.1\times 0.1$,
and copper/liquid-argon in the endcaps ($1.5<|\eta|<3.2$).
A dense forward calorimeter with granularity $0.2\times 0.2$ extends coverage to $|\eta|<5$.}
\item{A muon system consisting of both precision and fast (trigger) detection elements.
It is constructed within a toroidal magnet system and extends to transverse radii of about 10~m.}
\end{itemize}

\section{LHC reach for mAMSB}
\label{AMSB:SCAN}

While some of the phenomenological features are particular to AMSB,
one might expect to be able to distinguish AMSB from the SM
using the same types of cuts -- based on leptons, jets and missing energy 
-- that are applied in mSUGRA analyses.

The sensitivity of the LHC to mAMSB has been demonstrated\cite{Paige:1999ui}
for one point ($\mnought=200$~GeV, $\mthreehalfs=35$~TeV, 
$\tan\beta=3$, $\mu>0$) with relatively light sparticles,
where the sparticle spectrum was investigated in detail.
In \cite{Ghosh:2001xp,Ghosh:2000fv,Choudhury:Ghosh:Roy}, 
the signatures for AMSB at a future linear
$e^+ e^-$ or $e^-\gamma$ collider were investigated.
In \cite{Baer:2000bs} the reach of the LHC was investigated
using a simple generic detector simulation for 10~\invfb\ of integrated luminosity.
The production of charged and neutral Winos via vector boson fusion
was studied for AMSB in \cite{Datta:2001hv}.
In that paper the LHC's reach was investigated for a
signature consisting of two jets widely separated in pseudorapidity 
in association with missing transverse momentum.

In this section our the aim is to determine the reach of the LHC 
with a realistic detector simulator, using optimised but generic SUSY cuts
and for 100~\invfb\ of integrated luminosity.

\subsection{Event simulation}
\label{AMSB:SIM}

\FIGURE[tp]{
  \begin{minipage}[b]{.41\linewidth}
    \begin{center}
     \epsfig{file=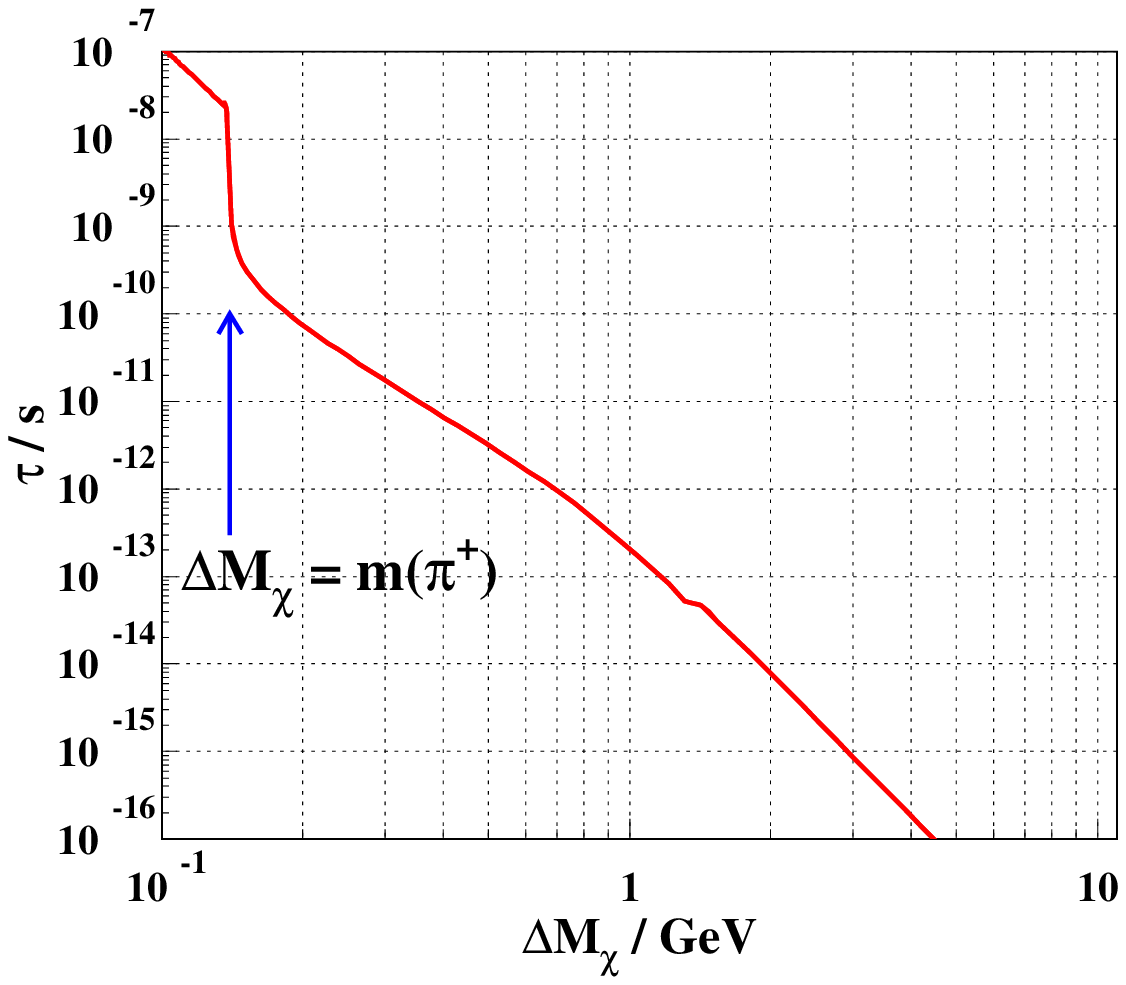, height=6.5cm}\\
      \hspace*{0.5cm}{\bf (a)}
    \end{center}
  \end{minipage}\hfill
  \begin{minipage}[b]{.45\linewidth}
    \begin{center}
     \epsfig{file=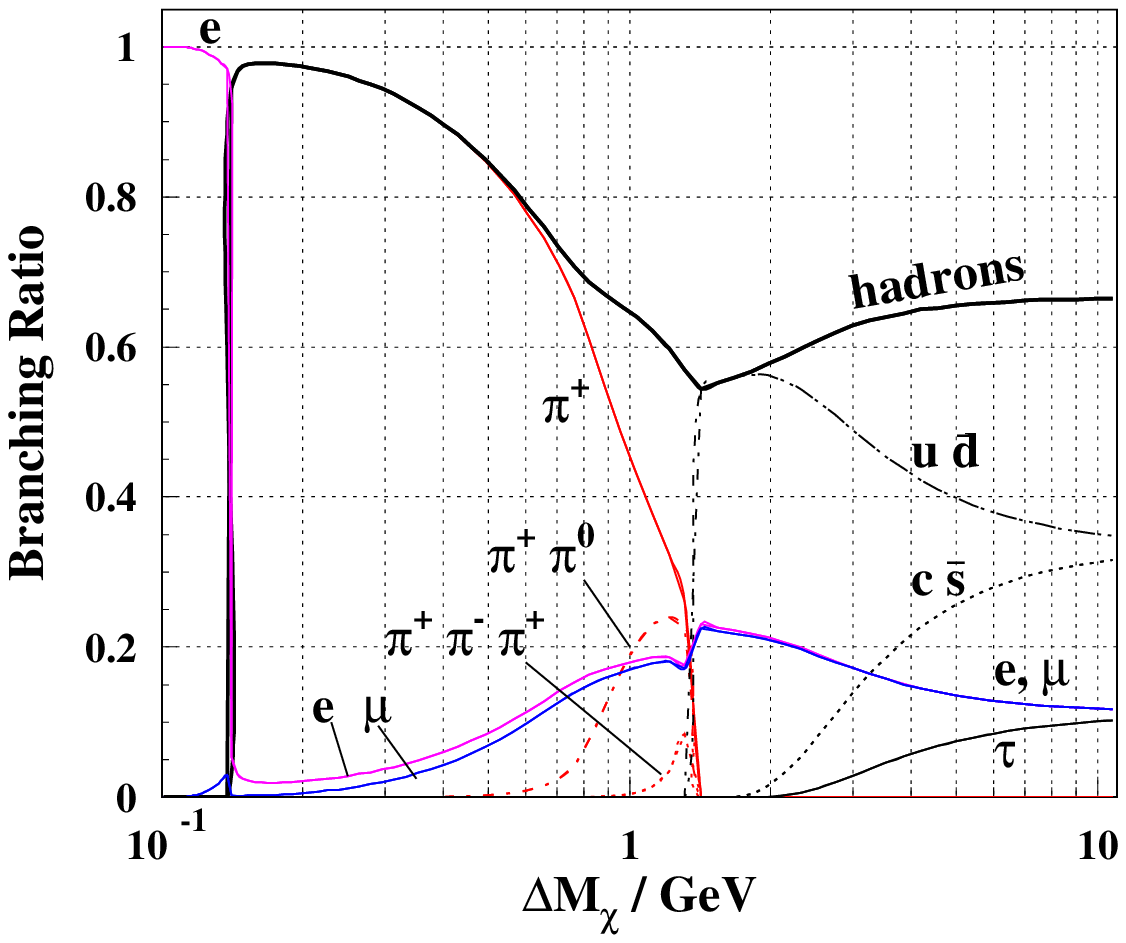,  height=6.3cm}\\
       \hspace*{0.5cm}{\bf (b)}
    \end{center}
  \end{minipage}\hfill
\caption{
The {\bf (a)} lifetime and {\bf (b)} branching ratios of the lightest chargino
as a function of the mass difference $\Delta M_\chi=M(\chgone)-M(\ntlone)$.
The rapid decrease in the lifetime occurs at $\Delta M_\chi=m_{\pi^+}$
where the single pion mode becomes available.
The discontinuity at $\Delta M_\chi=1.4$~GeV comes from the switch in the 
calculation from hadronic to partonic decay widths. 
The leptonic channels implicitly include the corresponding neutrino.
The branching ratio to $\pi^+\pi^0\pi^0$
was assumed to be equal to $\pi^+\pi^-\pi^+$.
After \cite{Chen:1997ap,Chen:1999yf,Kuhn:1990ad}.
\label{AMSB:CTAUBR}}
}

\afterpage{\clearpage}

The mAMSB spectra were generated using \isajetv{7.63}\cite{Paige:1998xm,Baer:1993ae}
on a grid 100~GeV $\times$ 5~TeV in the (\mnought, \mthreehalfs) plane.
In all cases the ratio of the Higgs vacuum expectation values
($\tan\beta$) was set equal to 10 and the sign of $\mu$ was positive.
The mass of the top quark (important for electroweak symmetry breaking)
was taken to be 175~GeV throughout.
The dependence of some of the key sparticle masses on the input values of 
\mnought\ and \mthreehalfs\ is shown in \figref{AMSB:MASSES}.

In the $\chgonepm\to\ntlone$ decays, \isajet\ does not include masses for the
leptons, and does not contain multi-pion decay modes.
Since the mass difference \DeltaMChi\ can be of the order of
the mass of the muon, the lepton mass effects can be important in AMSB.
To improve accuracy, the chargino decay modes calculated in \cite{Chen:1997ap,Chen:1999yf} 
were implemented with pion form factors from \cite{Kuhn:1990ad}
and massive leptons. The resulting chargino lifetime and
branching ratios are shown in \figref{AMSB:CTAUBR}.

\herwigv{6.3} was used to produce 200~\invfb\ of unweighted inclusive supersymmetry events for each point
(with a minimum of $5\times10^4$, and up to a maximum of $5\times10^5$ events).\footnote{
No significant differences were found on comparing with \herwigv{6.4} \cite{Corcella:2001wc} 
which includes spin correlations as described in \cite{Richardson:2001df}.}
This was then scaled to give the 
expected number of events for $\int\mathcal{L} = 100$~\invfb, 
which corresponds to the first year of `high luminosity'
$(10^{34}$~$\mathrm{cm}^{-2} \mathrm{s}^{-1})$ running of the LHC.
Background samples were generated with 
\herwig\ for the production of $W^{\pm}+\mathrm{jets}$, 
$Z^0+\mathrm{jets}$,  $t\tbar$, and QCD $2\to2$ (excluding $t\tbar$).
For the $W^{\pm}+\mathrm{jets}$ sample, the cross-sections were
multiplied by the factor
\begin{equation} 1.6\times\left[ \frac{m_W^2  + 
(p_t^\mathrm{thr})^2}{m_W^2}\right]^{2(N_\mathrm{jet}-1)} \end{equation}
where $N_\mathrm{jet}$ is the number of jets,
and $p_t^\mathrm{thr}$ is the jet transverse energy threshold which was set to 10~GeV.
This correction brings the \herwig\ cross-section into better 
agreement with tree-level matrix element calculations\cite{Giele:1990vh}.
A total of over 20~million background events were generated
in logarithmic intervals in the \herwig\ parameters {\tt PTMIN} and {\tt PTMAX} from 0 to 7000~GeV.
The background cross-sections are shown before and after preselection cuts in \figref{AMSB:BGXS}.

The events were passed through the ATLAS fast detector simulator, 
\atlfastv{2.50}\cite{Richter:1998at}, which gives a parameterised detector response
based on {\tt GEANT3} Monte-Carlo simulations\cite{buis}.
Jets were found using the \atlfast\ cone algorithm with a cone size $\Delta R=0.4$,
and a minimum \pt\ of 10~GeV.
The loss in resolution from pile-up was simulated.
Calorimeter cells with $E_T$ deposits below 1~GeV were not included in the
\etmiss\ calculation in order to more accurately model the expected resolution.
The cell sizes were $\Delta\eta\times\Delta\phi=0.1\times$0.1 for $|\eta|<3$ 
and 0.2$\times$0.2 for $3<|\eta|<5$.
Otherwise the default \atlfast\ parameters were applied.
The parameterised tracking simulation used is described in \appref{EIDENT}.

\subsection{Optimisation of cuts}

Cuts were applied to leptons (electrons and muons only), jets, and missing transverse energy
in a similar manner to those used in \cite{phystdr,Abdullin:1998pm,Bityukov:2001yf,Baer:2000bs,Tovey:2002}.
The variables to which cuts were applied were:
\begin{list}{\bfseries\upshape (\arabic{cut})}
{\newcounter{cut}
}\addtocounter{cut}{1}
\item {\etmiss, missing transverse energy;}\addtocounter{cut}{1}
\item {$p_{T(J1)}$, \pt\ of the hardest jet;}\addtocounter{cut}{1}
\item {$p_{T(J2)}$, \pt\ of the next-to-hardest jet;}\addtocounter{cut}{1}
\item {$\sum \pt$, scalar sum of the \pt\ of jets in the event;}\addtocounter{cut}{1}
\item {$N_\mathrm{jet}$, number of jets in the event;}\addtocounter{cut}{1}
\item {$S_\mathrm{T}$, transverse sphericity (circularity) of the event;}\addtocounter{cut}{1}
\item {$\Delta\phi_{(J1)}$, difference in azimuth between hardest jet and \Ptmiss\ vector;}\addtocounter{cut}{1}
\item {$p_{T(\ell 1)}$, \pt\ of hardest lepton (if any);}\addtocounter{cut}{1}
\item {$\Delta\phi_{(\ell 1)}$,  difference in azimuth between hardest lepton (if any)
and \Ptmiss\ vector;}\addtocounter{cut}{1}
\item {$M_T = \sqrt{2\ p_{T(\ell 1)}\ \etmiss\ ( 1 - \cos ( \Delta\phi_{(\ell 1)}))}$,
transverse mass of lepton and missing energy. 
Applied to single-lepton channel only, to reduce SM leptonic $W^\pm$ background.}
\end{list}

The preselection cuts which were applied to cut down the background, particularly at low \pt,
are shown in the third column of \tabref{AMSB:CUTVALS}.
The reduction in the background cross-section can be seen in \figref{AMSB:BGXS}. 
Hard preselection cuts were not placed on isolated electrons or muons since it is 
foreseen that one jet with $\pt>100$~GeV together with $\etmiss>100$~GeV will 
be sufficient to provide the on-line trigger.
Other trigger strategies which would improve the reach include:
\begin{itemize}
\item{Allowing a hard ($\approx 20$~GeV) isolated lepton as a trigger
could improve the selection for points for relatively low \mnought\
in which the cascade decay $\tilde{q}\to\cht^0_x\to\tilde{l}\to\cht^0_y$ 
is available;}
\item{Track vertexing which in ATLAS can be applied at the second trigger level 
could improve the selection for events containing heavy quarks;}
\item{If all of the coloured sparticles are heavy then the dominant 
production of SUSY particles will be to the lightest gauginos.
At leading order in perturbation theory, the production of 
\ntlone s and/or \chgonepm s would be difficult to trigger on, 
but gaugino production in association with a high \pt\ jet or 
photon and \etmiss\ might be observable\cite{Gunion:1999jr,Feng:1999fu}.
}
\end{itemize}
A fraction of the simulated gaugino production events
contain high \pt\ jets from initial state parton showers.
However the parton shower algorithm is based on the soft and co-linear
approximations so this fraction will be smaller than that which would 
result from higher order matrix element calculations. 
Our simulation of the trigger can therefore be considered conservative.

\EPSFIGURE[tp]{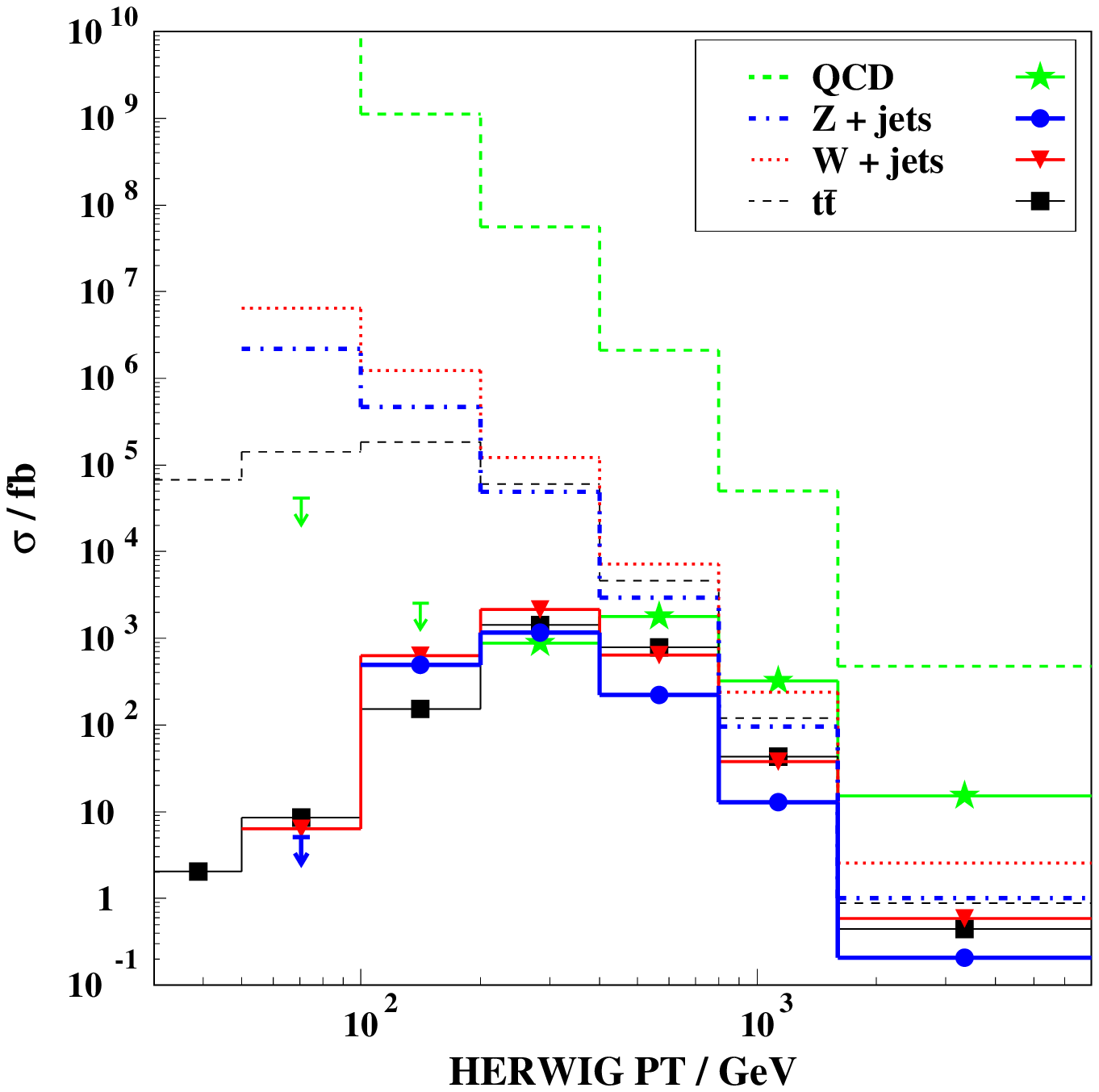, width=11cm}{
The cross-sections for the $t\tbar$, $W^{\pm}+\mathrm{Jets}$, and $Z^0+\mathrm{Jets}$
and QCD (excluding $t\tbar$)
backgrounds plotted as a function of the transverse momentum variable which
is compared in \herwig\ to {\tt PTMIN} and {\tt PTMAX}.
The production cross-sections are denoted with dashed lines, 
while the cross-section to pass the preselection cuts are solid lines.
Where no events passed the selection cuts (at low {\tt PT})
arrows indicate 90\% Poisson confidence limits on the selected cross-sections.
\label{AMSB:BGXS}
}

Different values of the cuts were applied to events with
zero, one, two or three leptons (electrons and muons only) as reconstructed by \atlfast,
as well as to an inclusive lepton (`ptmiss') analysis.
In the case of two-lepton events, different cuts were applied according
to whether the leptons were of the same sign~(SS) or opposite sign~(OS).

\renewcommand{\arraystretch}{1.2}
\TABULAR[bt]{|c|ll|c|c|c|c|c|c|c|c|c|c|c|cccc}{
\hline
{\bf } & \multicolumn{2}{c|}{\bf Variable} & \multicolumn{10}{c|}{\bf Allowed Values}\\
\hline\hline
1 &$\etmiss$ 	& $>$&    {\bf\textit{200}} & 250 & 300 & 400 & 500 & 600 & 800 & 1000 & 1500 & 2000\\
2 &$p_{T(J1)}$	& $>$&    {\bf\textit{100}} & 150 & 200 & 300 & 400 & 600 & 800 & 1000 & 1500 & 2000\\
3 &$p_{T(J2)}$	& $>$&    {\bf\textit{100}} & 150 & 200 & 300 & 400 & 600 & 800 & 1000 & 1500 & 2000\\
4 &$\sum \pt$ 	& $>$& {\bf\textit{200}} & 250 & 300 & 400 & 500 & 600 & 800 & 1000 & 1500 & 2000\\
5 &$N_\mathrm{jet}$ & $\ge$&    {\bf\textit{2}} & 3 & 4 & 5 & 6 & 7 & 8 & 9 & 10 & 11\\
\hline
6 &$S_\mathrm{T}$      & $>$&    0 & 0.1 & 0.2 & 0.3 & 0.4 & 0.5 & 0.6 & 0.7 & 0.8 & 0.9\\
7 &$\Delta\phi_{(J1)}$ & $>$&    0 & 0.3 & 0.6 & 0.9 & 1.2 & 1.5 & 1.8 & 2.1 & 2.4 & 2.7\\
8 &$p_{T(\ell 1)}$     & $>$&    10 & 15 & 20 & 40 & 60 & 100 & 200 & 500 & 1000 & 2000\\
9 &$\Delta\phi_{(\ell 1)}$&$>$&   0 & 0.3 & 0.6 & 0.9 & 1.2 & 1.5 & 1.8 & 2.1 & 2.4 & 2.7\\
10 & $M_T$               &$>$& 100 & 100 &  100 & 100 & 100 & 100 & 100 & 100 & 100 & 100\\
\hline
}{
Allowed values for each of the cuts described in the text. 
The preselection cuts (highlighted in {\bf\textit{bold italic}}) 
are shown in the third column (applies to cuts 1~to~5 only).
The units of the variables (1-4,8,10) are GeV.
\label{AMSB:CUTVALS}}

For each of these analyses the cuts were optimised in the order listed
in \tabref{AMSB:CUTVALS}, and allowed to take one of ten values shown. 
The significance of the signal was given by $S/\sqrt{B}$
where $S$ and $B$ are the number of signal and
background events expected respectively for 100~\invfb.
For each variable, the cut was chosen to maximise the significance
subject to the constraint that $S>10$.
The analysis was determined to be successful if significance greater than five 
was achieved with at least ten events passing for any set of cuts.

This technique will not necessarily generate the global maximum in 
$S/\sqrt{B}$ since the cut variables are not totally independent,
but it is sufficient for the purposes of a large parameter-space 
scan, and it decreases the chances of over-fitting to a sparse background.

\subsection{Results}
\label{AMSB:SCANRESULTS}

The $5 \sigma$ (and $\ge10$~event) discovery reaches for 100~\invfb\ of integrated luminosity
in \figref{AMSB:Lep} and \figref{AMSB:ThPt} show that the LHC will be able to 
distinguish mAMSB from the SM over a large range of parameter space. 
As expected, the results are similar to previous generic R-parity conserving SUSY 
searches \cite{phystdr,Baer:2000bs,Abdullin:1998pm,Bityukov:2001yf,Tovey:2002} with 
reach extending up to squark masses of about 2800~GeV, or to gluino masses of about 2100~GeV,
whichever is the lower.
The discovery reach is similar to, but greater than that found in 
\cite{Datta:2001hv} for Wino production via vector boson fusion.
\afterpage{\clearpage}

\FIGURE[tp]{
  \begin{minipage}[b]{.48\linewidth}
    \begin{center}
     \epsfig{file=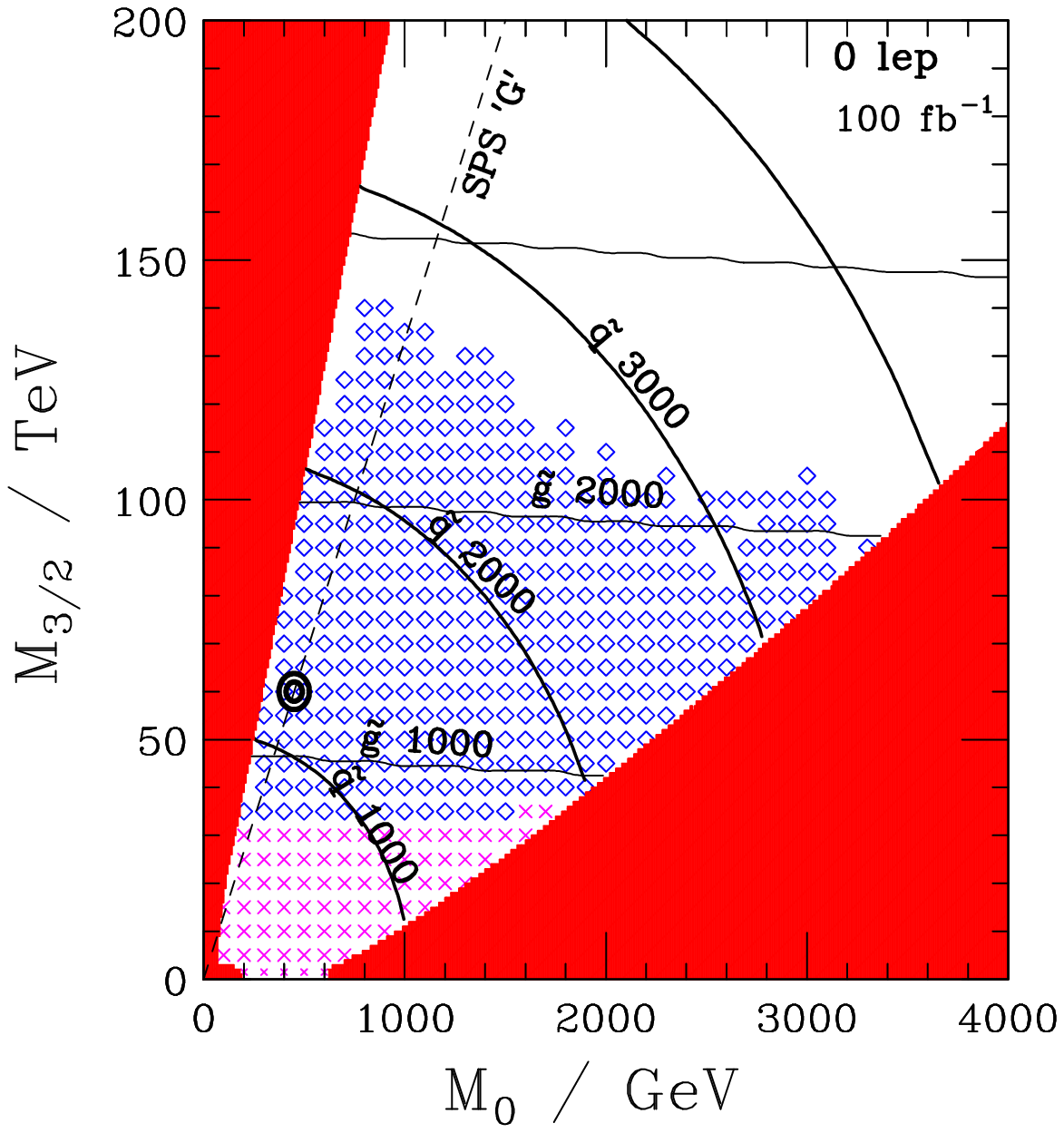, height=7.6cm,
      width=7cm, angle=90}\\
      \hspace*{0.2cm}{\bf (a)}\vspace*{0.5cm}\\
      \epsfig{file=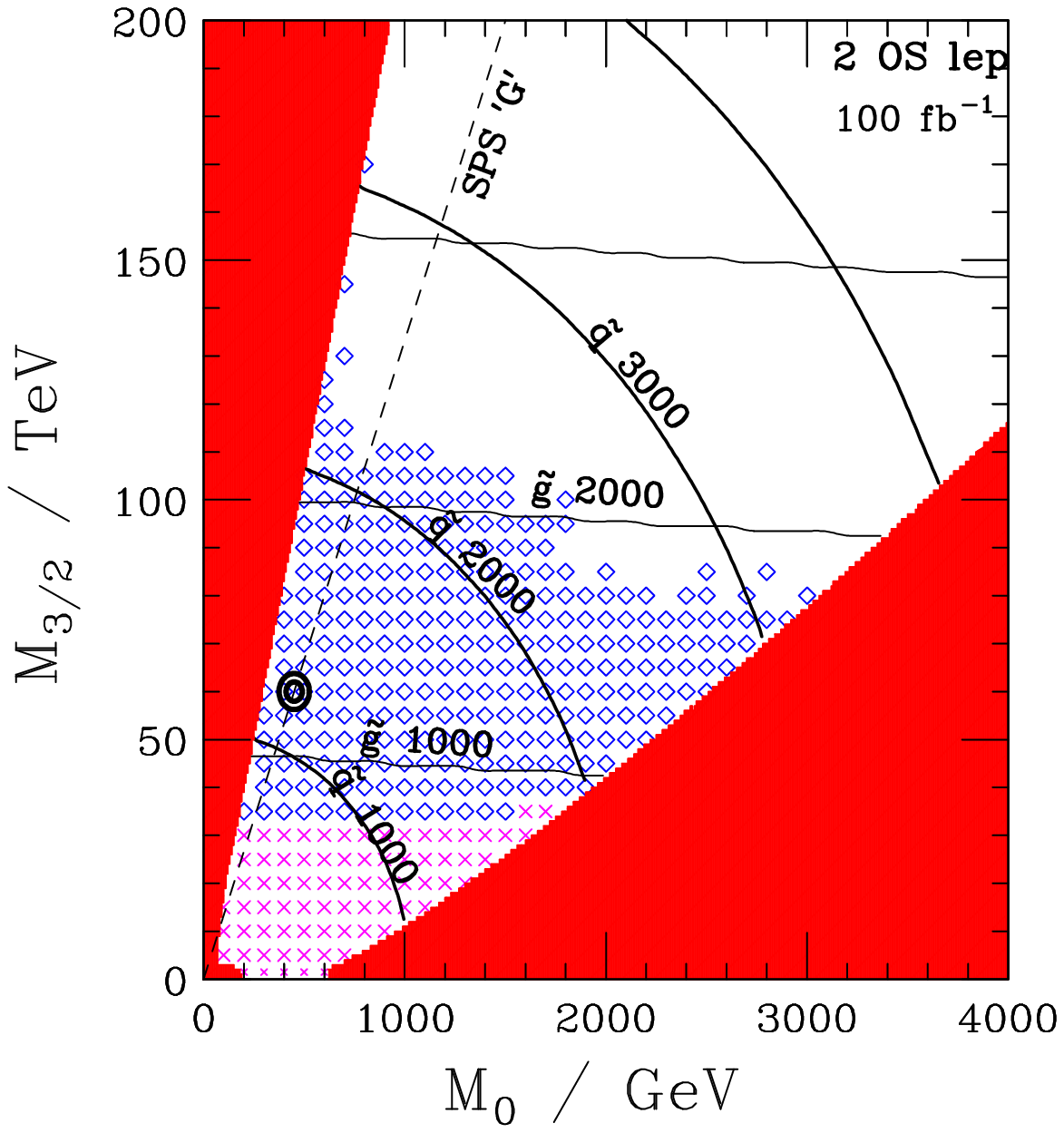, height=7.6cm,
       width=7cm, angle=90}\\
      \hspace*{0.2cm}{\bf (c)}\\
    \end{center}
  \end{minipage}\hfill
  \begin{minipage}[b]{.48\linewidth}
    \begin{center}
     \epsfig{file=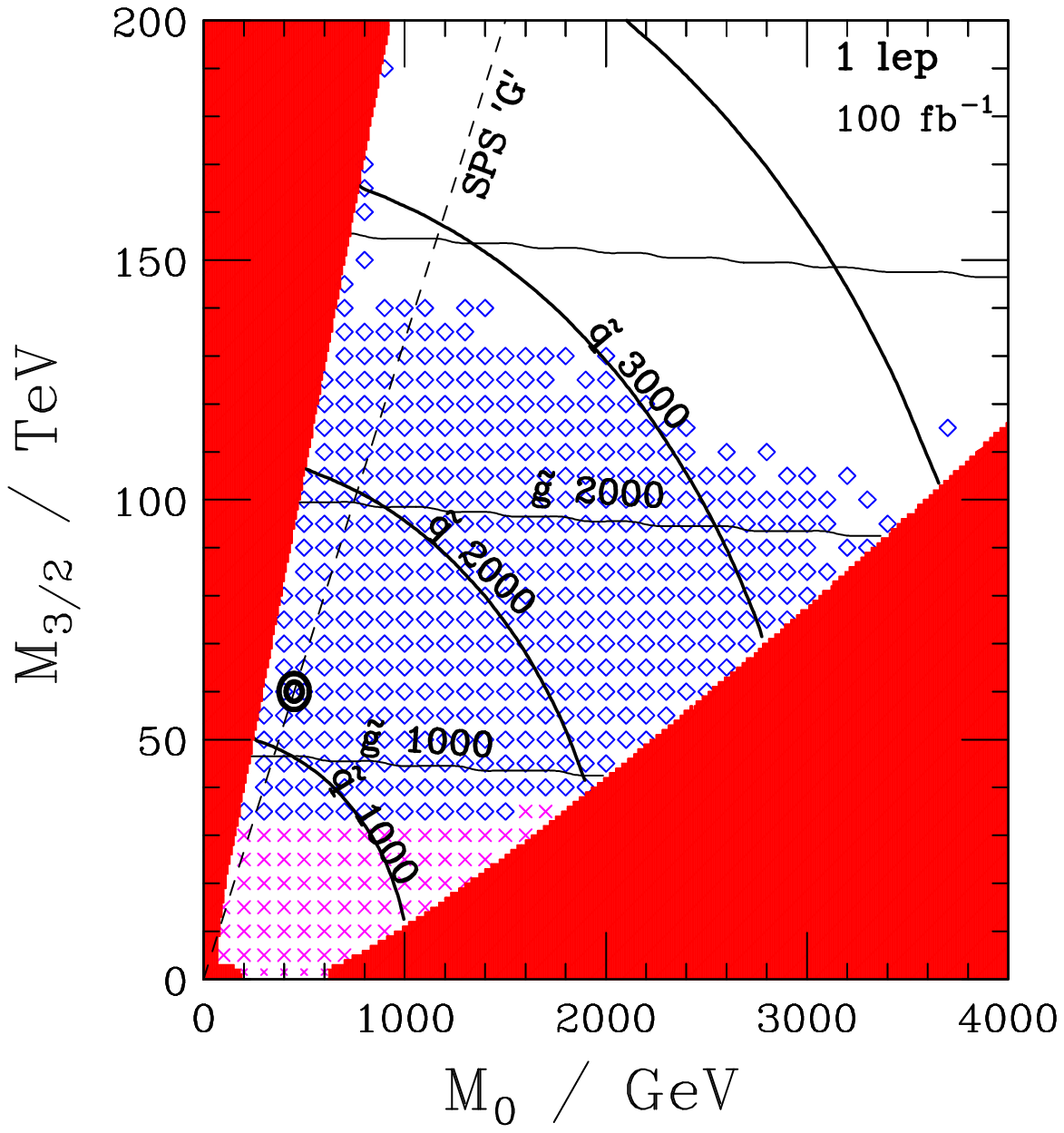, height=7.6cm,
       width=7cm, angle=90}\\
       \hspace*{0.5cm}{\bf (b)}\vspace{0.5cm}\\
      \epsfig{file=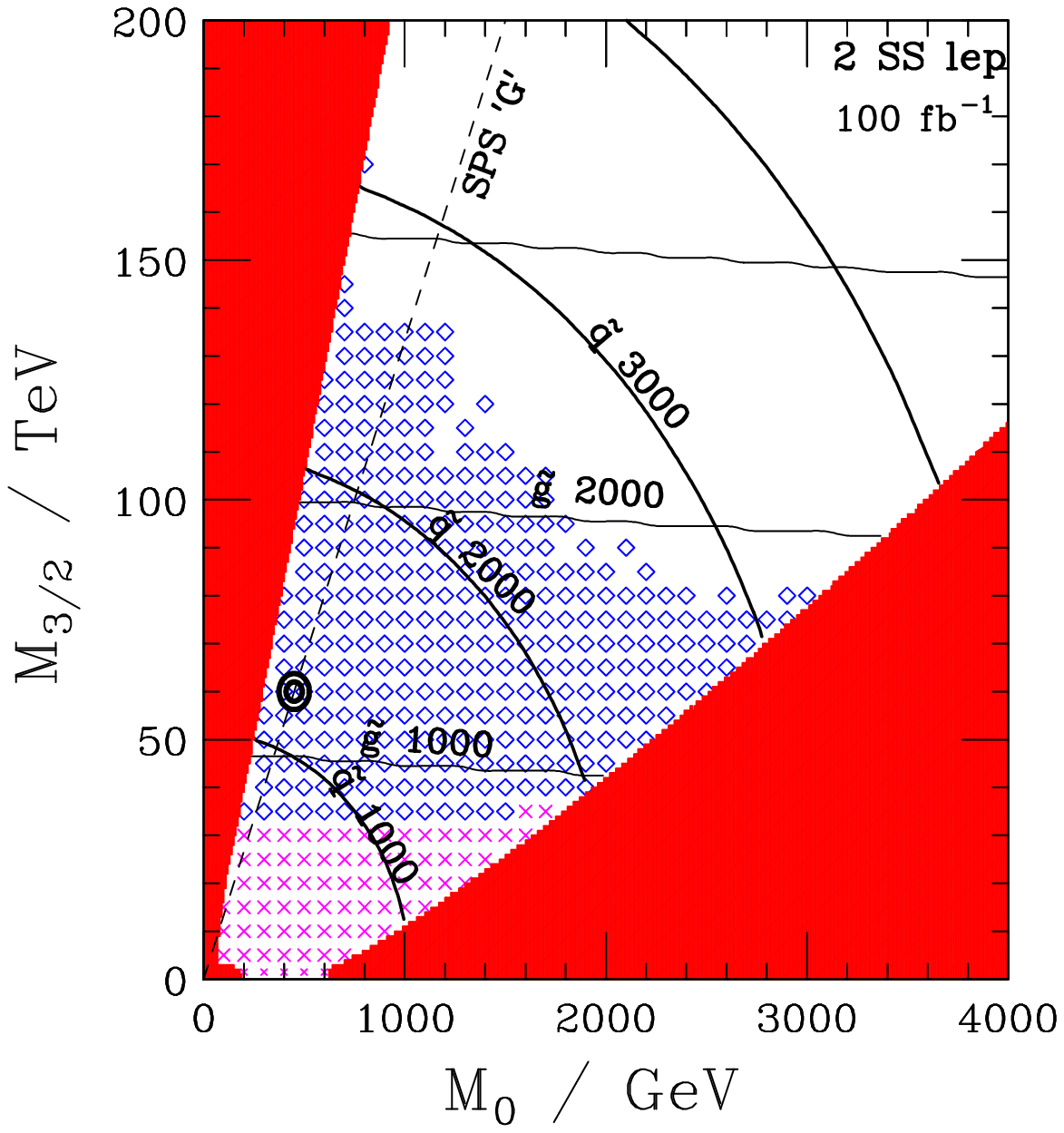, height=7.6cm,
       width=7cm, angle=90}\\
       \hspace*{0.5cm}{\bf (d)}\\
    \end{center}
  \end{minipage}\hfill
\caption{
The $5\ \sigma$~(and $\geq$~10 event) discovery region for mAMSB in the \mnought--\mthreehalfs\ plane,
is shown by the light blue diamonds, for the {\bf (a)} no lepton, {\bf (b)} single lepton, 
{\bf (c)} two opposite-sign leptons, {\bf (d)} two same-sign leptons channels.
The other parameters are $\tan\beta=10$ and $\mu>0$ in all cases.
The solid red regions are excluded because of lack of electroweak symmetry
breaking (bottom right) or because of charge- or colour-breaking
minima or non-\ntlone\ LSP (left).
The low \mthreehalfs\ region excluded by the LEP limit\cite{Heister:2002mn} 
on the chargino mass is indicated by purple crosses.
The Snowmass mAMSB point {\bf SPS~9} at $\mnought=450$~GeV,~$\mthreehalfs=60$~TeV is marked with a
pair of concentric circles,
through which passes the Snowmass slope `model line G' (dashed line).
Contours of \ssul\ and \gluino\ iso-mass are shown as solid lines.
\label{AMSB:Lep}} }

\FIGURE[pt]{
  \begin{minipage}[b]{.48\linewidth}
    \begin{center}
     \epsfig{file=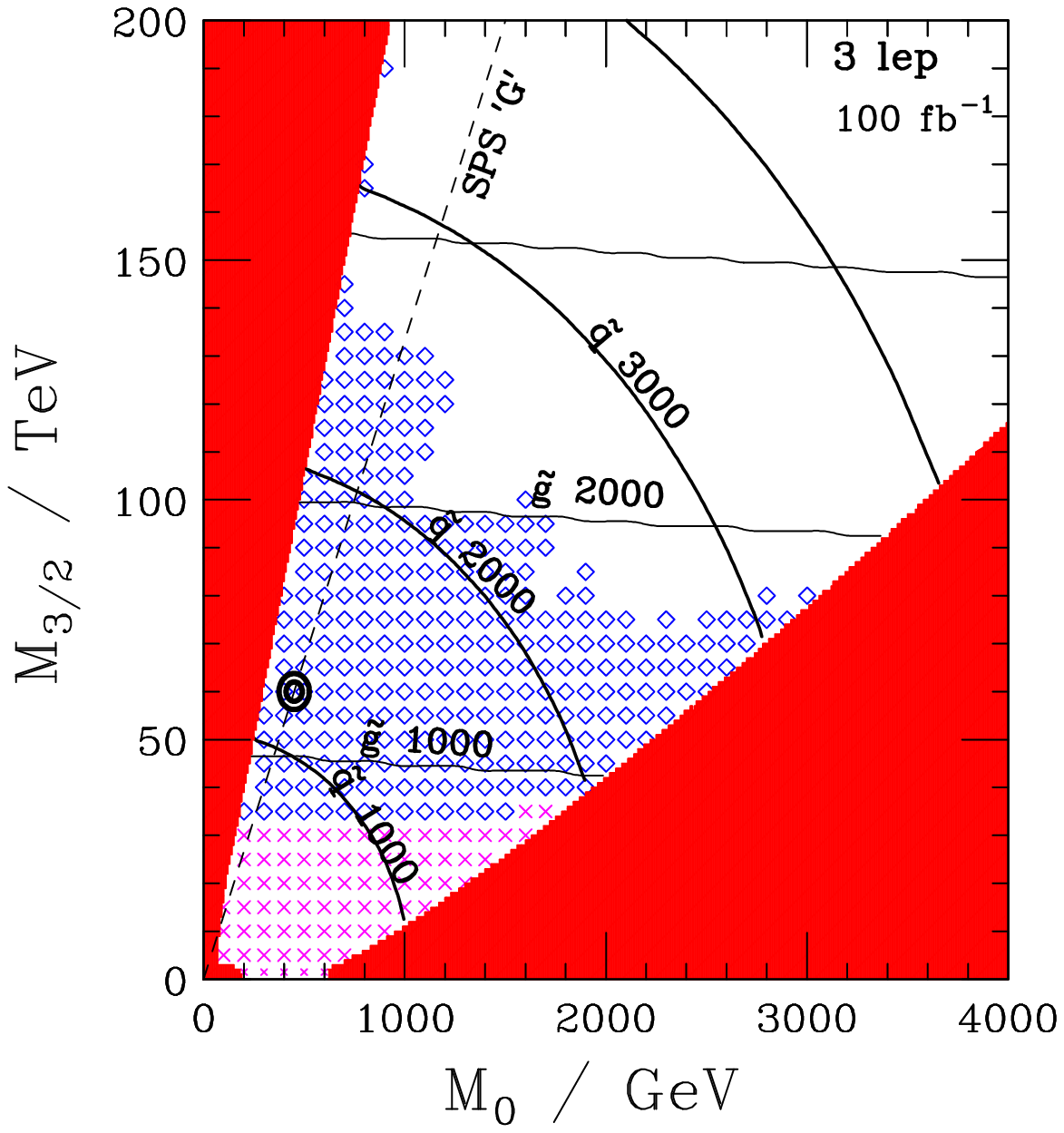, height=7.6cm,
      width=7cm, angle=90}\\
      \hspace*{0.5cm}{\bf (a)}
    \end{center}
  \end{minipage}\hfill
  \begin{minipage}[b]{.48\linewidth}
    \begin{center}
     \epsfig{file=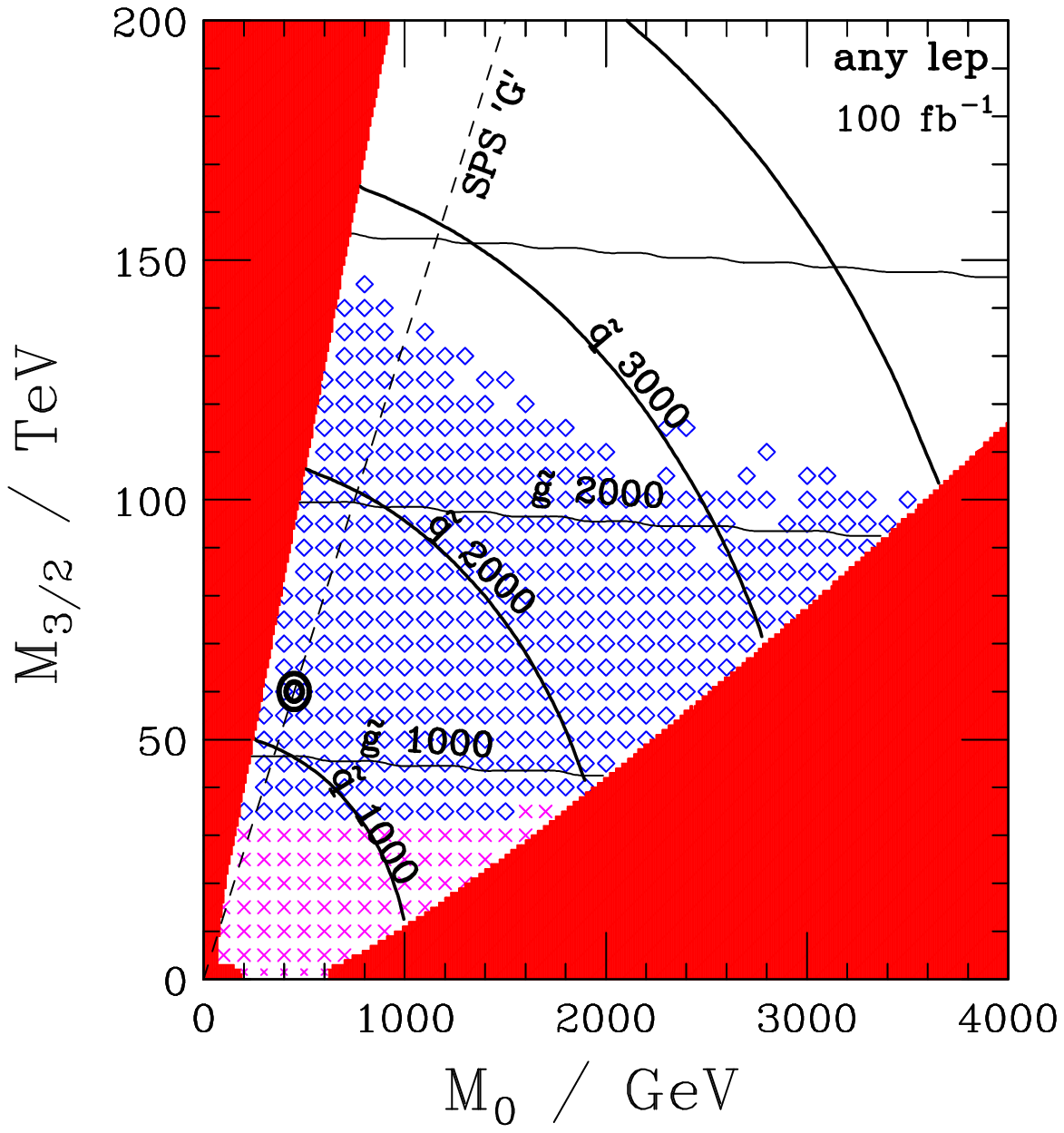, height=7.6cm,
       width=7cm, angle=90}\\
       \hspace*{0.5cm}{\bf (b)}
    \end{center}
  \end{minipage}\hfill
\caption{
As \figref{AMSB:Lep} but for the {\bf (a)} the three lepton channel
and {\bf (b)} the inclusive (`missing pt') channel.
\label{AMSB:ThPt} }}
\renewcommand{\arraystretch}{1.2}
\TABULAR[btp]{|l|c|c|c|c|c|c|c|c|c|c|}{ \hline
Analysis & $\etmiss$ & $p_{T(J1)}$ & $p_{T(J2)}$ & $\sum p_{T}$ & $N_\mathrm{jet}$ & $S_\mathrm{T}$
& $\Delta\phi_{(J1)}$ & $p_{T(\ell 1)}$ & $\Delta\phi_{(\ell 1)}$ \\
\hline
{\bf ptmiss} &  600 & 100 &  200 & 1000 & 9 & 0.2 & 0.3 & N/A & N/A \\ 
{\bf 0 Lep}  &  800 & 150 &  200 & 1000 & 10 & 0.1  & 0 & N/A & N/A\\
{\bf 1 Lep}  &  500 & 200 &  200 & 200  & 10 & 0.6  & 0 & 10 & 0 \\ 
{\bf 2 Lep OS}  & 800 & 100 &  150 & 200  & 9 & 0.7 & 0 & 10 & 0 \\
{\bf 2 Lep SS}  & 500 & 100 &  600 & 200  & 2 & 0.1 & 0 & 10 & 0\\
{\bf 3 Lep}   &  400 &  400 &  200 & 200  & 9 & 0   & 0 & 10 & 0\\ \hline
}{The value of the cuts which maximised the significance ($S/\sqrt{B}$)
for each of the analyses at the Snowmass mAMSB Point {\bf SPS~9} for 
$\int\mathcal{L}=100$~\invfb.
The $M_T$ cut applied only to the single lepton channel and
was kept at 100~GeV.\label{AMSB:SNOCUTS}}

In general the best reach is obtained in the single-lepton and the inclusive channel.
The same-sign dilepton signal is competitive at low \mnought,
when $m_{\tilde{\ell}}<m_\ntltwo$ -- a region which includes the model line {\bf SPS~G} -- 
but this sensitivity decreases as \mnought\ and $m_{\tilde{\ell}}$ increase.
Points with large \mnought\ have a spectrum qualitatively similar to that 
in \figref{AMSB:PDXSPEC}, with heavy squarks and sleptons.
Such points can still produce leptons, mostly from heavy chargino or neutralino decays.
Another source of leptons is from gluino decays to higgsinos 
along with a quark and anti-quark from the third generation,
followed by leptonic $t$ or $b$ decay.

\TABULAR[btp]{|l||c|c|c||c|c|c|}{ \hline
 & \multicolumn{3}{c||}{\bf SPS 9}&\multicolumn{3}{c|}{\bf Point d'Aix }\\ \hline
Analysis & $S$	& $B$	& $S/\sqrt{B}$ & $S$	& $B$	& $S/\sqrt{B}$ \\ \hline\hline
{\bf ptmiss}   & 3300 & 151  & 270 & 2650 & 0.25 & 5200 \\
{\bf 0 Lep }   & 950  & 21   & 206 & 7800 & 4.46 & 3700 \\
{\bf 1 Lep}    & 151  &  0.00005 & 21000  & 930 & 0.00005 & $10^5$ \\
{\bf 2 Lep OS} & 15  &  0.00005  & 2100  & 62 & 0.00005 & 8800 \\
{\bf 2 Lep SS} & 13   &  0.0005  & 1900 &  67 & 0.0048 & 970 \\
{\bf 3 Lep}    & 206  & 0.0001   & 20400  & 19 & 0.00010 & 1900 \\ 
\hline
}{The expected number of signal ($S$), and background ($B$) 
events and the significance ($S/\sqrt{B}$) for
for each of the analyses, at the Snowmass mAMSB Point {\bf SPS~9}
and the Point d'Aix,
for $\int\mathcal{L}=100$~\invfb.
For all six analyses and for both points the backgrounds are dominated by high 
\pt\ vector boson production in association with jets.
\label{AMSB:SNOSIG} }

Along the Snowmass model line `G' ($\mnought=0.0075\times\mthreehalfs$) the LHC will
be able to measure discrepancies from the Standard Model up to
\mnought=1050~GeV, \mthreehalfs=140~TeV, at which point the gluino mass is 2.76~TeV 
and  $m_\ssul=2.70$~TeV.
The Snowmass point {\bf SPS~9} lies well within the discovery region for
all of the six different analyses.
The optimised cuts and the resultant significance are shown in \tabref{AMSB:SNOCUTS}
and \ref{AMSB:SNOSIG}.
The high significance implies that it will be possible to extract
further information from the data, for example to determine sparticle masses.


\section{Distinguishing Wino-like LSPs}
\label{AMSB:DIST:ANAL}

\EPSFIGURE[tp]{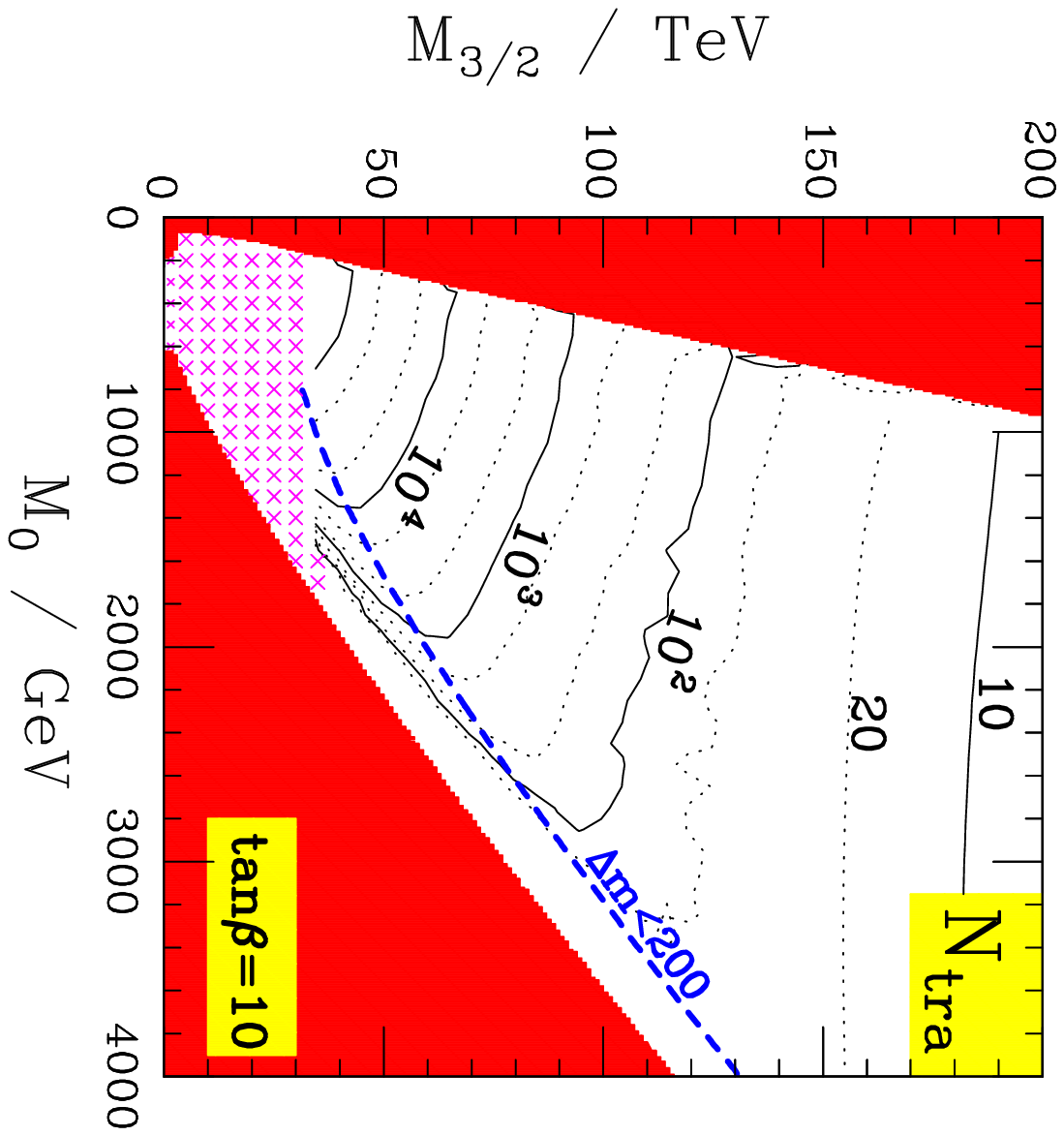, width=9cm, height=10cm, angle=90}{
The number of $\chgonepm\to\ntlone$ decays expected within the central
region of the detector ($|\eta|<2$) with transverse decay vertices between
100~mm and 800~mm from the interaction point for integrated luminosity of 
100~\invfb.
The initial track from the chargino is required have $\pt>10$~GeV.
All events are required to have passed the preselection cuts
(\tabref{AMSB:CUTVALS}) which ensure that the event will be triggered.
The 10~event and 20~event contours have been smoothed for clarity.
$\DeltaMChi$ is less than 200~MeV above and to the left of the 
dashed blue line on the bottom right-hand side of the plot. 
\label{AMSB:TRACK}}

Most of the mass-sum rules\cite{Huitu:2002fg} for mAMSB involve terms like
$M^2_\ssdl - M^2_\ssul$ which would be extremely difficult to determine
experimentally. The exception is the accidental near-degeneracy of the
\ssel\ and \sser, which might be observable at some points, 
as was noted in \secref{AMSB:BENCH}.
However points with large \mnought\ have relatively heavy sleptons
which would have a small production cross-section at a hadron collider.
The Wino-like LSP signature of AMSB, and the resultant near-degeneracy
of the \chgone\ and \ntlone\ will be the robust `smoking gun' for 
anomaly mediation, and is applicable beyond the minimal model.

The LHC, like LEP\cite{Heister:2002mn}, 
or the Tevatron\cite{Feng:1999fu,Gunion:1999jr}
or a future linear $e^+ e^-$ collider\cite{Gunion:2001fu}
can use a variety of analyses in the search for
long-lived charginos, each tailored to a
different regime in \DeltaMChi, 
according to the \chgone\ lifetime.
The various regimes are ordered by decreasing \chgone\ lifetime below:
\begin{list}{\bfseries\upshape (\arabic{anlctr})}
{\newcounter{anlctr}
}\addtocounter{anlctr}{1}
\item {$\DeltaMChi<m_{\pi^+}$.
If the pion decay mode is not available, the \chgone\ will be long lived ($c\tau\gsim 1$~m)
and can leave a track through the muon chambers.
Analysis of this type of signature was performed in \cite{Ambrosanio:2000ik}
for a GMSB model with long-lived $\tilde{\tau_1}$ NLSP, and in \cite{Allanach:2001sd} for 
an intermediate scale model with heavy stable leptons. The mass 
of the chargino can be measured using the muon detector as a time-of-flight system.
Additional discrimination can be obtained by considering $\frac{dE}{dx}$ information
from the transition radiation tracker.
Higher-mass sparticles can then be reconstructed by forming the
invariant mass of the stable particle with jets and leptons.
Provided the \chgone\ lifetime, although long, is short enough that a 
reasonable number of \chgone s decay within the inner detector,
the mass difference \DeltaMChi\ could also be determined from
that fraction using the techniques described below.
}\addtocounter{anlctr}{1}
\item {$m_{\pi^+}<\DeltaMChi\lsim 200$~MeV. This is the regime in which 
high \pt\ chargino tracks often decay within the body of the inner tracker to 
soft pions or leptons along with large amounts of missing transverse energy.
The details of the detector resolution and track reconstruction algorithm
are beyond the scope of this study, but such tracks 
should provide a striking signature provided they occur in sufficient numbers.
The number of \chgone s which would be produced at the LHC 
and decay within a fiducial volume in the active material of the ATLAS 
tracker is plotted in \figref{AMSB:TRACK} as a function of \mnought\ and \mthreehalfs.

One-prong chargino decays such as $\chgone\to\ntlone\pi^+$ and $\chgone\to\ntlone e^+\nu_e$
will produce `kinks' at the point where the charged SM particle is softly emitted.
In some cases the SM particle will go undetected and the high \pt\ chargino track
will seem to disappear. Such decays, producing track stubs,
could also be detected with a dedicated off-line analysis.

Even though rather strict triggering requirements have been made,
there are at least ten such events for 100~\invfb\ 
over almost the entire plotted parameter-space provided 
$\mthreehalfs\lsim185$~TeV.
This signature should be essentially free of physics background,
so `detector' backgrounds will dominate.
Since ATLAS should have three barrel pixel layers with $r<122$~mm
with noise occupancy of less than $10^{-5}$, and average physics occupancy of the order of $10^{-4}$
the instrumental backgrounds should also be under control\cite{pixtdr,Einsweiler}.
Such a search could therefore achieve a large reach and would be sufficient
to identify a near-degenerate LSP model.
}\addtocounter{anlctr}{1}
\item {200~MeV~$\lsim\DeltaMChi\lsim$a~few~GeV.
Chargino decays which occur before the first tracking layer can still be useful
provided that the soft track from the SM particle can be found, using \eg\ 
electron identification, track impact parameter, isolation from other tracks,
and the direction of the \Ptmiss\ 2-vector.
}
\end{list}

The maximum value of \DeltaMChi\ in minimal AMSB with $\tan\beta=10$ is less than about 200~MeV 
except close to the region of no electroweak symmetry breaking.
This leads to a decay distance $c\tau \gsim$ a~few centimeters, and corresponds to 
category~(2) above.
However, other models such as 0-II string models\cite{Brignole:1997dp,Chen:1997ap,Chen:1999yf}
have Wino-like \chgone\ and \ntlone\ with \DeltaMChi\ typically of the order of a GeV.
Such models fall into category~(3), which is potentially the most difficult at a 
hadron collider because the short-lived \chgone s will not be directly observable,
and their softly emitted SM daughters suffer from a
large background from other low-momentum tracks.
In \secref{AMSB:DIST:POINTS} we define a range of model points
which allows us to study this regime of shorter lifetimes.
Some of the methods which can enable the LHC to probe this region
are explored in \secref{AMSB:DIST:TRACK}.

\subsection{Definition of points}
\label{AMSB:DIST:POINTS}

\renewcommand{\arraystretch}{1.0}
The tree-level neutralino mass matrix is:
\begin{equation}\left( \begin{array}{cccc}
M_1		& 0	& -m_Z c_\beta s_W 	& m_Z s_\beta s_W \\
0		& M_2	& m_Z c_\beta c_W	& - m_Z s_\beta c_W \\
-m_Z c_\beta s_W& m_Z c_\beta c_W	& 0	& -\mu \\ 
m_Z s_\beta s_W & m_Z s_\beta c_W 	& -\mu	& 0 \\
\label{AMSB:NTLMATRIX}\end{array}\right) \end{equation}
and the tree-level chargino mass matrix is:
\begin{equation}\left( \begin{array}{cc}
M_2 & \sqrt{2}m_Ws_\beta \\ \sqrt{2}m_W c_\beta & \mu \\ \end{array}\right) \label{AMSB:CHGMATRIX}\ ,\end{equation}
where $s_W\equiv\sin\theta_W$, $c_W\equiv\cos\theta_W$, $s_\beta\equiv\sin\beta$, 
$c_\beta\equiv\cos\beta$ and we use the conventions of \cite{Haber:1985rc}.
The mass difference $m_\chgone-m_\chgone\equiv\DeltaMChi$ is highly suppressed at tree-level,
so the leading 1-loop correction can be important.
It takes the form\cite{Cheng:1998hc}:
\begin{equation}
\DeltaMChi^\mathrm{(1-loop)} = \frac{\alpha_2 M_2}{4\pi}
\left[ F \left( \frac{m_W}{M_2} \right) - c_W^2 F \left( \frac{m_Z}{M_2} \right) + 5 s_W^2 \right]\ ,
\label{AMSB:ONELOOP}\end{equation}
where $F(a)\equiv\int_0^1 dx (2+2x) \log[x^2+(1-x)a^2]$, and is included in \isajet. 

We are interested in the collider phenomenology of long-lived charginos
with various lifetimes. From \eqref{AMSB:NTLMATRIX} and \eqref{AMSB:CHGMATRIX},
the leading tree-level mass difference
term\cite{Cheng:1998hc} $\DeltaMChi\propto 1/\mu^2$.
In order to explore different regimes in \DeltaMChi,
we define new points based on {\bf SPS~9} and 
another minimal anomaly-mediated point with $\mnought=500$~GeV, $\mthreehalfs=36$~TeV, 
$\mu>0$ and $\tan\beta=10$. We then decreased the value of the 
$\mu$ parameter at the electroweak scale to the values shown in \tabref{AMSB:MUPOINTS}.

\afterpage{\clearpage}

In order to maintain consistent electroweak symmetry breaking when decreasing $\mu$,
the Higgs soft mass parameters, $m^2_{H_1}$ and $m^2_{H_2}$, must allowed to vary,
as can be seen from tree-level equation, 
\begin{equation} \mu^2=\frac{m^2_{H_1}-m^2_{H_2} \tan^2\beta}{\tan^2\beta-1}-\half m_Z^2 \nonumber\ .\end{equation}
The points with adjusted $\mu$ are then no longer in the minimal anomaly-mediated scenario,
since $m^2_{H_1}$ and $m^2_{H_2}$ are not those which would be predicted from 
the mAMSB parameters \mnought\ and \mthreehalfs.
Decreasing $\mu$  has the side-effect of decreasing the masses of the higgsinos,
and can change the phenomenology somewhat -- 
for example by opening chains such as $\squark\to\chgtwo q\to \chgone q Z^0$.
When followed by leptonic $Z^0$ decay these chains could, in principle, be used
to further constrain the dynamics. However the only requirement 
we make for our analysis is that the decay $\squark\to\chgone q$
occurs, producing highly boosted $\chgonepm$ pairs in association with jets.

\subsection{Identifying chargino decay products}
\label{AMSB:DIST:TRACK}

\renewcommand{\arraystretch}{1.0}
\TABULAR[Bt]{|l|c|c|c|c|c|c|}{\hline  &   & $m_\chgone$ & \DeltaMChi & $c\tau$  &  & \\
\raisebox{1.5ex}[0pt]{Point}    & \raisebox{1.5ex}[0pt]{$\mu$} &   (GeV)     &   (MeV)    &  
 (mm)    & \raisebox{1.5ex}[0pt]{$\chgone\to\ntlone e^+\nu_e$ } & 
\raisebox{1.5ex}[0pt]{$\chgone\to\ntlone \mu^+\nu_\mu$ }\\\hline
{\bf SPS~9} & 864  & 171 & 164 & 56 & 2.0 \% & 0.2 \%\\
{\bf SPS-300} & 300   & 165 & 886 & 0.11 & 17.0 \% & 15.9 \% \\
{\bf SPS-250} & 250   & 159 & 1798 & 0.004 & 21.9 \% & 21.5 \% \\\hline
{\bf A}       & 533   & 107 & 181  & 34     & 2.0 \%  & 0.3 \% \\
{\bf A-250}   & 250   & 101  & 766  & 0.20   & 15.4 \% & 13.9 \% \\
{\bf A-200}   & 200   & 97  & 1603 & 0.007 & 22.5 \% & 22.2 \% \\ \hline
}{The mass of the lightest chargino, its mass difference with the LSP, 
decay length and leptonic branching ratios for six points.
{\bf SPS~9} is the Snowmass mAMSB point, from which we define {\bf SPS~300}, 
and {\bf SPS~250} by adjusting $\mu$ at the electroweak scale to produce different $\DeltaMChi$.
Point {\bf A} is the mAMSB point with $\mnought=500$~GeV, $\mthreehalfs=36$~TeV, 
$\mu>0$ and $\tan\beta=10$, from which $\mu$ is adjusted at the electroweak scale to produce {\bf A-250} and {\bf A-200}.
\label{AMSB:MUPOINTS} } \renewcommand{\arraystretch}{1.5}

In this section we demonstrate how the low-\pt\ tracks from the charged SM
daughters of \chgone\ decays can be identified and used to constrain \DeltaMChi.
To do so we must distinguish these tracks from the large number of other low-\pt\ 
tracks from the particles generated in the proton remnant interactions,
and from any pile-up events.
To avoid having to deal with the contamination from multiple pile-up,
in this section we simulate the initial three years of `low luminosity' 
$(10^{34}$~$\mathrm{cm}^{-2} \mathrm{s}^{-1})$ running of the LHC, which is expected to
provide 30~\invfb\ of integrated luminosity with an average of one inelastic collision 
per bunch-crossing.
The loss of resolution from pile-up is therefore not simulated in this section.

\afterpage{\clearpage}

\EPSFIGURE[tp]{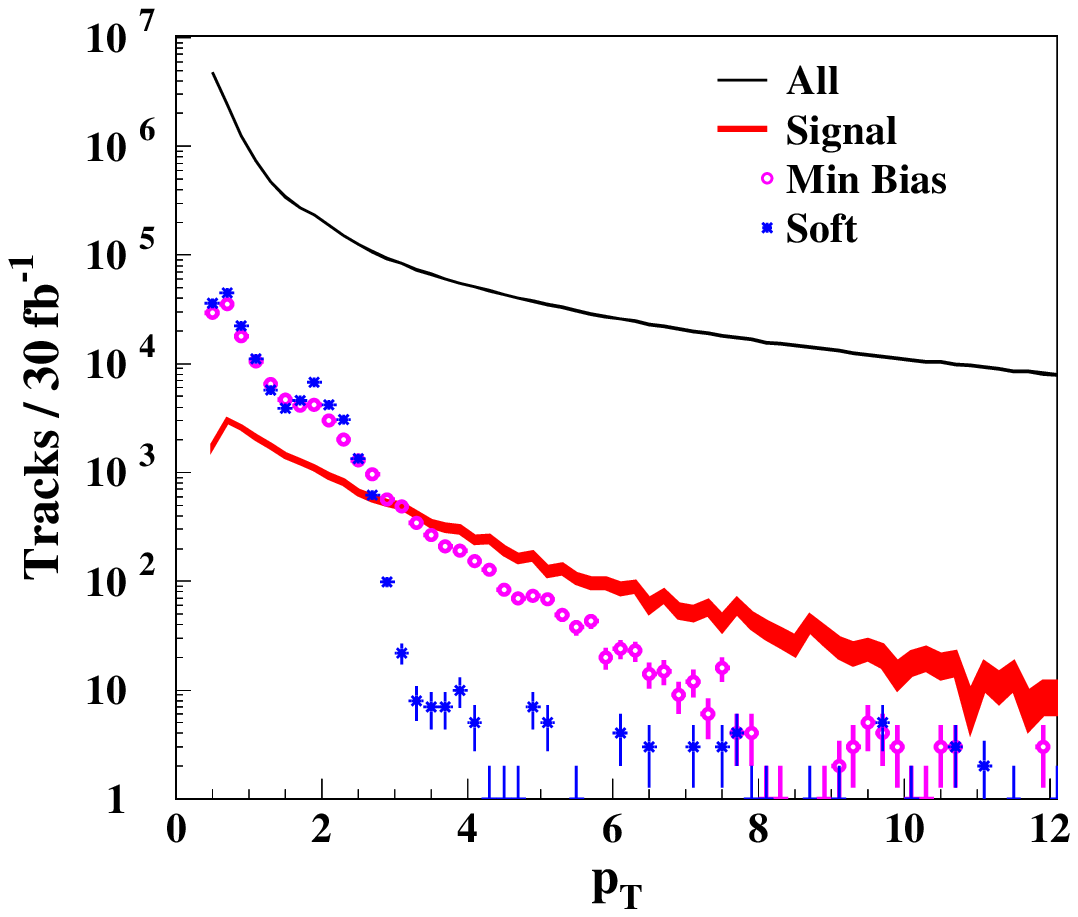, width=10cm, height=6.5cm}{
The transverse momentum distribution of reconstructed charged tracks for the point 
{\bf SPS-250} (defined in \tabref{AMSB:MUPOINTS}) after selecting events with $\etmiss>200$~GeV.
The thin black line shows all tracks; 
the distribution for particles originated from $\chgonepm$ decays (the signal)
is highlighted by the thick red line, where the thickness shows the statistical 
uncertainty; 
those from the minimum bias event and from the soft underlying event
are denoted by circles and asterisks respectively.
\label{AMSB:DIST:SNOC:PT} }

\EPSFIGURE[tbh]{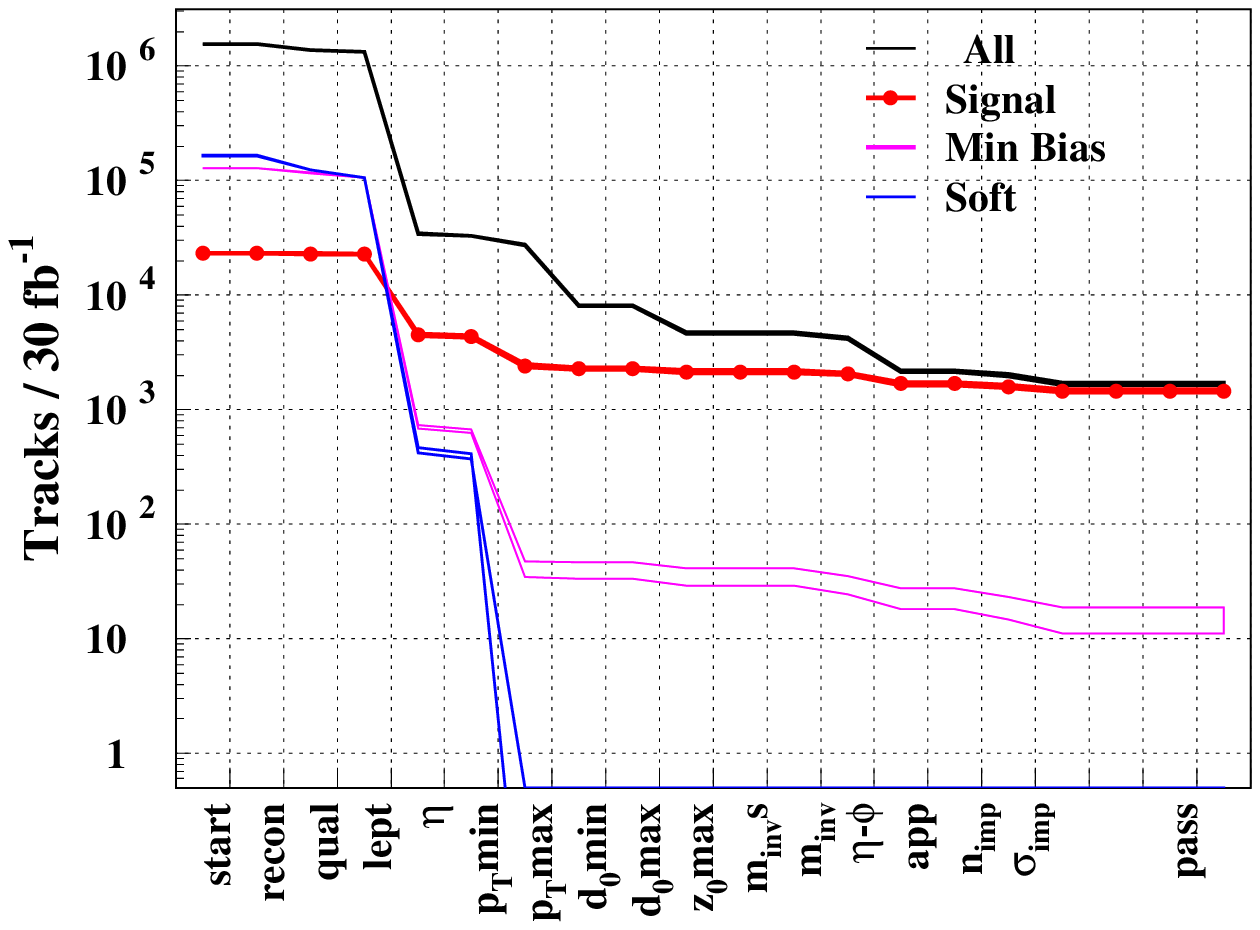, height=7.5cm, width=10cm}{
The number of tracks surviving subsequent cuts,
after a preselection cut $\etmiss>200$~GeV for the point {\bf SPS-250} 
(defined in \tabref{AMSB:MUPOINTS}).
The signal tracks (from \chgonepm\ decays) are also plotted separately, 
as are tracks from the \herwig\ soft underlying event and the additional minimum bias QCD event.
Not all cuts are used in any particular analysis.
For this point the most effective cuts are those on the $p_T$ of the track, and on 
selecting leptons (`lept'), while further cuts on the transverse impact parameter ($d_0$)
and isolation ($\eta-\phi$) improve the selection.
\label{AMSB:DIST:SNOC:WHYFAIL} }

The \herwig\ soft underlying event model, based on the UA5 collaboration\cite{Alner:1987is}
$p\bar{p}$ Monte-Carlo, is known to underestimate the number of tracks with $\pt>1$~GeV\cite{Tano:2001zc},
because it does not model the semi-hard physics of multiple parton interactions.
To ensure that our analysis was insensitive to contamination from such tracks, 
a sample of 5000 inclusive QCD events with ${\tt PTMIN}=4.5$~GeV was generated.
We refer to these as `minimum bias events' because they simulate 
(better than the default \herwig\ soft minimum bias option)
the expected minimum bias distribution of tracks, including a tail at 
higher \pt\ due to the onset of hard scattering. 
The particles from this minimum bias event were added to the SUSY event
in addition to \herwig's usual underlying event.

\TABULAR[bhtp]{|l|c|c|c|}{\hline
&Points & {\bf SPS-300} , {\bf A-250} & {\bf SPS-250}, {\bf A-200} \\\hline\hline
Event:&$\etmiss^\amin	$ & 500	& 500 \\
&$p_{T(J1)}^\amin$	& 400	&  400\\
&$S_T^\amin$	& 0.05 	& 0.05 \\
\hline
Track:&$\pt^\amin$  & 2.0	& 2.0\\
&$\pt^\amax$  & 5	& 10\\
&$d_0^\amin$ & 0.03	& 0\\
&$d_0^\amax$ & 0.3	& 0.1\\
&$|\eta|^\amax$ & 1.5 & 2\\
&$\Delta R_{ij}^\amin$ & 0.45	& 0.2\\
&$M_{ij}^\amin$ & 0.9	& 0.4\\
&$R_\mathrm{imp}$ & 0.4	& 0.4\\
&$\sigma_\mathrm{imp}$ 	& 3.0 	& 3.0\\
&Particle & any		& $\ell\in e, \mu$\\
\hline}{
The cuts applied to events and to tracks for the different points in \tabref{AMSB:MUPOINTS}.
The missing transverse energy (\etmiss), leading jet transverse momentum ($p_{T(J1)}$), 
and transverse sphericity ($S_T$) cuts were applied to the whole event, while the other cuts 
were applied track by track. The track transverse momentum was required to lie in the range $\pt^\amin \to \pt^\amax$;
its transverse impact parameter in the range $d_0^\amin \to d_0^\amax$;
and the absolute value of its pseudorapidity was required to be less than $|\eta|^\amax$.
The other cuts are described in the text.
Energy, momentum and mass units are GeV; the impact parameter, $d_0$, is measured in mm. 
\label{AMSB:DIST:CUTS}}

An initial selection requiring a high \pt\ jet and large missing transverse energy was made
to trigger the event, reduce the SM background, and select those SUSY events in which the 
neutralino was highly boosted.
In this section we are dealing with the short-lifetime regime ($c\tau<1$~mm),
so this boost will almost certainly {\em not} be sufficient to allow the chargino to live long
enough to decay in the body of the detector. 
However a boosted chargino means that even particles which are very softly 
emitted in the chargino rest frame will have a transverse momentum
\begin{equation}p_{Tx}\sim \frac{p_{T\ntlone}}{m_\ntlone} \times m_x
\nonumber\end{equation}
as measured in the detector, where $x$ is the charged SM daughter particle
and $p_{T\ntlone}$ is of the order of the missing transverse momentum.
This extra \pt\ assists the reconstruction of the SM daughters of charginos and
helps distinguish them from even softer particles coming from the underlying event (\figref{AMSB:DIST:SNOC:PT}).

\FIGURE[tp]{
  \begin{minipage}[b]{.48\linewidth}
    \begin{center}
     \epsfig{file=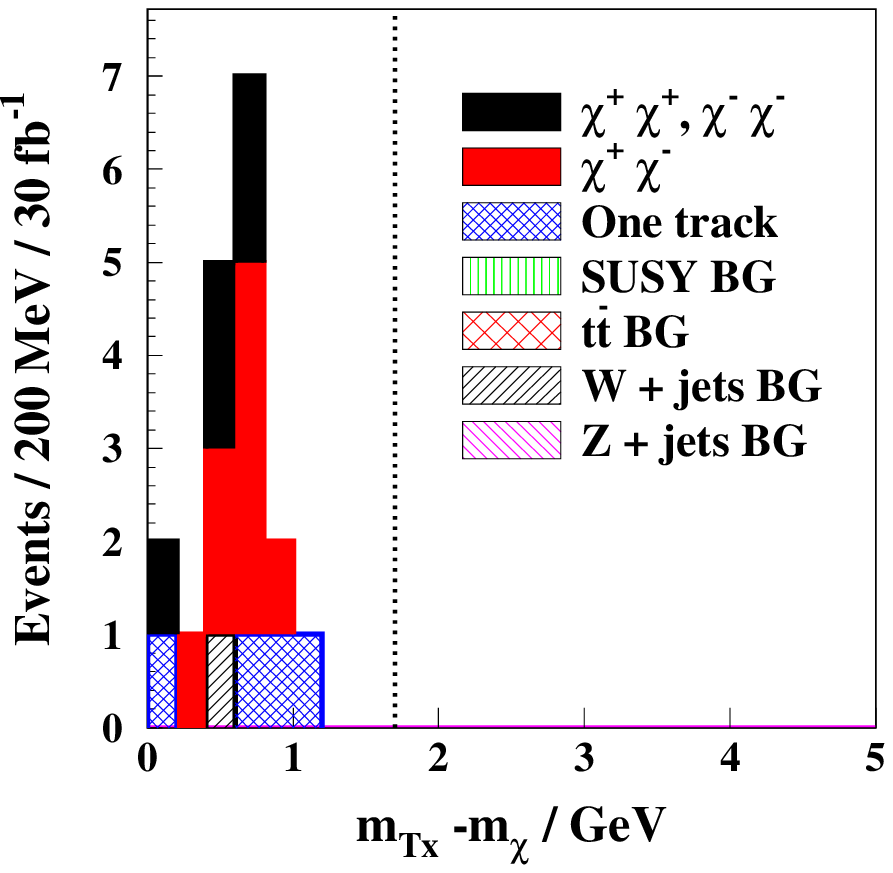, height=7cm}\\
      \hspace*{0.2cm}{\bf (a)}
    \end{center}
  \end{minipage}\hfill
  \begin{minipage}[b]{.48\linewidth}
    \begin{center}
     \epsfig{file=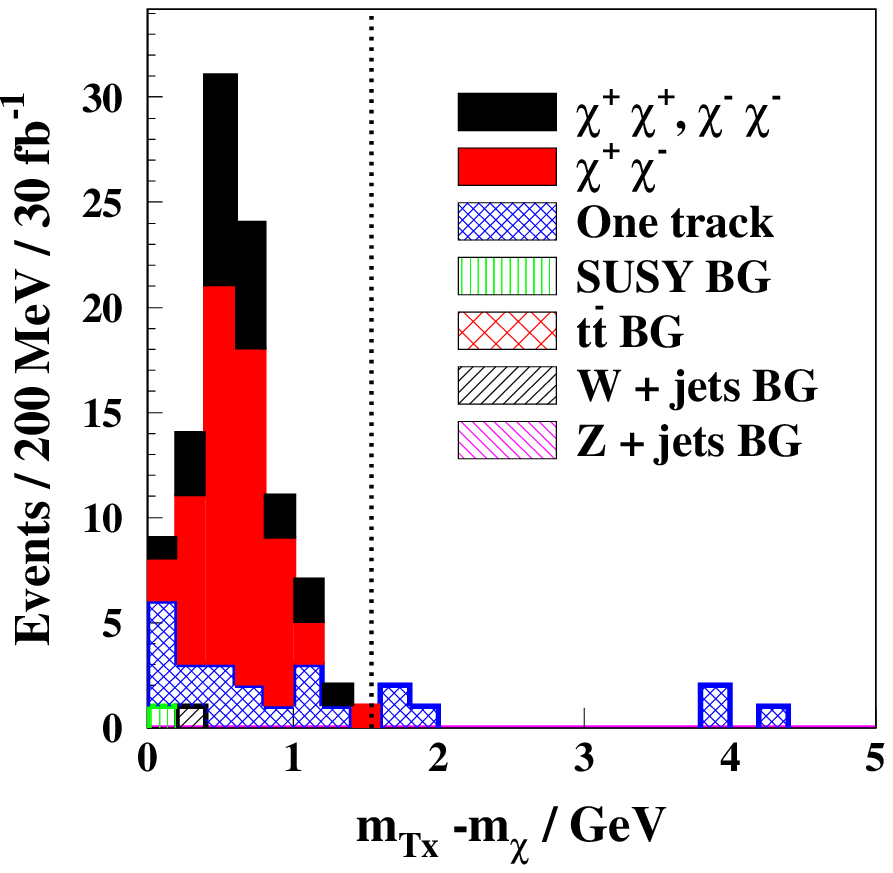,  height=7cm}\\
       \hspace*{0.2cm}{\bf (b)}
    \end{center}
  \end{minipage}\hfill
\caption{
The $m_{T4}-m_\ntlone$ distribution for the points {\bf (a)} {\bf SPS-250}, and {\bf (b)} {\bf A-200}
(defined in \tabref{AMSB:MUPOINTS}).
These two points have similar values of \DeltaMChi, and the same cuts were applied to both.
Only events in which exactly two leptons passed the selection cuts are plotted.
The signal tracks are those from \chgone\ and \chgonem\ decays.
Events where only one of the two tracks has been successfully identified are plotted separately.
The vertical dotted line indicates $\DeltaMChi=m_\chgone-m_\ntlone$.
\label{AMSB:DIST:SNOC:MTX}
}}

\FIGURE[tp]{
  \begin{minipage}[b]{.48\linewidth}
    \begin{center}
     \epsfig{file=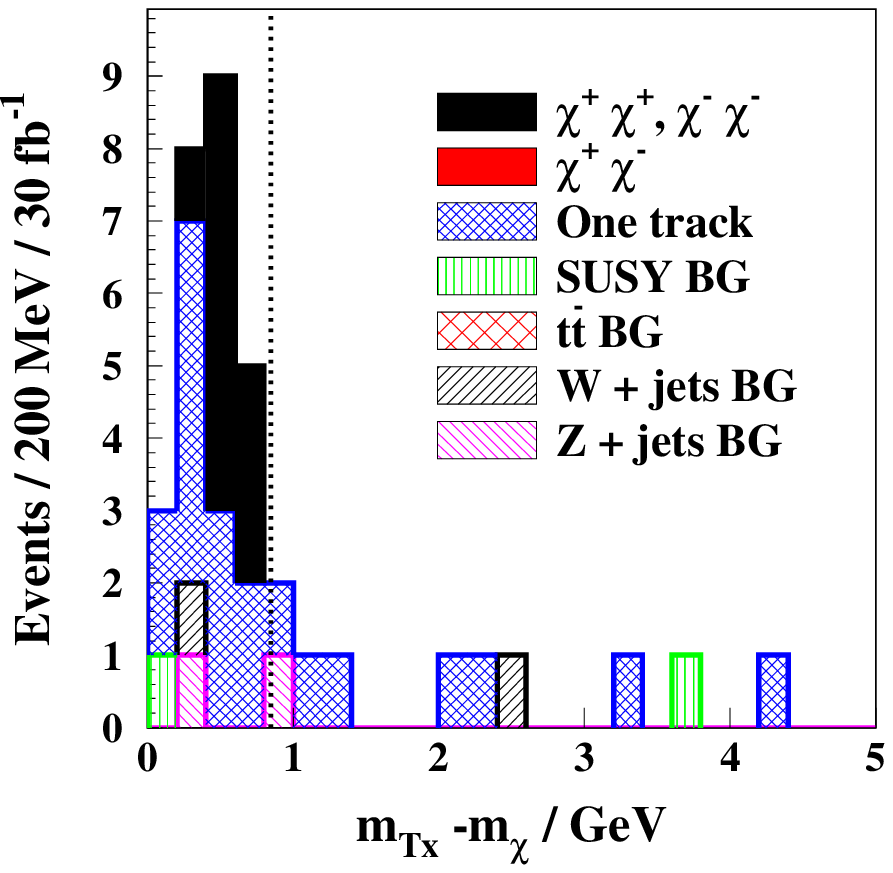, height=7cm}\\
      \hspace*{0.5cm}{\bf (a)}
    \end{center}
  \end{minipage}\hfill
  \begin{minipage}[b]{.48\linewidth}
    \begin{center}
     \epsfig{file=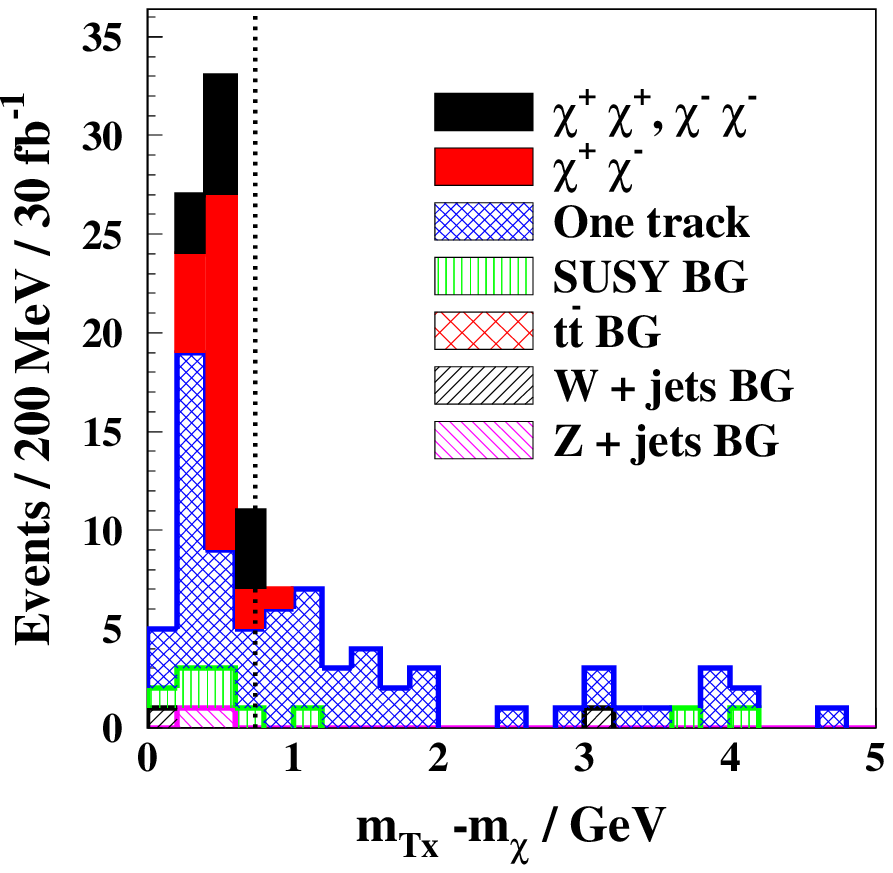,  height=7cm}\\
       \hspace*{0.5cm}{\bf (b)}
    \end{center}
  \end{minipage}\hfill
\caption{
The $\mttwo-m_\ntlone$ distribution for {\bf (a)} the point {\bf SPS-300}, and {\bf (b)} the point {\bf A-250}.
These two points have similar values of \DeltaMChi, and the same cuts were applied to both.
Only events in which exactly two particles (of any type) passed the selection cuts are plotted.
The peak is closer than in \figref{AMSB:DIST:SNOC:MTX} to the upper kinematic limit
at $\mttwo-m_\ntlone=\DeltaMChi$ (dotted line) since there are only two missing particles -- the two neutralinos. 
\label{AMSB:DIST:SNOB:MTX}}}

The tracks from SM daughters of chargino decays have other properties which can help in their identification. 
For larger \DeltaMChi\ the \chgone\ branching ratio to each of muon or electron and associated neutrino 
can be up to nearly twenty percent (\figref{AMSB:CTAUBR}b). 
In the ATLAS experiment electrons and muons will be distinguished from other tracks by transistion 
radiation and by penetration respectively (see \appref{EIDENT}).

Much of the background is associated with heavy-quark decay and can be removed by applying isolation cuts 
(\tabref{AMSB:DIST:CUTS}) which require that for any track, $i$, to be a candidate:
\begin{itemize}
\item{no other track is found with $\Delta R_{ij}=\sqrt{(\Delta\eta_{ij})^2+(\Delta\phi_{ij})^2}$ 
less than some value $\Delta R_{ij}^\amin$;}
\item{the invariant mass of the track with another track, $j$, is greater than $M_{ij}^\amin$ for all $j$, 
where it is assumed that $m_i=m_j=m_{\pi^+}$;}
\item{no other track with $R_{ij}<R_\mathrm{imp}$ has $d_0^{(j)}/\sigma(d_0^{(j)}) > 3$, 
where $d_0^{(j)}$ is the transverse impact parameter, and $\sigma(d_0^{(j)})$ is its uncertainty;}
\item{$\sum_j d_0^{(j)}/\sigma(d_0^{(j)})<\sigma_\mathrm{imp}$ where the sum is over all $j\ne i$ with $R_{ij}<R_\mathrm{imp}$.}
\end{itemize}

\afterpage{\clearpage}

\subsection{Results}
\label{AMSB:DMRESULTS}

We selected those events in which precisely two tracks satisfied the track cuts listed in
\tabref{AMSB:DIST:CUTS}. 
The numbers of tracks from chargino decays and from other sources which passed subsequent
cuts are shown in \figref{AMSB:DIST:SNOC:WHYFAIL} for the point {\bf SPS-250}.
We then considered the Cambridge \mtx\ variable which (as described in 
\appref{MTX}) remains sensitive to \DeltaMChi\ even when there are uncertainties in the
neutralino mass or the missing transverse momentum, providing 
an important handle on Wino-LSP physics at hadron colliders.
We assume that both the squark mass scale, $m_\squark$, and the $\ntlone$ mass have been previously determined
from the measurement of other kinematic edges as described in 
\cite{phystdr,Allanach:2000kt,Tovey:2002,Paige:1999ui},
and plot distributions of $\mtx-m_\ntlone$ which has the property: 
\begin{equation}
0\ < (\mtx-m_\ntlone)\ \leq\ (m_\chgone-m_\ntlone)\ \equiv\ \DeltaMChi.\end{equation}

For the two points with $\DeltaMChi\approx1.7$~GeV, we selected isolated leptons 
and plotted distributions of $m_{T4}-m_\ntlone$.
As can be seen in \figref{AMSB:DIST:SNOC:MTX},
{\bf A-200} has more signal events passing the cuts than {\bf SPS-250}, 
but the peak position is the same in both cases. 

As  \DeltaMChi\ decreases, the leptonic branching ratios decrease (see \figref{AMSB:CTAUBR}b)
and the detector's ability to distinguish the lower \pt\ leptons from hadrons is 
diminished (see \appref{EIDENT}). For this reason, at the two points 
with $\DeltaMChi\approx800$~MeV we select events with two tracks
of {\em any} type satisfying the cuts in the third column of \tabref{AMSB:DIST:CUTS}.
Most of these tracks will be pions because of the large branching ratio 
to $\pi^\pm$ or $\pi^\pm\pi^0$.
The $\mttwo-m_\ntlone$ distributions for these two points are plotted in 
\figref{AMSB:DIST:SNOB:MTX}.

\EPSFIGURE[t]{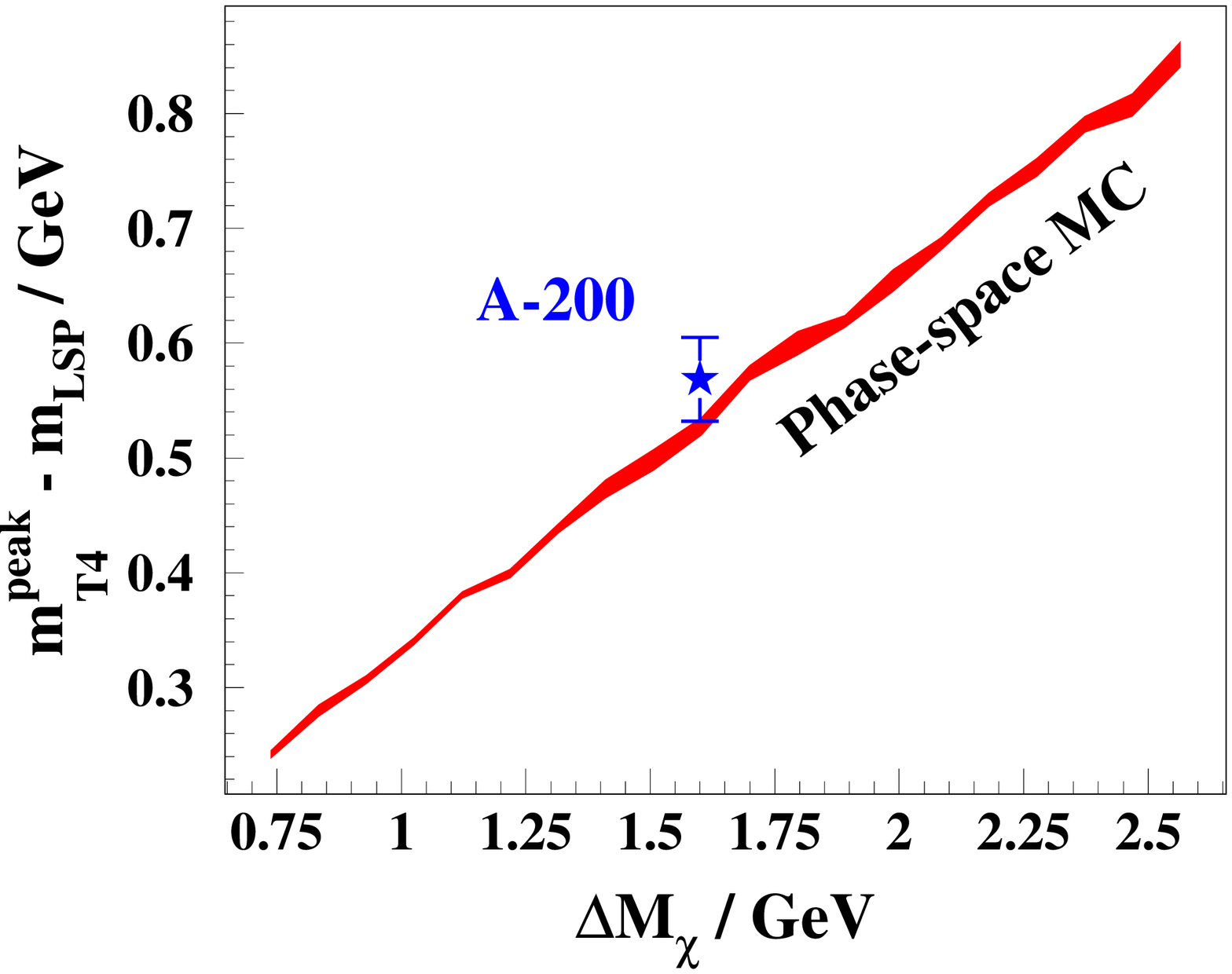, width=10cm}{
The diagonal line shows how the $m_{T4}-m_\ntlone$ peak value depends on \DeltaMChi.
The thickness of the line indicates the uncertainty 
in the peak position from a gaussian fit to the distribution 
combined in quadrature with a 10\% uncertainty in the LSP mass
and a 10 \% uncertainty in the squark mass scale.
The peak of the $m_{T4}-m_\ntlone$ distribution for the point 
{\bf A-200} is marked with a star
at its input value of \DeltaMChi; the error bar shows its uncertainty.
\label{AMSB:DIST:PEAKPLOT} }

For all four points the narrow peaks indicate that \DeltaMChi\ is of the order of 1~GeV. 
This confirms the Wino-like nature of the LSP, providing the `smoking gun' 
signature for anomaly mediation.

Each \mtx\ distribution depends principally on \DeltaMChi,
on the neutralino mass and the 
momentum distribution of the charginos. 
The latter depends largely on $m_\squark$ so if both $m_\squark$ and the lightest 
neutralino mass were already measured, then \DeltaMChi\ can be measured by 
fitting to each $\mtx-m_\ntlone$ distribution.

To demonstrate that $m_{T4}$ can indeed make a quantitative measurement of $\DeltaMChi$,
a phase-space Monte-Carlo program was used to generate
very simple `events' in which pairs of squarks 
decayed via the chain $\squark\to \chgone~q\to \ntlone~e~\nu_e~q$.
$m_{T4}-m_\ntlone$ distributions were produced for those events in which $\ptmiss>500$~GeV,
and the peak determined from a gaussian fit.
The correlation between the fitted peak positions and
the input values of \DeltaMChi\ is shown in \figref{AMSB:DIST:PEAKPLOT}.
The peak of the $m_{T4}-m_\ntlone$ distribution for point {\bf A-200}
(\figref{AMSB:DIST:SNOC:MTX}b), 
was likewise determined from a gaussian fit.
As can be seen from \figref{AMSB:DIST:PEAKPLOT}
the mass difference \DeltaMChi\ 
can be measured at that point with a statistical uncertainty of approximately 150~MeV (10\%).


\section{Other constraints}
\label{AMSB:OTHER}

\subsection{Cosmological relic density}
\FIGURE[p]{
  \begin{minipage}[b]{.48\linewidth}
    \begin{center}
     \epsfig{file=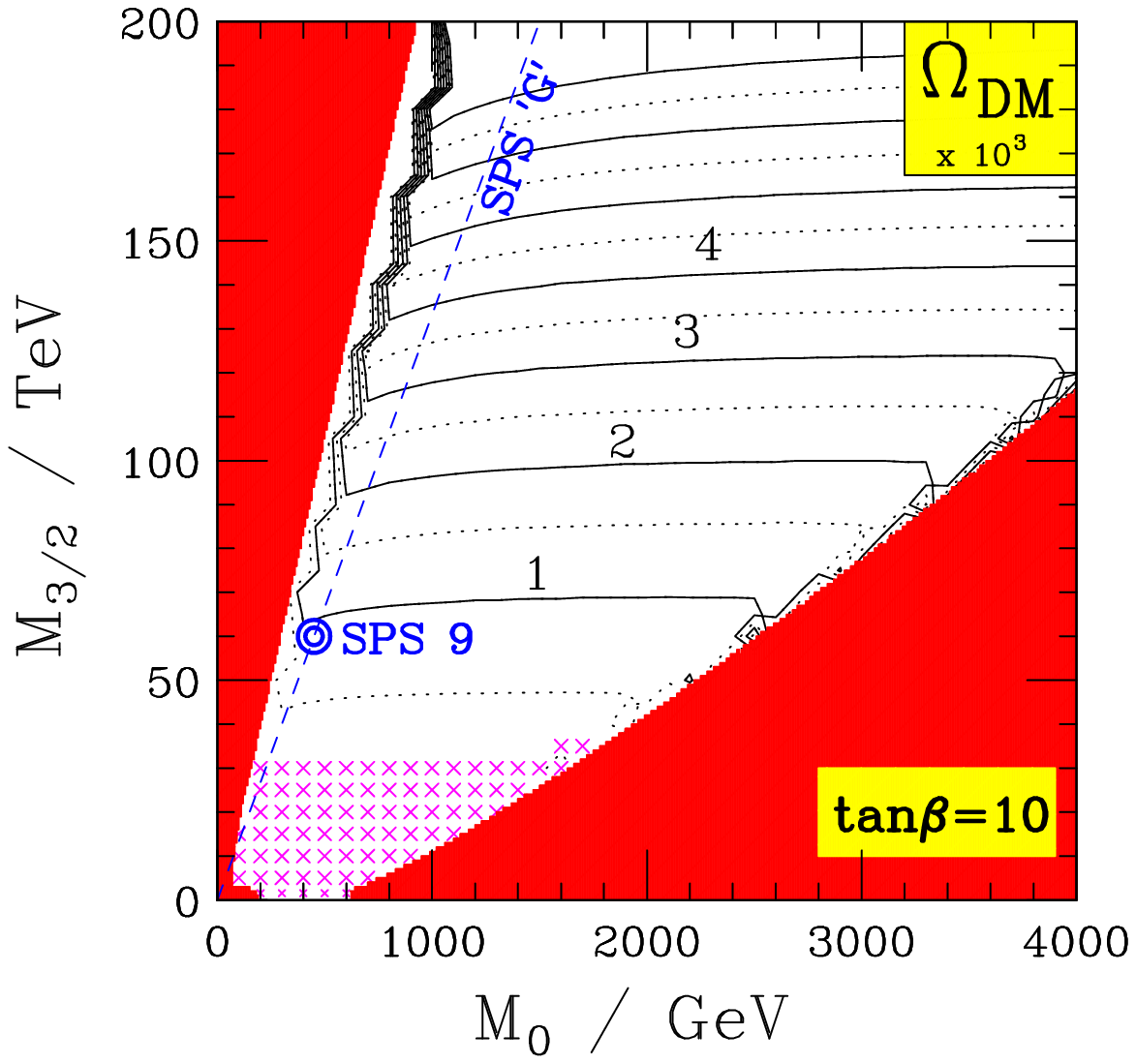, width=7cm, angle=90}\\
      \hspace*{0.5cm}{\bf (a)}
    \end{center}
  \end{minipage}\hfill
  \begin{minipage}[b]{.48\linewidth}
    \begin{center}
     \epsfig{file=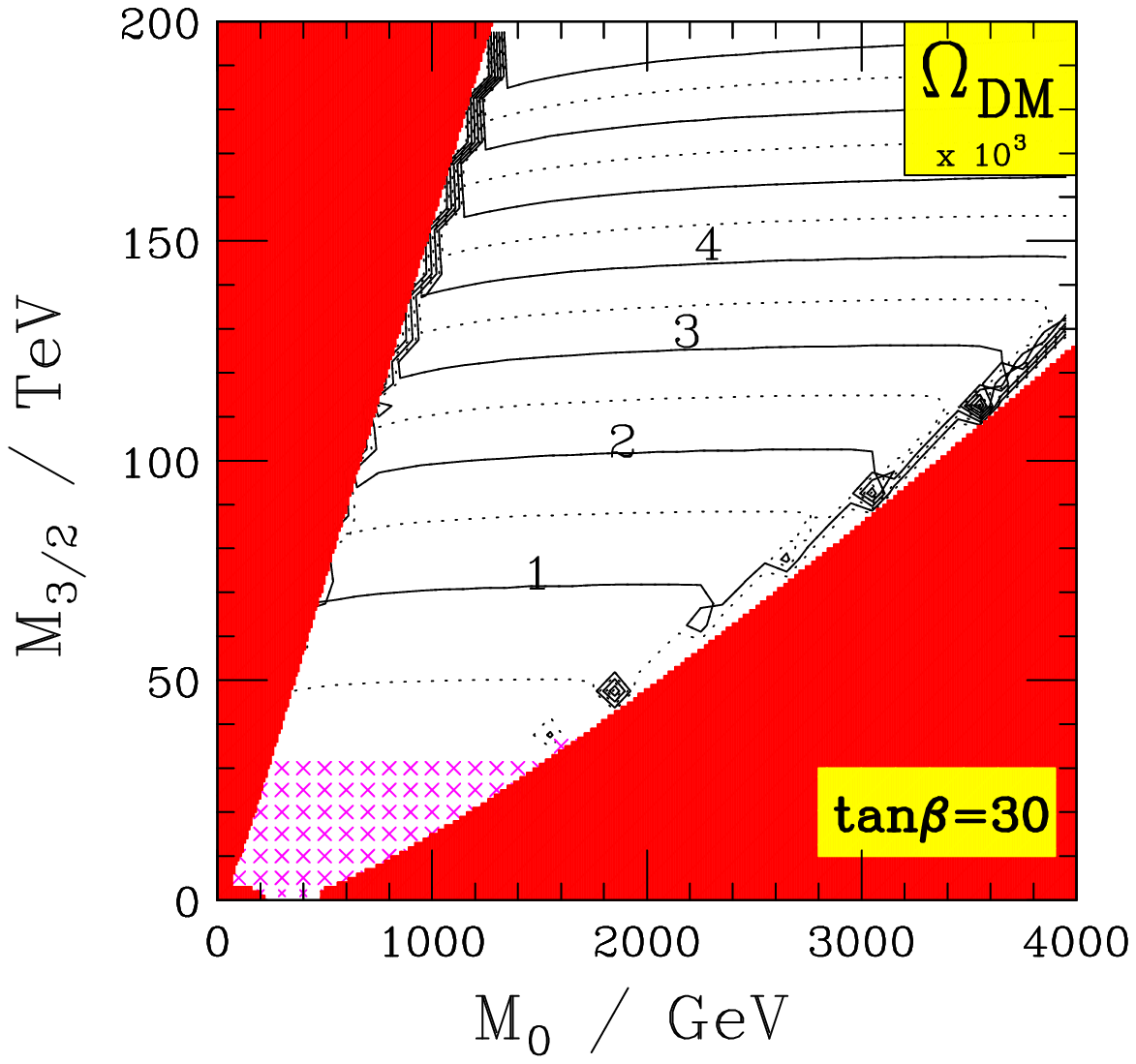,  width=7cm, angle=90}\\
       \hspace*{0.5cm}{\bf (b)}
    \end{center}
  \end{minipage}\hfill
\caption{
The relic density of the lightest neutralino, $\Omegadm h^2$
with $\mu>0$; 
{\bf (a)} for $\tan\beta=10$ and {\bf (b)} for $\tan\beta=30$.
The meaning of the symbols is explained in  \figref{AMSB:Lep}. 
The astrophysical limits on $\Omegadm h^2$ are discussed in the text.
The contours are in units of $10^{-3}$.
\label{AMSB:OMEGA}}}

\TABULAR[p]{|l|c||l|c|}{\hline
$\ntlone \ntlone \to W^+ W^-$ & 8 \% &  $\chgone \ntlone \to \mathrm{fermion\ pair}$ & 40 \% \\
$\chgone \ntlone \to W^+ + Z^0/h/A$ &11 \% & $\chgone \chgone \to W^+ W^+$ & 8 \% \\
$\chgone \chgonem \to \mathrm{fermion\ pair}$ &15 \% &
$\chgone \chgonem \to \mathrm{boson\ pair}$ &12 \% \\ \hline
}{
The main cosmological relic annihilation channels for the Snowmass point {\bf SPS~9}.\label{AMSB:OMG:CHANNELS}}

If R-Parity is conserved then the lightest supersymmetric particle can be a good candidate 
for the cold (non-relativistic) dark matter hypothesised by astrophysicists and cosmologists.
The cold dark matter contribution to critical density of 
the universe provided by the LSP is given by:
\begin{equation}
\omegadm\equiv\Omegadm h^2 = \frac{m_\ntlone n_\ntlone}{\rho_\mathrm{crit}/h^2}
\nonumber
\end{equation}
where $\rho_\mathrm{crit}=3H^2/8\pi G$ is the critical density, 
$n_\ntlone$ is the number density of LSPs,
and $H=h \times 100$~km~$\mathrm{s}^{-1}\ \mathrm{Mpc}^{-1}$ is
the Hubble constant.

Astronomical estimates from measurements of the acoustic power spectrum of the cosmic microwave background 
anisotropy\cite{Jaffe:2000tx,Pryke:2001yz,Pearson:2002,Rubino-Martin:2002rc,Lewis:2002} 
suggest that $\omegadm=0.106\pm0.010$.
Further constraints from the 2dF Galactic Redshift Survey\cite{Percival:2002} 
lead to $\omegadm=0.1151\pm0.0091$, assuming a flat universe $(\Sigma_i\Omega_\mathrm{i}=1)$.


In AMSB the Wino-like \ntlone\ and \chgone\ undergo rapid annihilation though reactions such as 
$\ntlone\ \ntlone\to W^+\ W^-$ and $\ntlone\ \chgone\to q\ q^\prime$.
The relic density of cold dark matter was calculated for AMSB 
using the program {\tt micrOmegas}\cite{Belanger:2001fz},
which includes the annihilation channels listed in \tabref{AMSB:OMG:CHANNELS}.
The results, in \figref{AMSB:OMEGA}, show that the value of 
$\omegadm$ is small throughout the region, and about a factor of two less 
than was estimated in \cite{Giudice:1998xp}:
\begin{equation}\omegadm^\mathrm{no-coanh}\approx 
5 \times 10^{-4}\times \left(\frac{m_\ntlone}{100\ \mathrm{GeV}}\right)\nonumber\ ,\end{equation}
which neglected the co-annihilation channels.

Thus in AMSB the LSP does not give any problems with relic overabundance, 
but suffers the opposite problem -- that sparticles in thermal equilibrium
in the early universe will not produce sufficient neutralino dark matter.
It has been suggested in  \cite{Moroi:1999zb} that decays of cosmological moduli --
which are low mass particles predicted by string theory when supersymmetry is broken --
could have produced Winos with sufficient abundance to be of astrophysical interest.

\subsection{Muon $g-2$}

\FIGURE[tp]{
  \begin{minipage}[b]{.48\linewidth}
    \begin{center}
     \epsfig{file=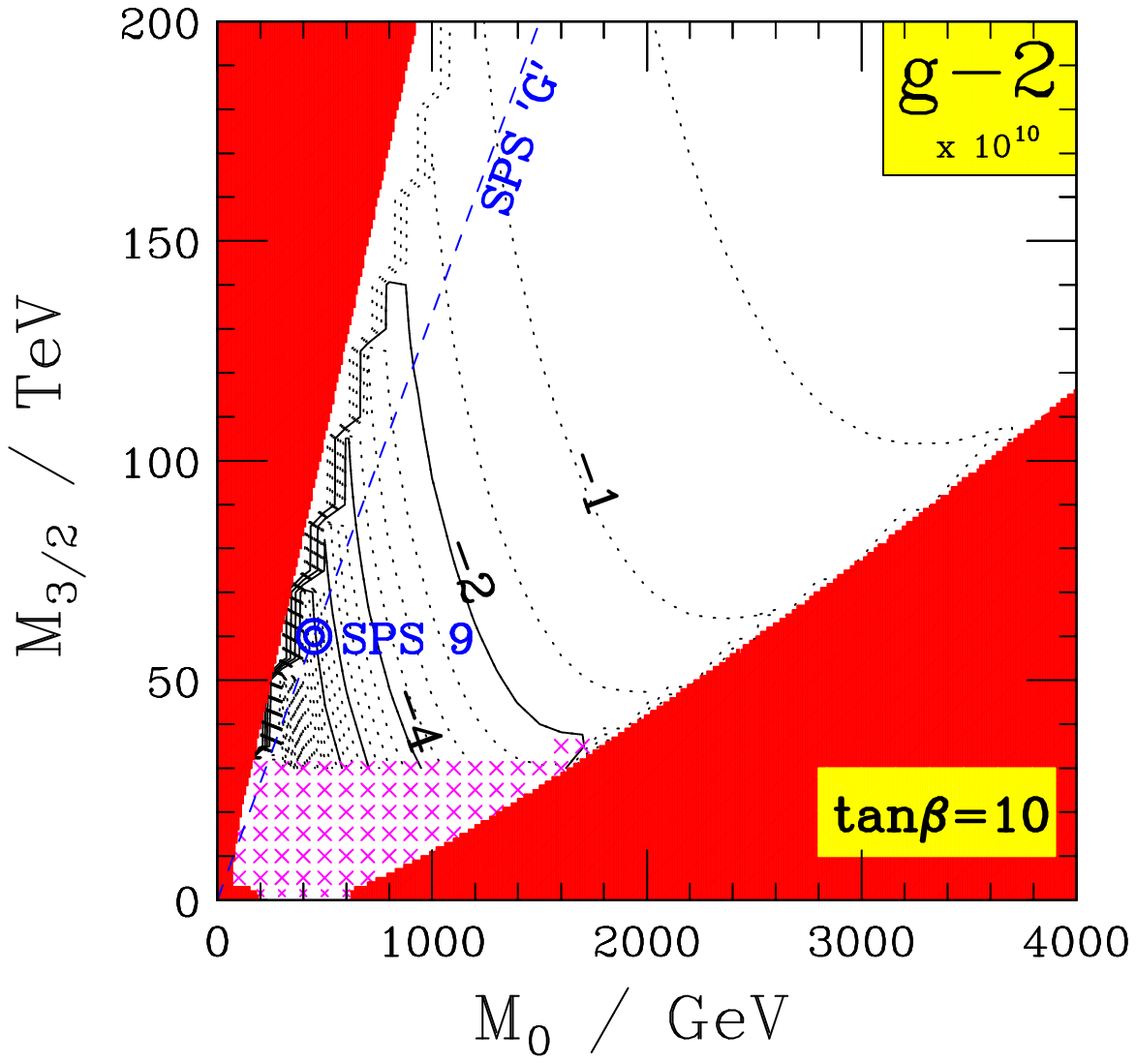, width=7cm, angle=90}\\
      \hspace*{0.5cm}{\bf (a)}
    \end{center}
  \end{minipage}\hfill
  \begin{minipage}[b]{.48\linewidth}
    \begin{center}
     \epsfig{file=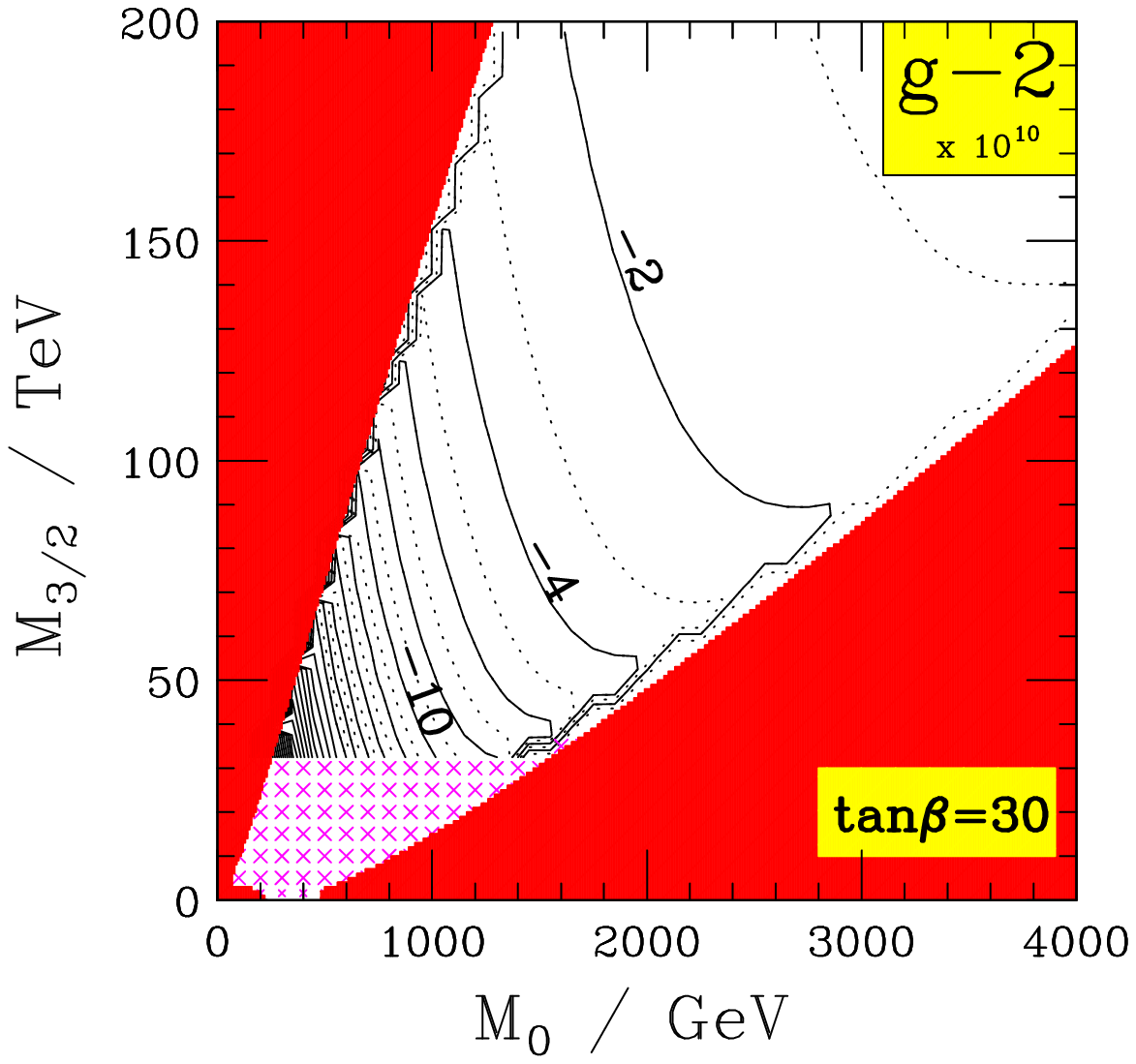,  width=7cm, angle=90}\\
       \hspace*{0.5cm}{\bf (b)}
    \end{center}
  \end{minipage}\hfill
\caption{
The SUSY contribution to the anomalous magnetic moment of the muon, (in units of $10^{-10}$)
with $\mu>0$; {\bf (a)} for $\tan\beta=10$ and {\bf (b)} for $\tan\beta=30$.
The meaning of the symbols is explained in  \figref{AMSB:Lep}. 
\label{AMSB:MUONG}}}

It is well know that the gyromagnetic moment of the muon, $g_\mu$ can sensitive to sparticle interactions
through loop corrections\cite{Chattopadhyay:1996ae,Martin:2001st,Baek:2002cc,Byrne:2002cw,Endo:2001ym}.
The recent BNL measurements\cite{Hertzog:2002ns,Bennett:2002jb} dominate the
world experimental average of the positive muon anomalous moment, 
\begin{equation}a_\mu^\mathrm{BNL} = (g_\mu-2)/2 = 11~659~203(8)\times 10^{-10}\ .\nonumber\end{equation}
Standard Model calculations of $a_\mu$ have been reviewed in \cite{Czarnecki:2001pv,Yndurain:2001qw}. 
Recent calculations \cite{Narison:2001jt,deTroconiz:2001jy},
which include a correction to the sign of the pion pole part of the hadronic light-by-light 
contribution\cite{Knecht:2001qf,Knecht:2001qg,Hayakawa:2001bb,Bijnens:2001cq,Blokland:2001pb}, 
were combined in \cite{Bennett:2002jb} to give:
\begin{equation}11~659~177(7)\times10^{-10}\quad \leq \quad a_\mu^\mathrm{SM}\quad
\leq\quad 11~659~186(8)\times10^{-10}\ .\nonumber\end{equation}

The SUSY contribution is typically dominated by loops involving charginos and 
neutralinos\cite{Lopez:1994vi,Czarnecki:2001pv,Chattopadhyay:2000ws}, 
and is proportional to $\tan\beta$ in the high $\tan\beta$ limit. 
The AMSB contributions to $a_\mu$ have been calculated in \cite{Feng:1999hg,Chattopadhyay:2000ws}.
In \figref{AMSB:MUONG} we plot the $a_\mu^\mathrm{AMSB}$ for $\tan\beta=10$ and 
30, where again we have made use of the program 
{\tt micrOmegas}.\footnote{Note that {\tt ISASUSY} produces AMSB points with negative $M_1$ and $M_2$
so the sign correlation between $\mu$ and $a_\mu^\mathrm{SUSY}$ is 
opposite to that in \cite{Chattopadhyay:2000ws}.}
The AMSB contribution is small compared to both the experimental and theoretical uncertainties, 
except at small \mnought,~\mthreehalfs.

\subsection{$B \to X_s \gamma$}

The inclusive radiative decay $B \to X_s \gamma$ is sensitive to sparticle properties
through radiative corrections involving charged higgs, chargino and $\tilde{t}$ loops 
(see for example \cite{Degrassi:2000qf,Ciuchini:1998xe,Gambino:2001ew,Feng:1999hg}).

\FIGURE[tpb]{
  \begin{minipage}[b]{.48\linewidth}
    \begin{center}
     \epsfig{file=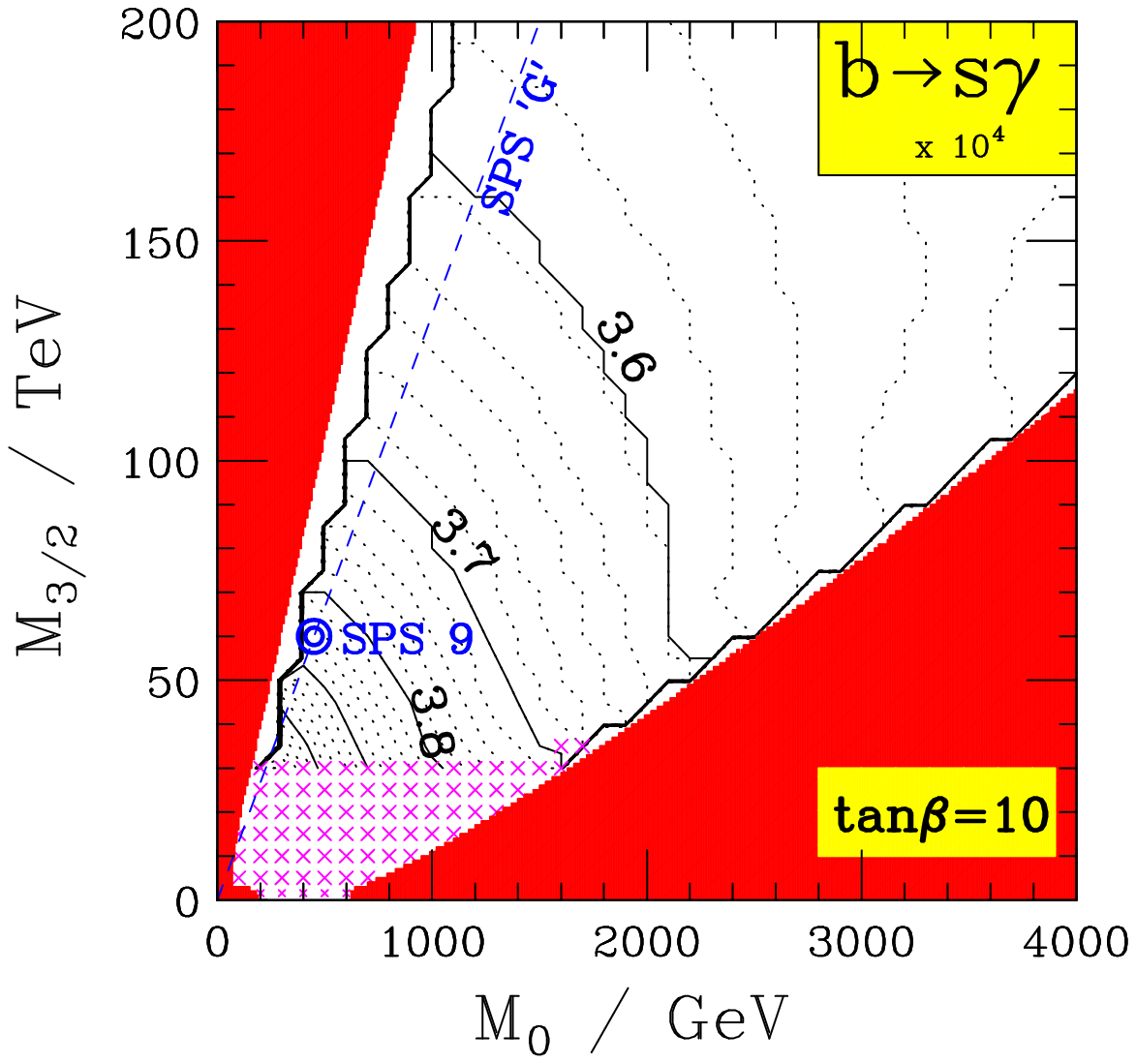, width=7cm, angle=90}\\
      \hspace*{0.5cm}{\bf (a)}
    \end{center}
  \end{minipage}\hfill
  \begin{minipage}[b]{.48\linewidth}
    \begin{center}
     \epsfig{file=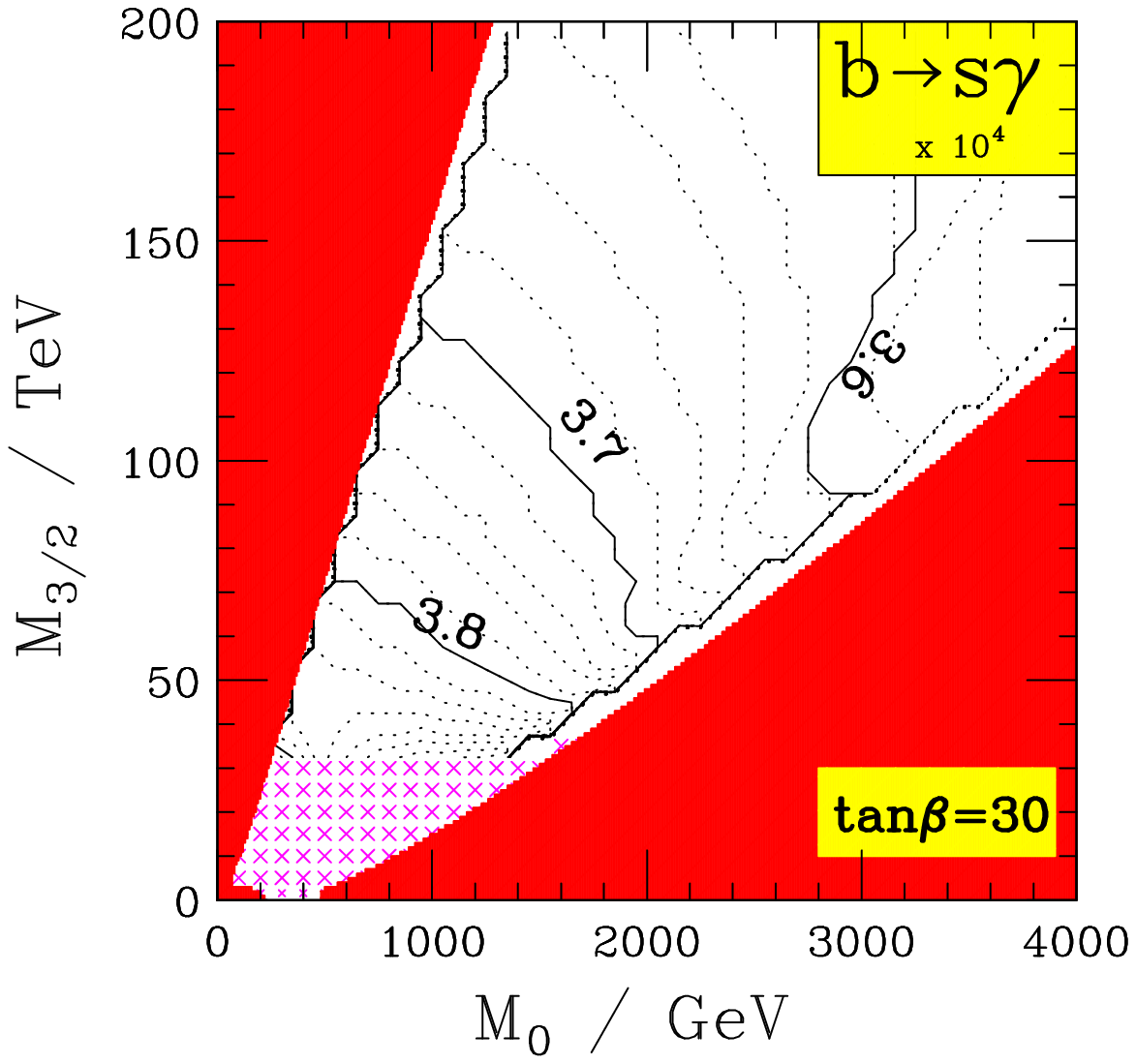,  width=7cm, angle=90}\\
       \hspace*{0.5cm}{\bf (b)}
    \end{center}
  \end{minipage}\hfill
\caption{
The branching ratio $\mathcal{B}(B\to\ X_s \gamma)$ with $\mu>0$; {\bf (a)} for $\tan\beta=10$ and 
{\bf (b)} for $\tan\beta=30$.
The contours are in units of $10^{-4}$.
The meaning of the symbols is explained in  \figref{AMSB:Lep}.
\label{AMSB:BSGAMMA}}}

Experimental measurements of the branching ratio from the CLEO\cite{Chen:2001fj}, 
BELLE\cite{Ushiroda:2001sb} and BaBar\cite{Aubert:2002pd} collaborations give: 
\begin{eqnarray}\mathcal{B}(B\to X_s \gamma)^\mathrm{CLEO} \quad &=& \quad
3.21\pm 0.43 \pm 0.27^{+0.18}_{-0.10}\times 10^{-4}\nonumber \\
\mathcal{B}(B\to X_s \gamma)^\mathrm{BELLE} \quad &=&\quad 
3.39\pm 0.53 \pm 0.42^{+0.51}_{-0.55}\times 10^{-4}\ ,\nonumber\\
\mathcal{B}(B\to X_s \gamma)^\mathrm{BaBar} \quad &=& \quad
3.88 \pm 0.36 \pm 0.37^{+0.43}_{-0.23} \times 10^{-4}\ ,\nonumber\\
\end{eqnarray}
where the errors are statistical, systematic, and from theory respectively.
Renormalisation scale uncertainties \cite{Gambino:2001au} lead to a 10\% ($\pm0.3\times10^{-4}$)
uncertainty in the SM prediction.
The mAMSB contributions were calculated using {\tt micrOmegas} and are plotted in \figref{AMSB:BSGAMMA}.
The combined experimental and theoretical uncertainty does not allow us to constrain 
the mAMSB parameter space at better than $2\sigma$ for either of $\tan\beta=10,30$.

The overall effect of the $B\to X_s \gamma$ and to a greater extent the $g_\mu-2$ constraint is to 
disfavour the low \mnought, \mthreehalfs\ region especially when $\tan\beta$ is large.

\section{Conclusions}
If anomaly mediated supersymmetry is present at the 1~to~2~TeV scale, the LHC
will observe excesses in various multi-lepton + \etmiss\ channels. 
We have used generic supersymmetry search procedures, and a realistic
detector simulation, to investigate the ability of the experiments at the 
LHC to discover AMSB scenarios.
By selecting events with very large missing energy and identifying 
tracks from chargino decays, the Wino-like nature of the LSP can be
determined, and $m_\chgone-m_\ntlone$ can be measured over a 
large range in parameter space. Careful study of the tracks left by \chgonepm s and 
their decay products can give clear evidence for this class of models even in cases 
where the chargino is shorter-lived than predicted in minimal AMSB.

\acknowledgments

The authors would like to thank Bryan Webber and Phil Stephens for providing constructive comments.
We are indebted to Carolina \"Odman and Phil Marshall of the Cavendish Astrophysics group
who provided many of the references in \secref{AMSB:OTHER}.
We have made use of the ATLAS physics analysis framework and tools which are
the result of collaboration-wide efforts. This work was partly funded by the UK
Particle Physics and Astronomy Research Council.

\appendix
\section{The variable \protect\mttwo\ and its generalisation}
\label{MTX}

The Cambridge \mttwo\ variable, proposed in \cite{Lester:1999tx}, 
can be used in analyses such as \cite{Allanach:2000kt},
where particles are pair-produced at hadronic colliders, and
decay semi-invisibly, as is the case in R-parity conserving SUSY.
This variable is used in symmetrical, two-body decays of supersymmetric particles, 
where the LSP is unobserved, and so must be inferred from missing energy.

\subsection{The properties of \protect\mttwo}

For events in which the decay of a heavy object produces an unseen particle, such as
\begin{equation} \chgone \to \ntlone \pi^+ \label{MTX:chgdecay} \end{equation}
one can write the Lorentz invariant
\begin{equation} m_\chgone^2 = m_\pi^2 + m_\ntlone^2 + 
2 \Bigl[  E_T^\pi E_T^\ntlone \cosh(\Delta\eta)-{\bf p}_T^\pi\cdot{\bf p}_T^\ntlone \Bigr]
\label{MT2:MCHIDEF}\end{equation}
where ${\bf p}_T^\pi$ and ${\bf p}_T^\ntlone$ indicate pion and neutralino 2-vectors in the transverse plane, and
the transverse energies are defined by
\begin{equation}E_T^{\pi} = { \sqrt { ({\bf p}_T^{\pi}) {^2} + m_\pi^2 }}
\qquad \hbox{ and } \qquad {E_T^\ntlone} = { \sqrt{ ({\bf p}_T^\ntlone){^2} + m_\ntlone^2 } }\ .
\label{MT2:ETDEF}\end{equation}
Also \begin{equation}\eta=\half \log\biggl[\frac{E+p_z}{E-p_z}\biggr]\end{equation} is the true rapidity, so that
\begin{equation}\tanh\eta=p_z/E\ ,\qquad \sinh\eta=p_z/E_T\ ,\qquad \cosh\eta=E/E_T .\end{equation}

In a hadron collider, only the transverse components of a missing particle's momentum 
can be inferred, so it is useful to define the transverse mass,
\begin{equation}
m_T^2 ( {\bf p}^{\pi}_T, {\bf p}_T^\ntlone) \equiv { m_{\pi^+}^2 + m_\ntlone^2 +
2 ( E_T^{\pi} E_T^\ntlone - {\bf p}_T^{\pi}\cdot{\bf p}_T^\ntlone ) }
\label{MT2:MTDEF}\end{equation}
which, because $\cosh(x)\geq 1$, is less than or equal to the mass of the lightest chargino, with equality
only when the rapidity difference between the neutralino and the pion, $\Delta\eta_{\ntlone\pi}$ is zero.
All other $\Delta\eta$ lead to $m_T<m_\chgone$, so if we knew the neutralino 
momentum we could use $m_T$ to give an event by event lower bound on the lightest chargino mass.
$m_T$ was used in this way by UA1 \cite{Arnison:1983rp} in the measurement of the $W^\pm$ mass.

In R-parity conserving SUSY events there are expected to be two unseen 
LSPs.\footnote{Though there may also be other unseen particles -- see \appref{MT2GEN}.}
Since only the {\em sum} of the missing transverse momentum of the two neutralinos is known,
the variable
\begin{eqnarray}
m_{T2}^2 &\equiv& { \min_{\slashchar{{\bf q}}_T^{(1)} + \slashchar{{\bf q}}_T^{(2)} = {\bf \pmiss}_T }}
{\Bigl[ \max{ \{ m_T^2({\bf p}_T^{\pi^{(1)}}, \slashchar{{\bf q}}_T^{(1)}) ,\ 
 m_T^2({\bf p}_T^{\pi^{(2)}} , \slashchar{{\bf q}}_T^{(2)}) \} } \Bigr]}
\qquad \label{MT2:MT2DEF}
\end{eqnarray}
is a lower bound on the transverse mass $m_T$
for events where two decays of the type \eqref{MTX:chgdecay} occur.
In \eqref{MT2:MT2DEF} we have been forced
to minimise over all consistent neutralino 2-momenta.
Note that $\slashchar{{\bf q}}_T^{(i)}$ is the hypothesised momentum
of the $i$th neutralino which need not be equal to its true momentum.

\EPSFIGURE[tbh]{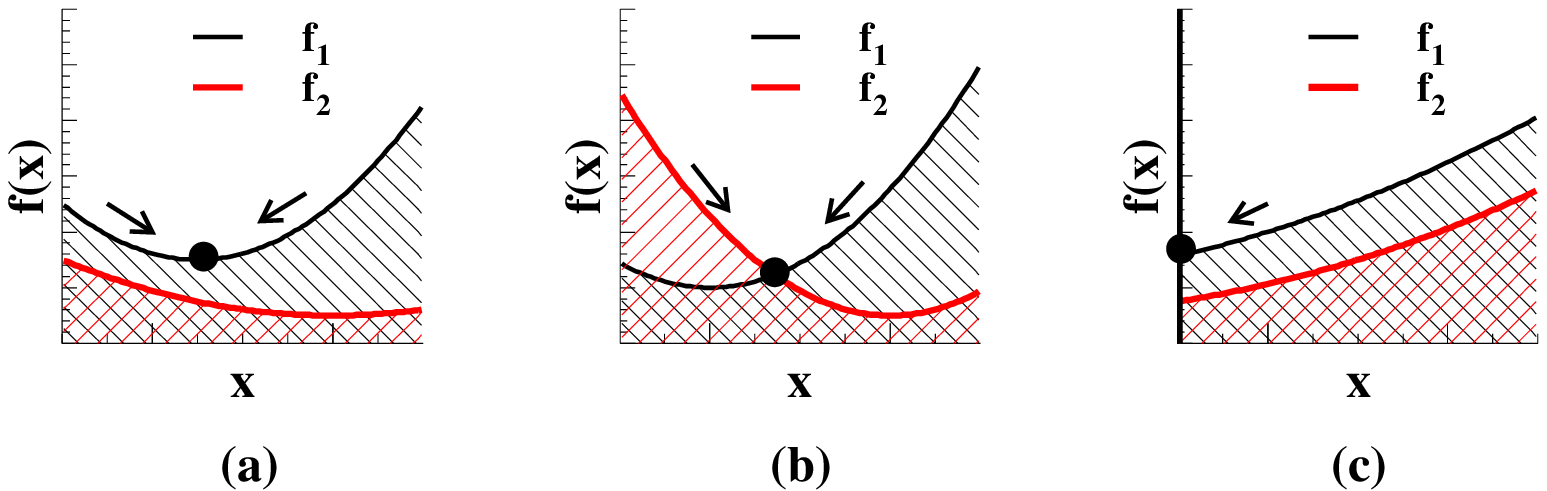, width=15cm}{
A diagram demonstrating that the minimisation over some parameter of the maximum of two well-behaved functions
may occur either at {\bf (a)} a minimum value of one of them, or {\bf (b)} when they are equal,
or {\bf (c)} at the boundary of the domain. 
\label{MTX:MINIMISING}
}

To find the range of values \mttwo\ may take we first
let $f_1=m_T^2({\bf p}_T^{\pi^{(1)}},~\slashchar{{\bf q}}_T^{(1)})$, 
and $f_2=m_T^2({\bf p}_T^{\pi^{(2)}},~\slashchar{{\bf q}}_T^{(2)})$.
We then note that the minimum over a parameter $x$ 
of the maximum of $f_1(x)$ and $f_2(x)$ can occur at a local minimum, $f_{1(2)}^\prime(x^*)=0$, 
provided $f_{1(2)}(x^*)>f_{2(1)}(x^*)$, as shown in \figref{MTX:MINIMISING}a. 
Alternatively the minimum can occur when the functions cross one another
when $f_1=f_2$ (\figref{MTX:MINIMISING}b) or at a boundary (\figref{MTX:MINIMISING}c).
The parameter $x$ corresponds to the fraction of the the missing momentum 
(in one of the transverse directions) which is assigned to each half of the event.
Since $f_1, f_2\to\infty$ as $x\to\pm\infty$ \figref{MTX:MINIMISING}c is not relevant to our minimisation problem. 

To see that case (a) cannot occur, consider the unconstrained minimisation 
over $\slashchar{{\bf q}}_T$, of $m_T^2 ( {\bf p}^{\pi}_T,\ \slashchar{{\bf q}}_T)$.
Using the relationship
\begin{equation}\frac{\partial \slashchar{E}_T}{\partial \slashchar{q}_k}=\frac{\slashchar{q}_k}{\slashchar{E}_T}\ ,
\end{equation}
where $\slashchar{E}_T^2=\slashchar{{\bf q}}_T^2+m_\ntlone^2$, it is straightforward to show that,
\begin{equation}\frac{\partial m_T^2}{\partial \slashchar{q}_k}=2\left(E_T^\pi\frac{\slashchar{q}_k}
{\slashchar{E}_T}-p_k^\pi\right) \qquad k=1,2\ .\end{equation}
This means that at the minimum
\begin{equation} {\bf v}_T^\pi = \slashchar{{\bf u}}_T\ , \label{MTX:GLOBMIN}\end{equation}
where we introduce the notation ${\bf v}_T \equiv {\bf p}_T / E_T$, $\slashchar{{\bf u}}_T \equiv \slashchar{{\bf q}}_T / \slashchar{E}_T$.
where ${\bf p}_T$ and ${\bf v}_T$ represent the true transverse momentum and velocity of a particle,
while $\slashchar{{\bf q}}_T$ and $\slashchar{{\bf u}}_T$ are assigned by the minimisation.

Using the basis ($t$,~$x$,~$y$) with the metric diag(1,-1,-1), one can write
\begin{equation} m_T^2 = (E_T^\mathrm{tot},{\bf p}_T^\mathrm{tot})\cdot(E_T^\mathrm{tot},{\bf p}_T^\mathrm{tot})\ ,
\label{MTX:MTLOR2} \end{equation} 
where $E_T^\mathrm{tot}=E_T^\pi+\slashchar{E}_T$ and 
${\bf p}_T^\mathrm{tot}={\bf p}_T^\pi+\slashchar{{\bf q}}_T$. 
This 1+2 dimensional Lorentz invariant can be evaluated in any frame boosted from the lab 
in the transverse plane. 
\eqref{MTX:GLOBMIN} has told us that at the unconstrained minimum the
transverse velocities ${\bf v}_T^\pi$ and $\slashchar{{\bf u}}_T$ are equal;
a statement necessarily true in all transverse frames, including the special one
in which both the transverse velocities (and associated momenta) are zero.
Evaluating \eqref{MTX:MTLOR2} in this frame, we find that the unconstrained minimum
of \eqref{MTX:MTLOR2} then becomes $(m_\pi+m_\ntlone,~0,~0)\cdot(m_\pi+m_\ntlone,~0,~0)$, 
and we recover the expected result \begin{equation} m_T^\amin = m_\pi+m_\ntlone\ .\label{MTX:MTMIN} \end{equation}
We therefore conclude that the function $m_T^2$ has only one stationary value and it is the global minimum, 
and is common to both sides of the event provided the same type of particles are emitted.
Thus when $f_1$ is minimum it cannot be greater than $f_2$, 
and so the minimisation in \eqref{MT2:MT2DEF} forces $f_1=f_2$.
This could of course occur when {\em both} $f_1$ and $f_2$ are at their global minima, 
in which case \mttwo\ takes its minimum value:
\begin{equation}\mttwo^\amin=m_\pi+m_\ntlone\label{MT2:MT2MIN}\ .\end{equation}

To summarise, \mttwo\ is the minimum of $m_T^{(1)}$ subject to the two constraints $m_T^{(1)}=m_T^{(2)}$, and 
${\bf \pmiss}_T ^{(1)} + {\bf \pmiss}_T ^{(2)} = {\bf \pmiss}_T$.
The condition for the minimisation can be calculated by Lagrange multiplier methods, the result of which is that 
the velocity vectors ${\bf \slashchar{u}}_T^{(1,2)}$ 
of the {\em assigned} neutralino momenta $\slashchar{{\bf q}}_T^{(1,2)}$ must satisfy
\begin{equation}
({\bf \slashchar{u}}_T^{(1)}-{\bf v}_T^{\pi^{(1)}})\ \propto\ ({\bf \slashchar{u}}_T^{(2)}-{\bf v}_T^{\pi^{(2)}})\ .
\label{MTX:MTTWOCOND} \end{equation}

To find the maximum of \mttwo\ over many events
we note that for each event the minimisation will select hypothesised momenta satisfying \eqref{MTX:MTTWOCOND}.
We now note events can occur in which the {\em true} transverse velocities of the neutralinos 
were exactly those which were assigned by the minimisation, \ie\ they can satisfy 
\begin{equation}
{\bf v}_T^{\ntlone(1)} = {\bf \slashchar{u}}_T^{(1)}, \qquad {\bf v}_T^{\ntlone(2)} = {\bf \slashchar{u}}_T^{(2)}\ .
\label{MT2:TRUEVEL}
\end{equation}
These events will have both hypothesised transverse masses equal not only to each other but also to 
true transverse masses which would have been calculated if the neutralino momenta had been known:
\begin{equation}
m_T^{(i)} \left( {\bf p}^{\pi^{(i)}}_T, \slashchar{{\bf p}}_T^{\ntlone(i)} \right)=
m_T^{(i)} \left( {\bf p}^{\pi^{(i)}}_T, \slashchar{{\bf q}}_T^{(i)}\right) 
\end{equation}
If events occur where, in addition to the transverse components of the neutralino momenta satisfying \eqref{MT2:TRUEVEL}, 
the rapidity differences satisfy $\eta_{\ntlone(1)}=\eta_{\pi(1)}$ and  $\eta_{\ntlone(2)}=\eta_{\pi(2)}$,
then by \eqref{MT2:MCHIDEF} \mttwo\ will equal the true mass of the chargino.
Combining this with \eqref{MT2:MT2MIN} and recalling that
\mttwo\ cannot be greater than the chargino mass by construction,
we can see that the event-by event distribution of \mttwo\ can span the range
\begin{equation} m_\ntlone+m_\pi\ \leq\ \mttwo\ \leq\ m_\chgone \,\end{equation}
showing that \mttwo\ is sensitive to the $m_\chgone-m_\ntlone\equiv\DeltaMChi$ mass difference.

The variable is equally applicable to two same-sign $\chgone$ decays so \mttwo\ signal events can be 
defined as those having two $\cht_1^\pm \to \ntlone\ \pi^{\pm}$ decays with any combination of charges.

\subsection{Generalisations of \protect\mttwo}
\label{MT2GEN}

In our analysis we also wish to make use of the leptonic decays $\chgone\to\ntlone~\ell~\nu_\ell$ where $\ell\in e,\mu$.
We therefore generalise \mttwo\ to cases where more than two particles go undetected.

\FIGURE[t]{
  \begin{minipage}[b]{.41\linewidth}
    \begin{center}
     \epsfig{file=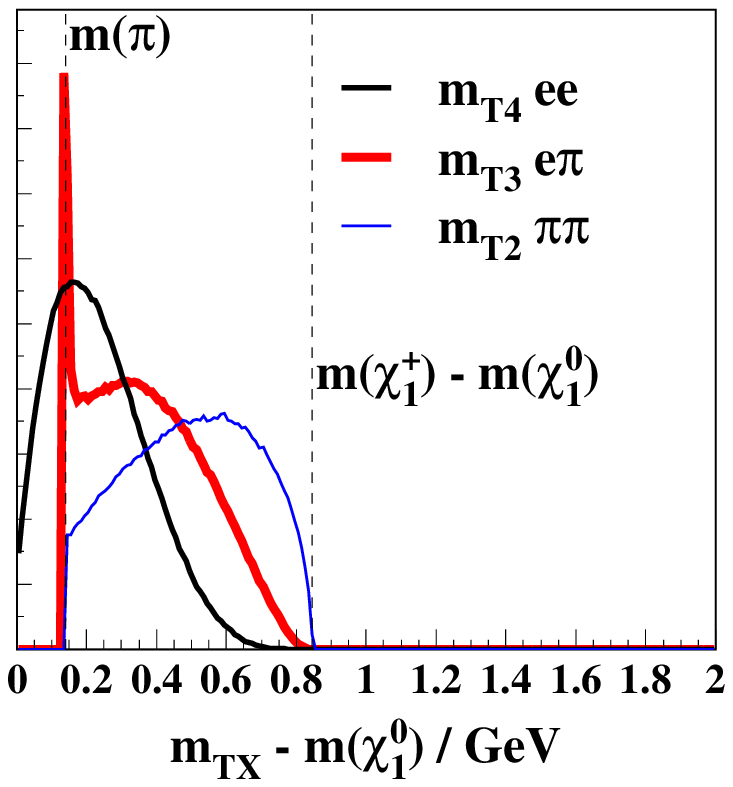, height=7cm}\\
      \hspace*{0.5cm}{\bf (a)}
    \end{center}
  \end{minipage}\hfill
  \begin{minipage}[b]{.57\linewidth}
    \begin{center}
     \epsfig{file=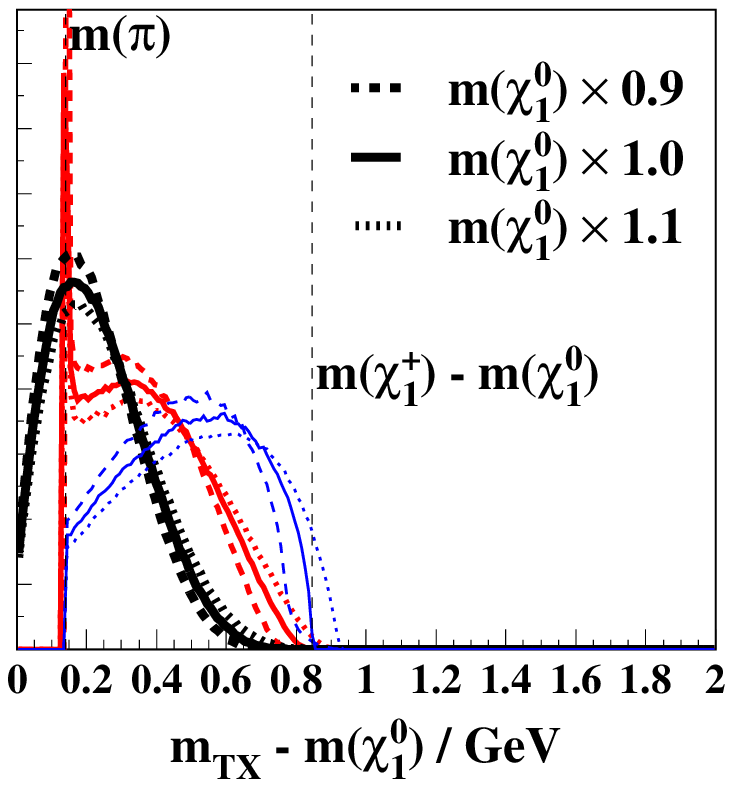,  height=7cm}\\
       \hspace*{0.5cm}{\bf (b)}
    \end{center}
  \end{minipage}\hfill
\caption{
{\bf (a)} Simulations of $\mtx-m_\ntlone$ for $X=2,3,4$ using a simple
phase-space Monte-Carlo generator program for a pair of decays 
$\squark\to\chgone q$ followed by $\chgone\to\ntlone~\pi$ or $\chgone\to\ntlone~e~\nu_e$.
As the number of invisible particles increases the proportion of events near the upper
limit decreases. The peak in $m_{T3}-m_\ntlone$ near the pion mass is explained in the text.
{\bf (b)} The distortion of $\mtx-m_\ntlone$ when the LSP mass is varied by $\pm$~10\%,
showing that $M_{TX}-m_\ntlone$ remains sensitive to the mass difference 
$\DeltaMChi=m_\chgone-m_\ntlone$.
In this simulation $\DeltaMChi=0.845$~GeV, $m_\ntlone=161.6$~GeV, and the electron and neutrino mass
were neglected.
The normalisation is arbitrary.
\label{MTX:MTXPLOTS}
}}

Consider events in which a chargino is produced and then decays to $\ntlone\ e\ \nu_e$.
If we expand the Lorentz invariant\begin{equation}
(m_\chgone)^2  = (p_\ntlone + p_e + p_\nu)^2
\end{equation}
we obtain three mass-squared terms for each of the decay particles and three cross-terms. 
The cross-terms can each be written in the form
\begin{equation}
p_a \cdot p_b = E_T^{(a)}E_T^{(b)}\cosh(\Delta\eta_{ab})-{\bf p}_T^{(a)}\cdot{\bf p}_T^{(b)}\ .
\end{equation}
If the neutralino and neutrino transverse momenta were individually known
we could evaluate the transverse mass,
\begin{equation} m_T^2 = m_\ntlone^2 + m_e^2 +
2\ \Bigl[ (E_T^e E_T^\chi - {\bf p}_T^e \cdot {\bf p}_T^\chi ) + 
(E_T^\nu E_T^\chi - {\bf p}_T^\nu \cdot {\bf p}_T^\chi ) + 
(E_T^e E_T^\nu - {\bf p}_T^e \cdot {\bf p}_T^\nu ) \Bigr]\ ,\label{MTX:MT3PART}\end{equation}
where the neutrino mass is assumed to be negligible.
$m_T$ will be equal to the \chgone\ mass in events where $\Delta\eta_{ab}=0$ for all pairs of 
$e$, $\nu_e$, and \ntlone.

In events with two leptonic chargino decays a variable like \mttwo\ can be defined
as in \eqref{MT2:MT2DEF} but using the three-particle definition of $m_T$ from \eqref{MTX:MT3PART}
and with the modified constraint,
\begin{equation}
{\bf q}_T^{\nu(1)} + {\bf q}_T^{\chi(1)} +{\bf q}_T^{\nu(2)} + {\bf q}_T^{\chi(2)} = \Ptmiss\ ,
\end{equation}
where the labels (1) and (2) indicate which chargino the particles were emitted from.
We call this variable $m_{T4}$ (or indeed \mtx\ where $X$ is the number of undetected particles).

The conditions for the minimisation required to calculate $m_{T4}$ can be calculated
just as for \mttwo.
The Euler-Lagrange equations involving 
\begin{equation}\frac{\partial (m_T^{(i)})^2}{\partial {\bf q}_T^{\nu(i)} }\qquad \mathrm{and} \qquad
\frac{\partial (m_T^{(i)})^2}{\partial {\bf q}_T^{\ntlone(i)} }\nonumber\end{equation}
show that the minimisation will select the invisible particles' momenta such that 
${\bf u}_T^{\ntlone(i)}={\bf u}_T^{\nu(i)}$.
The other E-L equations reproduce \eqref{MTX:MTTWOCOND} but with electrons replacing pions.

This means that when calculating $m_{T4}$ one can replace the missing particles
from each chargino decay with a pseudo-particle with mass equal to the sum of the 
masses of those invisible particles and proceed as for \mttwo.
In the case of leptonic chargino decay the mass of the neutrino can be safely neglected
in comparison to that of the \ntlone, and the constraint ${\bf u}_T^{\chi(i)}={\bf u}_T^{\nu(i)}$
is equivalent to ${\bf q}^{\nu(i)}_T=(0,0)$.

The generalisation to \mtx\ for other values of $X$ is straightforward.
The distribution of $m_{T3}$ is shown in \figref{MTX:MTXPLOTS}a for events in which 
one chargino decays to $\ntlone,e,\nu$ and another to $\ntlone,{\pi^+}$.
Unlike $m_{T2}$ and $m_{T4}$ it has a sharp peak at $m_{T3}=m_\ntlone+m_\pi$. 
This occurs because the visible particles on each side of the event are different
and so the unconstrained minimum of the values of $m_T$ on each side of the event
are not equal as they are in the case of \mttwo\ and $m_{T4}$:
\begin{equation}
\min_{\slashchar{\bf q}_T^{(1)}} \left(m_T^{(1)}( {\bf p}^{\pi}_T, \slashchar{{\bf q}}_T^{(1)})\right) 
\ =\ m_\pi + m_\ntlone\quad \ne\quad m_e + m_\ntlone\ =\ 
\min_{\slashchar{\bf q}_T^{(2)}} \left(m_T^{(2)}( {\bf p}^{e}_T, \slashchar{{\bf q}}_T^{(2)})\right)
\end{equation}
Some of the events can then fall into the category shown in \figref{MTX:MINIMISING}a, 
producing a peak of events with $m_{T}=m_\ntlone+m_\pi$.

The distribution over events of $m_{T4}$ will have fewer entries near the upper kinematic limit
($m_{T4}=m_\chgone$) because when more particles go undetected an event at that limit must satisfy 
a larger number of constraints.
For fully leptonic chargino decay, there are six constraints of the type $\Delta\eta=0$,
two ${\bf p}_T^{\nu(i)}=0$ and finally the modified constraint from \eqref{MTX:MTTWOCOND}.
This effect can be seen in \figref{MTX:MTXPLOTS}a for events where a total of two, 
three and four invisible particles are produced.

A further generalisation which we do not require here might be relevant in cases where 
more than one {\em visible} particle is emitted from each mother. 
For such decays, one would sum the full 4-momenta of the visible particles
from each decay as well as summing the masses of the invisible particles 
from each side of the event and proceed as for \mttwo.

\subsection{Uncertainties in \Ptmiss\ and $m_{\protect\ntlone}$}

The sensitivity of \mtx\ to the estimated mass of the neutralino is shown in \figref{MTX:MTXPLOTS}b,
where 10\% (16~GeV) errors in $\chi$ result in similar {\em fractional} errors in \DeltaMChi\ \ie\ 
of a few tens of MeV. \mtx\ shows similar insensitivity to measurement uncertainties in the missing
transverse momentum vector.
This behaviour can be (at least partially) understood from the non-relativistic limit of \mttwo, 
when the proportionality in \eqref{MTX:MTTWOCOND} becomes an equality and
\begin{equation} \mttwo^2-(m_\pi+m_\ntlone)^2 = \frac{1}{4 m_\pi m_\ntlone}
\left(m_\pi \Ptmiss - m_\ntlone {\bf p}_T^{\pi_1} - m_\ntlone {\bf p}_T^{\pi_2}\right)^2 
+ \mathcal{O}\left(({\bf v}_T\cdot {\bf v}_T)^2\right)\ . \label{AMSB:NONREL}\end{equation}
One can see that in \eqref{AMSB:NONREL} $m_\pi$ multiplies the missing momentum, 
while $m_\ntlone$ multiplies the pion transverse momenta.
\section{Simulation of particle tracks}
\label{EIDENT}

Part of the \atlfast\ \cite{Richter:1998at} software provides fast parameterised
simulation of the \atlas\ inner detector performance.
In particular it calculates the expected track reconstruction efficiency,
and smears the five track helix parameters 
and according to particle type.
The parameterisation used was based on {\tt GEANT3} Monte-Carlo simulations\cite{buis}
using a Kalman-filter algorithm for track reconstruction\cite{xkalman}.
A parameterisation from a large statistics Monte-Carlo sample\cite{nairz}, 
was used for the hadronic track smearing for this study.

\subsection{Low \pt\ electron identification}
In \atlas\ the transition radiation tracker allows the identification 
of low-energy electrons because they emit more transition radiation than 
more massive particles with the same momentum.
The issue of particle misidentification is not dealt with in the standard 
version of \atlfast, and so has been implemented independently in our analysis.

\TABULAR[bth]{|l|c|c|c|c|c|c|}{\hline
$\pt$		& 0.5	& 1	&   2	&  5	& 10	& $\geq 20$ \\
\hline
$\epsilon_{\pi}$& 0.01	& 0.003	& 0.005	& 0.007	& 0.008	& 0.012 \\ \hline
}{
Pion misidentification efficiency at $|\eta|=0.3$ as a function of transverse momentum
for an electron identification efficiency of 90\%.
Interpolations were made linearly in \pt\ and logarithmically in $\epsilon_{\pi}$.
Particles with $\pt<0.5$~GeV are assumed not to be reconstructed into tracks. 
\label{EIDENT:EPI}
}

\TABULAR[bth]{|cc|c|c|c|c|c|c|c|c|c|}{\hline
 	&$|\eta|$	& $\leq 0.2$	& 0.3	& 0.65	& 0.8	& 1.15	& 1.4	& 1.9	& 2.15	& $\geq 2.4$	\\
\hline
\pt 	& 2~GeV	& 1.2	&  	1.0 	& 0.88	& 1. 	& 0.6	& 0.4	& 0.028	& 0.88	& 2	\\
\hline
\pt	& 20~GeV& 1.1	& 	1.0	& 1.	& 2.0	& 1.4	& 0.78	& 0.22	& 1.4	& 2.3 \\ \hline
}{
$\eta$ dependent correction factor applied to $\epsilon_\pi$,
the hadron misidentification efficiency, for two different values of transverse momentum.
Interpolations were made logarithmically in $\epsilon_\pi$ and linearly in \pt\ and $\eta$.
Tracks beyond $|\eta|=2.5$ are not reconstructed.
\label{EIDENT:CORRETA}
}
Test beam performance was compared to Monte-Carlo simulations in \cite{trttot}
using a combined discriminator which considered both the number of high
threshold clusters and the time for which the charge deposited exceeded a lower threshold.
In that paper the pion misidentification probability was calculated as a function of transverse
momentum at $\eta=0.3$ for an electron efficiency of 90\% (\tabref{EIDENT:EPI}).
For our simulation these efficiencies were extrapolated to other values of $\eta$
by comparison with the {\tt GEANT3} simulations performed in \cite{phystdr}.
The \pt\ and $\eta$ dependent correction factors are shown in \tabref{EIDENT:CORRETA}.

\subsection{Low \pt\ muon identification}

High energy muons are easily identified because of their high penetration.
Identification efficiency is drastically reduced when muons have an insufficient 
transverse momentum to extend their track helix into the dedicated muon detectors.
The applied  identification efficiencies for muons were based on \cite{phystdr}
and are shown in \tabref{EIDENT:MU}. 

\TABULAR[bth]{|l|c|c|c|c|c|c|c|}{\hline
\pt 	& $\leq 2$	& 3	&  4	&  5	&  6	&  8	&  $\geq 10$	\\
\hline
Efficiency & ~0~& 0.34	& 0.66	& 0.89	& 0.95 	& 0.96	& 0.98  \\\hline
}{Applied muon identification efficiency as a function of transverse momentum.
Efficiencies were interpolated linearly in $\epsilon_\mu$ and \pt.
\label{EIDENT:MU}
}

\bibliography{thesis}

\end{document}